\documentclass{siamltex}

\usepackage{amsfonts} 
\usepackage{amsmath} 
\usepackage{xspace} 
 
\usepackage{cite} 
 
\newif\ifPDF 
\PDFfalse 
 
\ifPDF 
  \usepackage[T1]{fontenc} 
  \usepackage{aeguill} 
  \pdfpagewidth=\paperwidth 
  \pdfpageheight=\paperheight 
  \usepackage[pdftex]{color,graphicx} 
\else 
  \usepackage{color,graphicx} 
\fi 
 
\newcommand{\Rset}{\mathbb{R}}

\def\kdiff{k_{\rm diff}} 
 
\def\bfk{\mbox{\mathversion{bold}${k}$}} 
\def\bfkone{\mbox{\mathversion{bold}${k_1}$}} 
\def\bfktwo{\mbox{\mathversion{bold}${k_2}$}} 
\def\bfkthr{\mbox{\mathversion{bold}${k_3}$}} 
\def\bfkfou{\mbox{\mathversion{bold}${k_4}$}} 
\def\bfknin{\mbox{\mathversion{bold}${k_9}$}} 
\def\bfkele{\mbox{\mathversion{bold}${k_{11}}$}} 
\def\bfktwe{\mbox{\mathversion{bold}${k_{12}}$}} 
 
\def\bfm{\mbox{\mathversion{bold}${m}$}} 
 
\def\bfkm{{\mbox{\mathversion{bold}${k}$}_{\mbox{\mathversion{bold}$\scriptstyle{m}$}}}} 
 
\def\bfx{\mbox{\mathversion{bold}${x}$}} 
 
\def\bfsk{\mbox{\mathversion{bold}$\scriptstyle{k}$}} 
\def\bfskone{\mbox{\mathversion{bold}$\scriptstyle{k_1}$}} 
\def\bfsktwo{\mbox{\mathversion{bold}$\scriptstyle{k_2}$}} 
\def\bfskthr{\mbox{\mathversion{bold}$\scriptstyle{k_3}$}} 
 
\def\bfsx{\mbox{\mathversion{bold}$\scriptstyle{x}$}} 
 
\def\eikdotx{e^{i\bfsk\cdot\bfsx}} 
\def\eikonedotx{e^{i\bfskone\cdot\bfsx}} 
\def\eiktwodotx{e^{i\bfsktwo\cdot\bfsx}} 
\def\eikthrdotx{e^{i\bfskthr\cdot\bfsx}} 
 
\def\etwoikonedotx{e^{2i\bfskone\cdot\bfsx}} 
\def\etwoiktwodotx{e^{2i\bfsktwo\cdot\bfsx}} 
\def\etwoikthrdotx{e^{2i\bfskthr\cdot\bfsx}} 
 
\def\eikonepktwodotx{e^{i(\bfskone+\bfsktwo)\cdot\bfsx}} 
 
\def\eikonemktwodotx{e^{i(\bfskone-\bfsktwo)\cdot\bfsx}} 
\def\eiktwomkthrdotx{e^{i(\bfsktwo-\bfskthr)\cdot\bfsx}} 
\def\eikthrmkonedotx{e^{i(\bfskthr-\bfskone)\cdot\bfsx}} 
 
\def\cc{\mbox{$c.c.$}} 
 
\def\semidirectprodsymbol{ %
    \mathbin{\vrule height1.20ex depth 0.00ex width0.35pt 
    \mkern-2.9mu \mathchar"0202}} 
 
\def\sdp{%
       \mskip-\medmuskip \mkern5mu 
       \mathbin{\semidirectprodsymbol} 
       \penalty 900 
       \mkern5mu \mskip-\medmuskip} 
 
\title{Design of parametrically forced patterns and quasipatterns} 
 
\author{A.M.~Rucklidge\thanks{Department of Applied Mathematics, 
University of Leeds, Leeds LS2 9JT, UK} 
 \and 
 M.~Silber\thanks{Department of Engineering Sciences and Applied Mathematics,
and Northwestern Institute on Complex Systems, Northwestern University,
Evanston, IL 60208, USA}}
 
\begin{document} 
 
\maketitle 
 
 \begin{abstract} 
The Faraday wave experiment is a classic example of a system driven by
parametric forcing, and it produces a wide range of complex patterns, including
superlattice patterns and quasipatterns. Nonlinear three-wave interactions
between driven and weakly damped modes play a key role in determining which
patterns are favoured. We use this idea to design single and multi-frequency
forcing functions that produce examples of superlattice patterns and
quasipatterns in a new model PDE with parametric forcing. We make quantitative
comparisons between the predicted patterns and the solutions of the \hbox{PDE}.
Unexpectedly, the agreement is good only for parameter values very close to
onset. The reason that the range of validity is limited is that the theory
requires strong damping of all modes apart from the driven pattern-forming
modes. This is in conflict with the requirement for weak damping if three-wave
coupling is to influence pattern selection effectively. We distinguish the two
different ways that three-wave interactions can be used to stabilise
quasipatterns, and present examples of 12-, 14- and 20-fold approximate
quasipatterns. We identify which computational domains provide the most
accurate approximations to 12-fold quasipatterns, and systematically
investigate the Fourier spectra of the most accurate approximations.
 \end{abstract}



\pagestyle{myheadings}
\thispagestyle{plain}
\markboth{A.M. Rucklidge and M. Silber}{Design of parametrically forced
patterns and quasipatterns}


\section{Introduction}
\label{sec:Introduction}

The classic Faraday wave experiment consists of a horizontal layer of fluid
that spontaneously develops a pattern of standing waves on its surface as it is
driven by vertical oscillation with amplitude exceeding a critical value;
see~\cite{Arbell2002,Kudrolli1996a,Muller1998a} for recent reviews and surveys.
Many other experimental, biological and environmental systems also form
patterns~\cite{Cross1993,Hoyle2006}, but Faraday wave experiments have
consistently produced patterns with remarkably high degrees of symmetry. One
consequence of this is that, over the years, Faraday wave experiments have
repeatedly produced new patterns of behaviour that went beyond the existing
range of theoretical understanding and required the development of new ideas
for their explanation. For example, in the early 1990's, {\em quasipatterns}
were discovered in two different Faraday wave experiments, one with a
low-viscosity deep layer of fluid with single-frequency
forcing~\cite{Christiansen1992,Binks1997,Binks1997b}, and the other with a
high-viscosity shallow layer of fluid and forcing with two commensurate
temporal frequencies~\cite{Edwards1994}. These patterns are periodic in time
but are quasiperiodic in any spatial direction, that is, the amplitude of the
pattern (taken along any direction in the plane) can be regarded as the sum of
waves with incommensurate spatial frequencies. In spite of this, the spatial
Fourier transforms of quasipatterns have 8, 10 or 12-fold rotational order.
Quasipatterns are of course related to quasicrystals~\cite{Shechtman1984a}, and
quasipatterns have been found in nonlinear optical systems~\cite{Herrero1999},
in shaken convection~\cite{Volmar1997,Rogers2005} and in liquid
crystals~\cite{Lifshitz2007a} as well as the Faraday wave
experiment~\cite{Christiansen1992,Binks1997,Binks1997b,Edwards1994,
Kudrolli1998,Arbell2002}. There is, as yet, no satisfactory theoretical
understanding of the formation of quasipatterns owing to the problem of small
divisors~\cite{Rucklidge2003}.

Theoretical efforts aimed at understanding the pattern selection problem have
centered around {\em weakly nonlinear theory}. The calculations for the
real Faraday wave problem, with finite depth and non-zero viscosity, are
difficult (these involve solving the Navier--Stokes equations with a free
surface boundary condition~\cite{Skeldon2007}). Most calculations aimed at
producing superlattice patterns and quasipatterns have focussed on simpler
equations, such as the Zhang--Vi\~nals~\cite{Zhang1996} equations, which model
Navier--Stokes when the depth is infinite and the viscosity is small, or on
model equations, such as variations on the Swift--Hohenberg
equation~\cite{Lifshitz1997,Muller1994,Frisch1995} or the Fitzhugh--Nagumo
equations~\cite{Dewel2001}.

Nonlinear three-wave resonant interactions have long been recognised as playing
a key role in pattern selection in Faraday wave experiments, or other
situations where complex patterns are
found~\cite{Mermin1985,Newell1993,Edwards1994,Zhang1997}. A series of
papers~\cite{Silber1999,Silber2000,Porter2002,Topaz2002,
Porter2004a,Topaz2004,Porter2004} has developed this idea, using symmetry
considerations to understand pattern selection in Faraday wave experiments with
two-frequency forcing, exploiting the three-wave resonant interactions in the
context of weakly broken Hamiltonian structure. This approach was able to
explain several of the experimentally observed superlattice patterns, and
suggested ways of designing multi-frequency forcing functions that could be
used to control which patterns would
emerge~\cite{Topaz2002,Porter2004a,Topaz2004}. The approach is in principle
predictive, but has only been used to determine which additional frequencies to
add to the forcing function in order to make observed patterns more
robust~\cite{Arbell2002,Ding2006,Epstein2006}. The theory has not been tested
quantitatively against solutions of a pattern-forming system, though weakly
nonlinear coefficients have been computed for the Zhang--Vi\~nals
equations~\cite{Topaz2004,Porter2004} and, more recently, for the
Navier--Stokes equations in the infinite depth case~\cite{Skeldon2007}.

The main goal of this paper is to come to a greater understanding of some of
the complex patterns that are found in high-precision large aspect ratio
Faraday wave experiments. Much of the complexity arises from using two or more
frequencies in the forcing of the experiment, and recent
work~\cite{Porter2004,Topaz2004} explains in principle how to connect the
amplitudes and phases of the various components of the forcing frequencies to
the nonlinear pattern selection problem. The existing theory provides rules of
thumb for designing forcing functions that should encourage the appearance of
particular patterns, and it seems to work well, at least qualitatively, and at
least in some circumstances~\cite{Arbell2002,Ding2006,Epstein2006}.

We take the point of view that in order to claim convincingly that we
understand the pattern selection process in these problems with multi-frequency
forcing, we should be able to predict {\em in advance} which patterns will be
found for different parameter values, we should be able to predict the
amplitudes and range of stability of the patterns, and and we should test
against a pattern forming system that is not constrained to produce only a
limited range of patterns. In order to have a flexible framework for testing
predictions, we have devised a partial differential equation (PDE) with
multi-frequency forcing~(\ref{eq:pde}) that shares many of the characteristics
of the real Faraday wave experiment, but that has easily controllable
dissipation and dispersion relations, and simple nonlinear terms. The linear
behaviour of the PDE reduces to the damped Mathieu equation, with subharmonic
and harmonic tongues, and the simple quadratic and cubic nonlinearities allow
three-wave interactions. The PDE has a Hamiltonian limit, but it differs from
the real situation in the details of its dispersion relation, and the lack of
any coupling to a large-scale mean flow. Notwithstanding these differences, the
PDE allows us to explore in detail some of the generic issues surrounding
pattern selection in very large aspect pattern forming systems with parametric
forcing. A preliminary discussion of the question of quasipattern selection in
the new model PDE can be found in~\cite{Rucklidge2007}.

In section~\ref{sec:Background}, we review the details of how resonant
three-wave interactions influence pattern selection. The main idea is that two
pattern-forming modes, with wavevectors separated by an angle~$\theta$, are
coupled to a weakly damped mode, and this coupling can lead to the
angle~$\theta$ either featuring in the resulting pattern or being eliminated
from the resulting pattern~\cite{Mermin1985,Newell1993,Edwards1994,Zhang1997}.
We introduce the model PDE in section~\ref{sec:thePDE}, and describe its linear
and nonlinear features in sections~\ref{sec:Linear} and~\ref{sec:WNLT}.
Appendix~\ref{app:WNLT} gives full details of the weakly nonlinear
calculations.

In section~\ref{sec:Numericssuperlattice} we devise a forcing function that
stabilises the $22^\circ$ superlattice patterns that have been observed in
large aspect ratio Faraday wave
experiments~\cite{Epstein2006,Kudrolli1998,Arbell2002}. We compute fully
nonlinear solutions of the PDE and compare the computed pattern amplitudes with
those predicted by weakly nonlinear theory. This demonstrates that the
agreement is quantitatively accurate only very close to onset (within~0.1\%).
We show that the limit on the range of validity is because of the presence of
the weakly damped modes that are required for the superlattice pattern to be
stabilised. Thus parameter regimes that are likely to produce the most
interesting patterns, arising from coupling to weakly damped modes, are also
parameter regimes where weakly nonlinear theory has the most restricted
validity.

Two mechanisms have been proposed for quasipattern formation, both building on
ideas of Newell and Pomeau~\cite{Newell1993}, and one aim of this paper is to
demonstrate that both proposed mechanisms for quasipattern formation are viable
(preliminary work is reported in~\cite{Rucklidge2007}). One mechanism applies
to single frequency forced Faraday waves~\cite{Zhang1996} and has been tested
experimentally~\cite{Westra2003}. Another was developed to explain the origin
of the two length scales in superlattice patterns~\cite{Topaz2002,Porter2004}
found in two-frequency experiments~\cite{Kudrolli1998}. The ideas have not been
tested quantitatively, but have been used qualitatively to control
quasipattern~\cite{Arbell2002,Ding2006} and superlattice
pattern~\cite{Epstein2006} formation in two and three-frequency experiments. We
explore the two mechanisms of quasipattern formation in the model PDE in
sections~\ref{sec:Numericsquasipatterns}
and~\ref{sec:Numericsturbulentcrytals}.

Also in section~\ref{sec:Numericsquasipatterns}, we address the distinction
between true and approximate quasipatterns, as found in numerical experiments
with periodic boundary conditions. Owing to the problem of small divisors,
there is as yet no satisfactory mathematical treatment of
quasipatterns~\cite{Rucklidge2003}. In spite of this, the weakly nonlinear
stability calculations, which are in the framework of a 12-mode amplitude
expansion truncated at cubic order, prove to be a reliable guide to finding
parameter values where approximate quasipatterns are stable. The fact that
stable 12-fold quasipatterns are found where they are expected demonstrates
that this approach provides useful information, in spite of the reservations 
expressed in~\cite{Rucklidge2003}. We explore the effect of domain size on the
accuracy of the approximation to a true quasipattern, and show how certain
domains yield particularly accurate approximations.

In section~\ref{sec:Numericsturbulentcrytals}, we present examples of
{\em turbulent crystals}~\cite{Newell1993}: situations in which Fourier modes
oriented more than about $20^\circ$ apart do not affect each other, at the
level of a cubic truncation. In this context, we find examples of 12-, 14- and
20-fold quasipatterns. These are the first examples of quasipatterns of order
greater than 12 found as stable solutions of a~\hbox{PDE} (a preliminary
presentation of the 14-fold example is in~\cite{Rucklidge2007}).

We summarise our result in section~\ref{sec:Conclusions}.


\section{Theoretical background}
\label{sec:Background}

\begin{figure}
\hbox to \hsize{\hfil
 \hbox to 0.3\hsize{\hfil (a)\hfil}\hfil
 \hbox to 0.3\hsize{\hfil (b)\hfil}\hfil
 \hbox to 0.3\hsize{\hfil (c)\hfil}\hfil}
\vspace{0.5ex}
\hbox to \hsize{\hfil
  \mbox{\includegraphics[width=0.3\hsize]{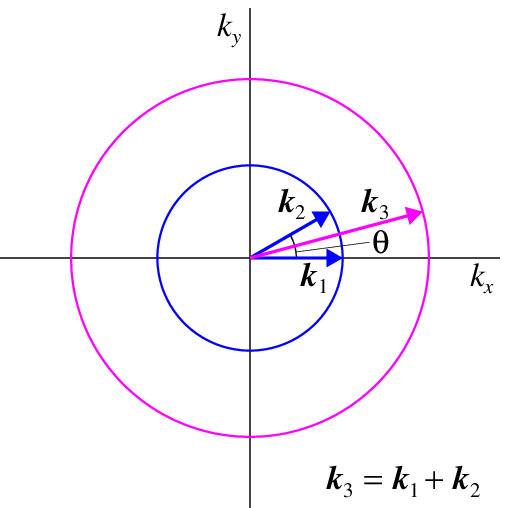}}\hfil
  \mbox{\includegraphics[width=0.3\hsize]{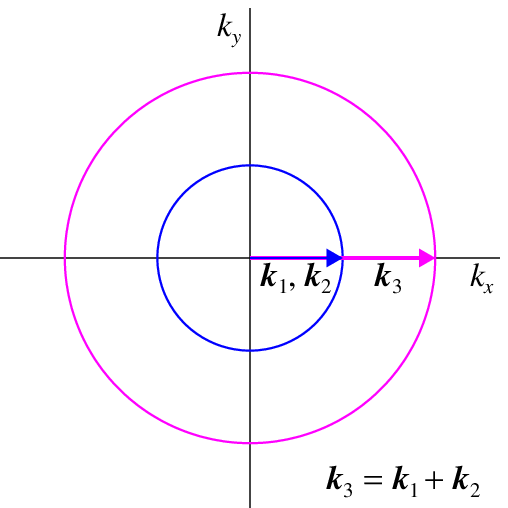}}\hfil
  \mbox{\includegraphics[width=0.3\hsize]{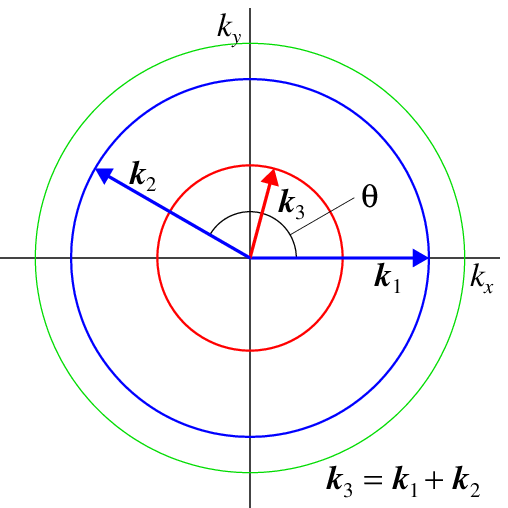}}
\hfil}
\caption{(a)~$1:2$ resonance occurs between two modes with wavevectors
$\bfkone$ and $\bfktwo$ (blue), with the same wavenumber~$k_1$
and separated by an angle~$\theta$, and a third mode
with wavevector $\bfkthr$ (magenta), provided $\bfkthr=\bfkone+\bfktwo$ and
$\Omega(k_3)=2\Omega(k_1)$.
 (b)~A special case of $1:2$ resonance in space and time occurs for $\theta=0$
when the dispersion relation satisfies $\Omega(2k_1)=2\Omega(k_1)$.
 (c)~With two-frequency $m:n$ forcing, the nonlinear combination of two modes
with wavevectors $\bfkone$ and $\bfktwo$, and with $\Omega(k_1)=m/2$ (blue),
can, in the presence of a second component of the forcing at frequency~$n$,
interact with a mode with wavevector~$\bfkthr$ (red), provided
$\bfkthr=\bfkone+\bfktwo$ and $\Omega(k_3)=|m-n|$. Waves driven by frequency~$n$
(green) do not enter the resonance condition.}
 \label{fig:resonances}
 \end{figure}

Resonant triads have played a key role in our understanding of pattern
formation~\cite{Mermin1985,Newell1993,Edwards1994,Zhang1997}.  This section
reviews, somewhat heuristically, the basic selection mechanisms in
the case of wave patterns that are parametrically pumped by a two (or more)
frequency forcing function.  The details behind this summary can be found
in~\cite{Porter2004a,Topaz2004}.  We write the forcing function as
 \begin{equation}\label{eq:ft}
 f(t)=f_m\cos(mt+\phi_m) + f_n\cos(nt+\phi_n) + ...,
 \end{equation}
where $m$ and $n$ are integers, $f_m$ and $f_n$ are real amplitudes,
and $\phi_m$ and $\phi_n$ are phases. (We could, of course, set
$\phi_m=0$ without loss of generality.)  Here we consider $m$~to be
the dominant driving frequency, and focus on a pair of waves, each
with wavenumber $k_m$, which satisfies the dispersion relation
$\Omega(k_m)=m/2$ associated with the linearized problem. In other
words, these waves naturally oscillate at a frequency that is
subharmonic to the dominant driving frequency $m$, and are typically
the easiest to excite parametrically.  We write the critical modes in
the form $z_1 e^{i\bfskone\cdot\bfsx+imt/2}$ and
$z_2 e^{i\bfsktwo\cdot\bfsx+imt/2}$ (together with their complex conjugates),
neglecting the higher temporal frequency contributions to the waves.
These waves will interact nonlinearly with waves
$we^{i\bfskthr\cdot\bfsx+i\Omega(k_3) t}$, where $w$~is a complex amplitude,
$\bfkthr=\bfkone+\bfktwo$ and $\Omega(k_3)$ is the frequency
associated with~$k_3$, provided that either (1)~the same resonance
condition is met with the temporal frequencies, {\em i.e.},
$\Omega(k_3) =\frac{m}{2}+\frac{m}{2}$, as in
figure~\ref{fig:resonances}(a,b), or (2)~any mismatch
$\Delta=|\Omega(k_3)-\frac{m}{2}-\frac{m}{2}|$ in this temporal
resonance condition can be compensated for by the forcing
function~$f(t)$ (figure~\ref{fig:resonances}c). Case~(1) corresponds
to $1:2$ resonance, which occurs even for single frequency forcing
($f_n=0$), and case~(2) applies, for example, to two-frequency forcing
with the third wave oscillating at the difference frequency:
$\Omega(k_3)=m-n$ and $\Delta=n$. Other cases analogous to~(2), such
as $\Omega(k_3)=m+n$, are discussed in~\cite{Topaz2004}, where the
special significance of the difference frequency case is explained.
Note that in both cases (1) and~(2), the temporal frequency
$\Omega(k_3)$ determines the angle $\theta$ between the wave-vectors
$\bfkone$ and $\bfktwo$ via the dispersion relation, and therefore
provides a possible selection mechanism for certain preferred angles
appearing in the power spectrum associated with the
wavepatterns. Selecting an angle of $0^\circ$, with $1:2$ resonance in space
and time (figure~\ref{fig:resonances}b), is a special case.

The nonlinear interactions of the modes can be understood by considering
resonant triad equations describing small amplitude standing wave patterns,
which take the form
 \begin{align}\label{eq:resonanttriad}
 \dot{z}_1&= \lambda z_1+q_1 \bar{z}_2w+(a|z_1|^2+b|z_2|^2)z_1+\cdots\nonumber\\
 \dot{z}_2&= \lambda z_2+q_1 \bar{z}_1w+(a|z_2|^2+b|z_1|^2)z_2+\cdots\\
 \dot{w}&= \nu w+q_3z_1z_2+\cdots,\nonumber
 \end{align}
where all coefficients are real, and the dot refers to timescales long compared
to the forcing period.  Here the quadratic coupling coefficients $q_j$
are {\rm O}(1) in the forcing in the $1:2$ resonance case, and {\rm O}$(|f_n|)$
in the difference frequency case~\cite{Porter2004}.
For other angles $\theta$ between the
wavevectors $\bfkone$ and $\bfktwo$ we expect $q_1\approx q_3\approx 0$
because the temporal resonance condition for the triad of waves is not met.

Since the $m$ frequency is dominant, the $z_1$ and $z_2$~modes will be excited
($\lambda$~increases through zero) while the $w$~mode is damped
($\nu<0$ in~(\ref{eq:resonanttriad})). In this case, $w$~can be
eliminated via center manifold reduction~\cite{Guckenheimer1983b} near the bifurcation point
($w\approx \frac{q_3z_1z_2}{|\nu|}$), resulting in the bifurcation
problem
 \begin{align}\label{eq:ampsrhombs}
 \dot{z}_1&= \lambda z_1+(a|z_1|^2+\tilde{b}|z_2|^2)z_1\nonumber\\
 \dot{z}_2&= \lambda z_2+(a|z_2|^2+\tilde{b}|z_1|^2)z_2\ .
 \end{align}
These equations describe the competition between standing waves
separated by an angle $\theta$, where $\tilde{b}\equiv
b+\frac{q_1q_3}{|\nu|}$ explicitly includes the contribution
from the slaved mode~$w$, and hence depends on the angle between the two
wavevectors $\bfkone$ and~$\bfktwo$.

The contribution of the damped~$w$ mode is significant
whenever $q_1q_3$ is non-negligible and the damping $|\nu|$ is
not too great, and can be made more important by increasing $q_1q_3$
and/or by decreasing the damping~$|\nu|$. For instance, in the
$1:2$ resonance case, for which $q_1q_3$ is ${\rm O}(1)$, the damping
can be decreased by taking $n=2m$ in~(\ref{eq:ft}), since the $w$~mode,
with frequency~$m$,
will then
be driven
subharmonically by the $n$~component of the forcing (as well as
harmonically by the $m$~component).

In the difference
frequency case,
the quadratic interactions rely on the presence of the $n$~component of the
forcing to allow the temporal resonance condition to be met, so
$q_1q_3$ is ${\rm O}(|f_n|^2)$. Thus
the contribution
of the damped~$w$ mode
to~$\tilde{b}$
can be made more important in two ways:
first, by increasing~$f_n$, or second, by parametrically
driving the damped mode so that $|\nu|$ is decreased.
This requires a third
driving frequency $p=2|n-m|$ to be added to the forcing~(\ref{eq:ft}).
In both of these
instances, the relative phases of the components of the forcing matter,
since the
modes are being generated nonlinearly with a preferred phase. These
ideas are developed systematically in~\cite{Porter2004,Topaz2004},
where it is also shown that if there is an underlying Hamiltonian structure, then
$q_1q_3<0$ for the $1:2$ resonance, and $q_1q_3>0$ in the
difference frequency case provided $n>m$. Note that when $q_1q_3>0$
($q_1q_3<0$) then the $\bfkthr$-mode makes a positive (negative)
contribution to the growth rate of the mode $z_2$ when $z_1$ is
present, and vice versa.

In order to determine more precisely whether the resonant contribution to
$\tilde{b}$ is significant enough to lead to a qualitative change in the
resulting pattern, it is useful to rescale the amplitudes $z_1$ and $z_2$
in~(\ref{eq:ampsrhombs}) by a factor of $1/\sqrt{|a|}$. Then we obtain, for
$a<0$, the rescaled equations
 \begin{align}\label{eq:ampsrhombsnorm}
 \dot{z}_1&= \lambda z_1-(|z_1|^2+B_{\theta}|z_2|^2)z_1\nonumber\\
 \dot{z}_2&= \lambda z_2-(|z_2|^2+B_{\theta}|z_1|^2)z_2\ ,
 \end{align}
where $B_{\theta}\equiv \frac{\tilde{b}}{a}$. Here the $\theta$~subscript
indicates that the cross-coupling coefficient between the $\bfkone$ and
$\bfktwo$ modes depends on the angle~$\theta$ between them.

The $B_{\theta}$ function has important consequences for the stability of
regular patterns. As a simple example, note that stripes
($|z_1|=\sqrt{\lambda}$, $z_2=0$) are stable if $B_{\theta}>1$, while rhombs
associated with a given angle $\theta$
($|z_1|=|z_2|=\sqrt{\lambda/(1+B_{\theta})}$) are preferred if
$|B_{\theta}|<1$. Moreover, if $|B_{\theta}|<1$ for any angle $\theta$, then
stripes will necessarily be unstable near onset.  Since, by judicious choice of
forcing frequencies we have at least some ability to control both the magnitude
and sign of $B_\theta$ over some range of angles~$\theta$, we have a mechanism
for enhancing or suppressing certain combinations of wavevectors in the
resulting weakly nonlinear patterns. Alternatively, as suggested
by~\cite{Zhang1996}, if we choose forcing frequencies that lead to a large $1:2$
resonant contribution at $\theta=0$ (figure~\ref{fig:resonances}b), then the
coefficient $a$ can become large compared to $b$, which in turn will cause the
rescaled cross-coupling coefficient $B_{\theta}$ to be small over a broad range
of angles away from $\theta=0$.  (As $\theta\to 0$, it can be shown that
$B_\theta\to 2$.)  These two cases are contrasted in
sections~\ref{sec:Numericsquasipatterns}
and~\ref{sec:Numericsturbulentcrytals}, with preliminary work described
in~\cite{Rucklidge2007}.

Before continuing, we reiterate the constraint on using this analysis to
design forcing functions that will stabilise a given pattern. In eliminating
the damped mode~$w$, we performed a center manifold reduction, which is valid
provided that all non-neutral modes are linearly damped with decay rates
that are bounded away from zero. The domain of validity of the reduced
equations depends on the extent to which $w$ is damped near the bifurcation
point. If that damping is very weak, then the reduced equations will only be
quantitatively predictive in a correspondingly small neighborhood of the
bifurcation point.  We point this out since, from the discussion above, it is
clear that the two ways of influencing the magnitude of~$\tilde{b}$, namely
increasing the driving force~$f_n$ to control~$q_1q_3$, or driving the
difference frequency to reduce the damping~$|\nu|$, both can lead to situations
where the center manifold reduction is no longer valid: the $m$ and~$n$ modes
could set in together, or the $m$ and difference frequency modes could set in
together. In either case, a codimension-two analysis could be performed, but
this is beyond the scope of this paper. In practice, the severity of this
constraint can only be seen by comparing predictions for the amplitudes and
stability of patterns with solutions of the problem at hand (which we do
systematically in section~\ref{sec:Numericssuperlattice}).


\section{The model PDE}
\label{sec:thePDE}

In order to explore these issues in detail, we have devised a
phenomenological PDE for which the leading nonlinear coefficients in
bifurcation problems such as~(\ref{eq:ampsrhombs}), can be calculated
relatively easily, and which is reasonably simple to integrate numerically. The
equation is:
 \begin{eqnarray}\label{eq:pde}
 \frac{\partial U}{\partial t} &=& (\mu+i\omega) U
          + (\alpha+i\beta)\nabla^2 U
          + (\gamma+i\delta)\nabla^4U  \nonumber\\
     & & {} + Q_1 U^2 + Q_2 |U|^2 + C |U|^2U
          + i\hbox{Re}(U) f(t),
 \end{eqnarray}
where
$f(t)$ is a real $2\pi$-periodic function,  $U(x,y,t)$ is a
complex-valued function, with $(x,y)\in\Rset^2$, and $\mu<0$, $\omega$, $\alpha$,
$\beta$, $\gamma$ and $\delta$ are real parameters, and $Q_1=Q_{1r}+iQ_{1i}$,
$Q_2=Q_{2r}+iQ_{2i}$ and $C=C_{r}+iC_{i}$ are complex parameters.

The way the forcing function enters the PDE was chosen so that the linearised
problem reduces to the damped Mathieu equation (in much the same way that
hydrodynamic models of the Faraday instability  reduce to this
equation~\cite{Benjamin1954a}). The PDE has the advantage that the dispersion
relation can be controlled easily, and weakly nonlinear theory is relatively
straightforward to compute. The linear terms are diagonal in Fourier space, so
the PDE is readily amenable to the Exponential Time Differencing numerical
methods of~\cite{Cox2002}. In addition, the nonlinear terms are simple (they do
not involve any derivatives), and so numerical solutions are relatively
inexpensive.

One of the special features of parametric systems is that, even though pattern
selection is a nonlinear process, the position of the linear stability curves
determines which resonant triad interactions are dominant. In turn, it is the
resonant triad interactions rather than the details of the particular form of
nonlinearity in the equation that drives the pattern selection process. For
these reasons, the model PDE is a useful testing ground for results derived
from symmetric bifurcation theory.

The model PDE is similar to the complex Ginzburg--Landau equation -- but we
point out that $U(x,y,t)$ is itself the pattern-forming field, and is not the
amplitude of some other underlying pattern. With $\mu<0$, all waves are damped
in the absence of driving. Note also that the parametric forcing $f(t)$ is
explicitly a function of time so that we are resolving dynamics on the fast
time-scale set by the periodic forcing. In contrast, other
authors~\cite{Coullet1992c,Conway2007,Conway2007a,Halloy2007} have investigated
Ginzburg--Landau equations that describe the slow, large spatial-scale
evolution of the amplitude of an otherwise spatially homogeneous oscillatory
mode arising through Hopf bifurcation. Instead of resolving the fast
oscillations of the subharmonic response to the time dependent forcing, a
constant-coefficient $\bar U$ term is introduced into the equation,
proportional to the amplitude of the parametric forcing~\cite{Coullet1992c}.
With multi-frequency forcing, other terms, such as~$\bar U^2$, are also
introduced~\cite{Conway2007,Conway2007a,Halloy2007}.

There are important qualitative differences, of course, between the model PDE
and the Faraday wave experiment. One difference is the role of the $k=0$
mode. In the PDE, this mode is damped ($\mu<0$) and has a non-zero
frequency~$\omega$; moreover the $k=0$ mode can be nonlinearly excited. In the
Faraday wave problem, owing to mass conservation, the $k=0$ mode is neutral and
cannot be excited, and this may have important consequences in the
dynamics~\cite{Matthews2000,Cox2003}. (In the Zhang-Vi\~nals
model~\cite{Zhang1996} this  requirement is met since all nonlinear terms
appear with an overall spatial derivative that prevents the excitation of the
$k=0$ mode.) Another important difference between the model PDE and the Faraday
wave experiment is that the dispersion relations have a different structure: in
the model PDE, the frequency is a polynomial function of the square of the
wavenumber, but the dispersion relation for Faraday waves is more
complicated~\cite{Benjamin1954a,Kumar1994,Besson1996}. Nonetheless, the
marginal stability curves of the model PDE, especially with multi-frequency
forcing, are similar to those that are observed in large aspect ratio Faraday
wave experiments.

The theory developed by Porter, Topaz and Silber~\cite{Porter2004,Topaz2004}
applies in the weakly damped, weakly forced regime, and certain of their
results also require that the undamped problem have a Hamiltonian structure.
This limit can be realized for our model~(\ref{eq:pde}) by setting
$\mu=\alpha=\gamma=C_r=0$ and $Q_2=-2{\bar Q_1}$. In this case, the Hamiltonian
is:
 \begin{align}\label{eq:hamiltonian}
   H(U,{\bar U}) &= \int\!\!\!\int_R \big[\omega |U|^2 -\beta |\nabla U|^2
 +\delta |\nabla^2 U|^2+f(t)(\hbox{Re}(U))^2\nonumber\\
         &\qquad\qquad\qquad\qquad {}-i Q_1 U^2 {\bar U}
        +  i {\bar Q}_1{\bar U}^2 U + \frac{C_i}{2} |U|^4\big]\ dxdy,
 \end{align}
and $U$ evolves according to
 \begin{equation}
    \frac{\partial U}{\partial t} = i \frac{\delta H}{\delta{\bar U}}\ .
 \end{equation}
The region $R$ corresponds to the domain of integration of the
PDE~(\ref{eq:pde}), where we have assumed periodic boundary conditions
apply on $\partial R$. In the examples presented below, some of the parameter
choices are nearly Hamiltonian (sections~\ref{sec:Numericsquasipatterns}
and~\ref{sec:Numericsturbulentcrytals}) and some are not
(section~\ref{sec:Numericssuperlattice}).


\section{Linear theory}
\label{sec:Linear}

The linear problem associated with~(\ref{eq:pde}) takes the form of
a damped Mathieu equation for each Fourier mode $\eikdotx$. Specifically, if
we set
$U(\bfx,t)=\eikdotx (p(t)+iq(t))$ in~(\ref{eq:pde}) linearized about
$U=0$, then we obtain
 \begin{equation}
 {\ddot p} + {\hat\gamma} {\dot p} + \left(\Omega^2 + {\hat\Omega} f(t)\right) p=0\ ,
 \end{equation}
where
\begin{equation}
 {\hat\gamma}=2\left(-\mu+\alpha k^2-\gamma k^4\right),
 \qquad
 {\hat\Omega}=\omega-\beta k^2+\delta k^4,
 \qquad
 \Omega^2=\frac{{\hat\gamma}^2}{4} + {\hat\Omega}^2\ .
 \end{equation}
We use the method of~\cite{Besson1996} to solve this linear problem for
multi-frequency forcing~$f(t)$, which determines the critical forcing
amplitude. Further details can be found in Appendix~\ref{app:WNLT}.

We require that the damping $\hat{\gamma}$ be positive for all~$k$, so $\mu<0$,
$\gamma\le0$ and $\alpha>-\sqrt{4\gamma\mu}$. However, we do not necessarily
insist that it be monotonic, which would require $\alpha\ge0$. Non-monotonic
damping is possible in the Faraday wave experiment in shallow layers: if the
viscous boundary layers extend from top to bottom of the experimental
container, modes with long wavelengths can be more heavily damped than
short-wave modes, and indeed the first harmonic mode can be unstable at lower
forcing that the subharmonic mode~\cite{Muller1997,Kumar1996a}. We use
non-monotonic damping for the example in
section~\ref{sec:Numericsquasipatterns}. Likewise, we restrict ourselves to the
parameter regime where the dispersion relation $\hat\Omega(k)$ is non-negative
and an increasing function of $k^2$; thus we assume $\omega\ge0$, $\delta\ge0$
and $\beta\le0$.

We will use both $1:2$ and difference frequency resonances to control
how modes interact, so it is important to understand that these
resonances impose further constraints on the parameters in the
dispersion relation. In the case of single frequency forcing
$f(t)=a\cos(t+\phi)$, we expect a subharmonic instability to set in
first with increasing $a$ at a critical wavenumber $k_1$, which can be
estimated by solving the equation
${\hat\Omega}(k_1)=\frac{1}{2}$. This estimate of $k_1$ is good
provided the damping is not too large, and that the damping is a
(nearly) monotonic function of~$k$. The wavenumber associated with the
first harmonic instability is determined by solving
${\hat\Omega}(k_2)=1$. These calculations, together with simple
trigonometry, determine that the angle $\theta$ associated with a
$1:2$ resonant triad satisfies the equation
$k_2=2k_1\cos\left(\frac{\theta}{2}\right)$, which has a solution
provided $k_2\le2k_1$. Choosing a scaling of $\bfx$ so that $k_1=1$,
then we find that $\omega$, $\beta$ and $\delta$ are related to
$\theta$ by
 \begin{align}
  {\hat\Omega}(k=1)&= \omega-\beta+\delta=\frac{1}{2}\nonumber\\
  {\hat\Omega}\left(k=2\cos\left(\frac{\theta}{2}\right)\right)&=
 \omega-4\beta\cos^2\left(\frac{\theta}{2}\right)
 +16\delta\cos^4\left(\frac{\theta}{2}\right)=1\ .
 \end{align}
In particular, the $1:2$ resonance will be at $\theta= 0^\circ$ ($k_2=2$)
if we choose
 \begin{equation} \omega= \frac{1}{3}+4\delta\ ,\qquad
                  \beta=-\frac{1}{6}+5\delta\ ,
 \end{equation}
 where $\delta\in[0,\frac{1}{30}]$
ensures $\beta\le 0$ and hence a monotonic dispersion relation. The $1:2$
resonance moves to $\theta= 90^\circ$ ($k_2=\sqrt{2}$) if we choose
 \begin{equation}
 \omega= 2\delta\ , \qquad
 \beta= -\frac{1}{2}+3\delta\ ,
 \end{equation}
where we require $\delta\in[0,\frac{1}{6}]$.

Next we consider the case of two-frequency forcing
 \begin{equation}
 f(t)=F\left(a_m\cos(mt+\phi_m) + a_n\cos(nt+\phi_n)\right)\ ,
 \end{equation}
where $m$ and $n$ are coprime integers, $(a_m,a_n)$ are relative amplitudes
scaled by an overall amplitude~$F$, and $(\phi_m,\phi_n)$ are phases. We
focus on a resonant triad involving two critical modes with dominant frequency
$\frac{m}{2}$ and a damped difference frequency mode with dominant frequency
$n-m$. We will typically take $m$ even and $n$ odd with $n>m$ and
$n-m<\frac{m}{2}$ ({\em i.e.}, $\frac{n}{m}\in(1,\frac{3}{2}))$. These
conditions imply that the initial instability is expected to be harmonic, that the
difference frequency mode decreases $B_\theta$ at the resonance angle in the
Hamiltonian limit, and that the difference frequency mode has a wavenumber
($\kdiff$) that is smaller than the critical wavenumber and hence it is not too
strongly damped (at least in the case of monotonic dissipation)~\cite{Porter2004}.  We estimate
the critical wavenumber of instability as $k\approx k_1$, where we assume a
scaling such that $k_1=1$. We then have that $k_1$ and $\kdiff$ satisfy
\begin{align}
   {\hat\Omega}(k=1)&=\omega-\beta+\delta=\frac{m}{2}\nonumber\\
   {\hat\Omega}(k=\kdiff)&=\omega-\beta \kdiff^2+\delta \kdiff^4=n-m\ ,
 \end{align}
which we can solve for $\omega$ and $\beta$:
 \begin{equation}
   \omega = \frac{-\kdiff^2\frac{m}{2}+(n-m)}{1-\kdiff^2}
                  + \delta \kdiff^2
\quad\hbox{and}\quad
   \beta = \frac{-\frac{m}{2} + (n-m)}{1-\kdiff^2}
                  + \delta (1+\kdiff^2)\ .
 \end{equation}
Setting $\kdiff=2 \cos\left(\frac{\theta}{2}\right)$, we can
relate an angle in the power spectrum of the pattern with the
wavenumber $\kdiff$ of the damped mode associated with the
resonant triad. For instance, for $\theta=150^\circ$ (the
complementary angle to $30^\circ$, which appears in the 12-fold
quasipatterns), we have $\kdiff=\sqrt{2-\sqrt{3}}$. Alternatively, if
$\theta=158.2^\circ$ (the
complementary angle to $21.8^\circ$, which appears in the simplest hexagonal
superlattice patterns), we have $\kdiff=\frac{1}{\sqrt{7}}$.


\section{Weakly nonlinear theory}
\label{sec:WNLT}

Our weakly nonlinear calculations are aimed at determining the coefficients of
the leading nonlinear terms in finite-dimensional bifurcation problems
associated with certain families of patterns in the plane, and the
corresponding lattices of wavevectors.  These finite-dimensional bifurcation
problems allow us to rigorously compute the relative stability of various
simple planforms, {\em e.g.}, stripes \hbox{\em vs.} squares, rhombs, hexagons,
and also to calculate relative stability of superlattice patterns and hexagons,
stripes, and certain rhomb patterns. Moreover these calculations lead to
quantitative predictions of the amplitude of the standing wave patterns as a
function of the distance $\lambda$ from the bifurcation point, where
$\lambda\equiv (F-F_c)/F_c$ and $F_c$ is the critical value of the overall
forcing amplitude.

\begin{figure}
\hbox to \hsize{\hfil
 \hbox to 0.24\hsize{\hfil (a)\hfil}\hfil
 \hbox to 0.24\hsize{\hfil (b)\hfil}\hfil
 \hbox to 0.24\hsize{\hfil (c)\hfil}\hfil
 \hbox to 0.24\hsize{\hfil (d)\hfil}\hfil}
\vspace{0.5ex}
\hbox to \hsize{\hfil
  \mbox{\includegraphics[width=0.24\hsize]{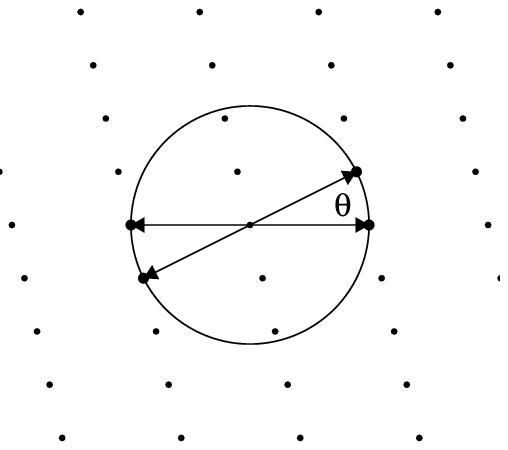}}\hfil
  \mbox{\includegraphics[width=0.24\hsize]{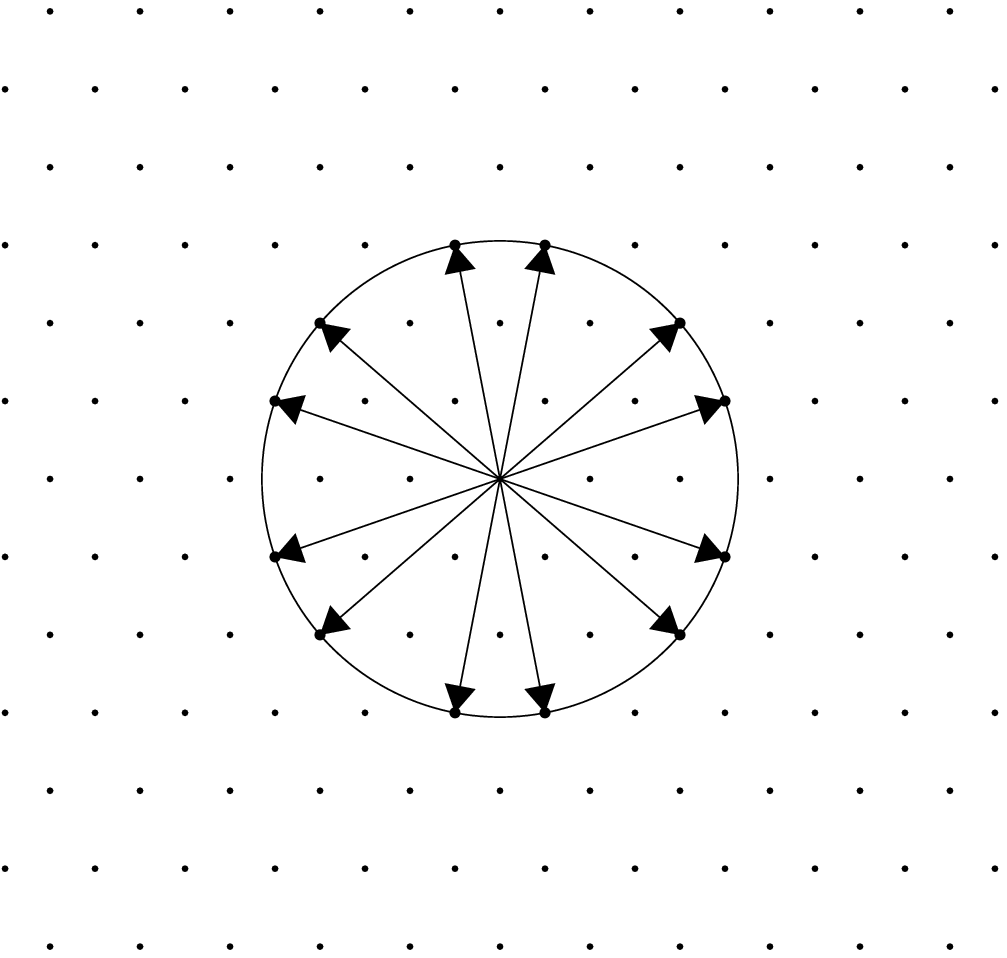}}\hfil
  \mbox{\includegraphics[width=0.24\hsize]{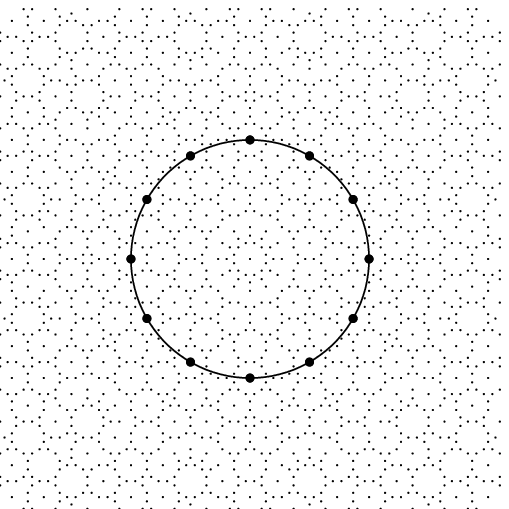}}\hfil
  \mbox{\includegraphics[width=0.24\hsize]{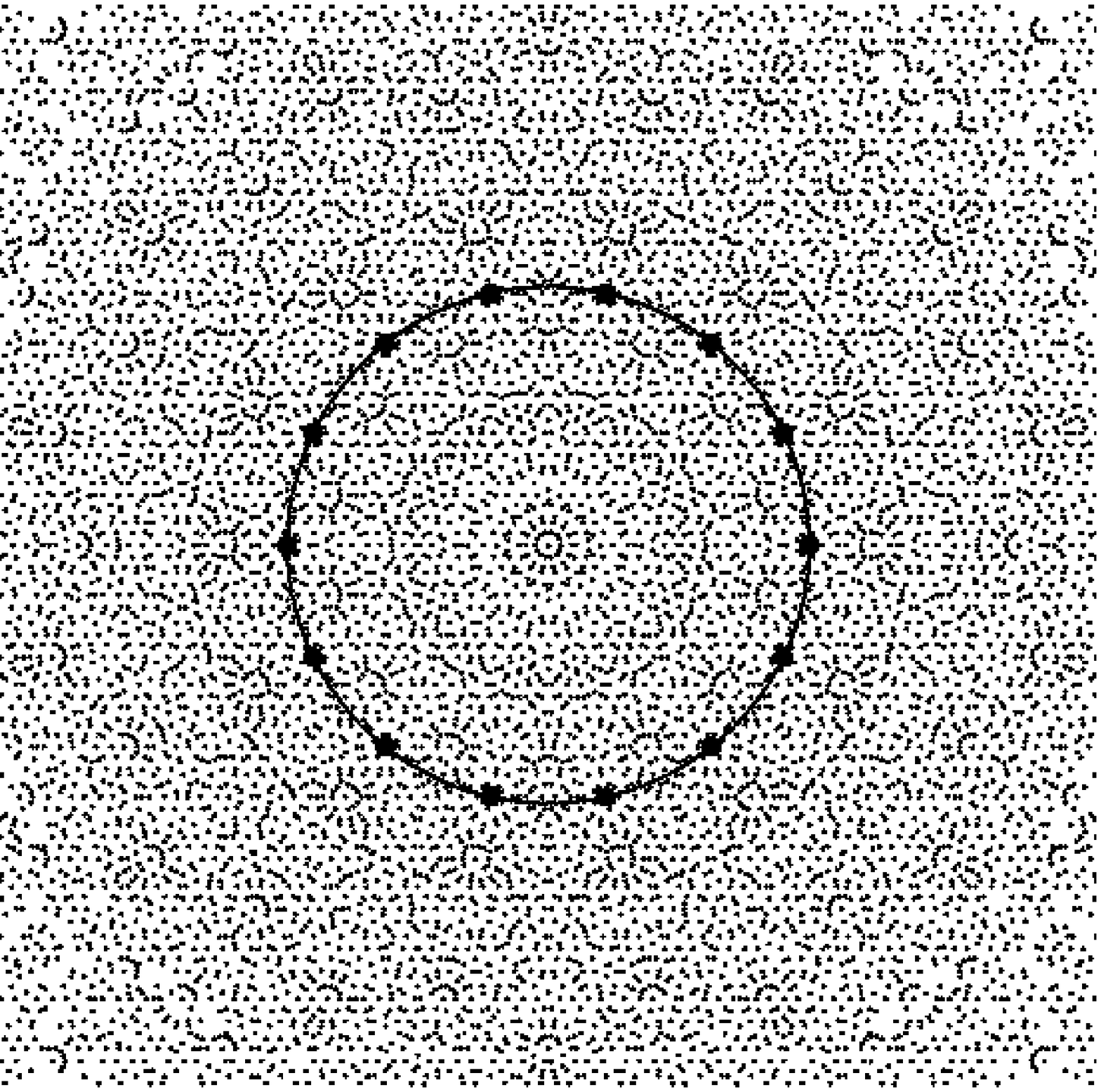}}
\hfil}
\caption{(a)~Rhombic lattice with an angle~$\theta$ between the primary
wavevectors. (b)~Hexagonal superlattice, with an angle of $21.8^\circ$ between
the most closely spaced wavevectors. (c,d)~12-fold and 14-fold quasilattices,
up to 11th order and 7th order respectively~\cite{Rucklidge2003}. See
section~\ref{sec:Numericsquasipatterns} for a discussion of how these
quasilattices are generated.}
 \label{fig:lattices}
\end{figure}

A simple example of such a reduction to a finite-dimensional problem was
presented in Section~\ref{sec:Background}, where we described the bifurcation
problem associated with a pair of standing waves oriented at an angle~$\theta$
relative to each other, where $\theta\in(0,\pi/2]$ was bounded away
from~$\pi/3$. An example rhombic lattice is shown in
figure~\ref{fig:lattices}(a). In that case, the bifurcation problem consisted
of a pair of amplitude equations given by~(\ref{eq:ampsrhombsnorm}), and, after
appropriate scaling, there was just a single nonlinear coefficient $B_\theta$.
The details of the (numerical) computation of this coefficient from the
governing PDE~(\ref{eq:pde}) is given in Appendix~\ref{app:WNLT}. In subsequent
sections of this paper we present plots of $B_\theta$ as a function of $\theta$
for certain parameter sets and forcing functions $f(t)$ used in our numerical
simulations of the model PDE.

As an additional, specific example we consider patterns associated with the
hexagonal superlattice that has been observed in several Faraday wave
experiments~\cite{Kudrolli1998,Arbell2002,Epstein2006}. Equivariant bifurcation
theory~\cite{Golubitsky1988} was used to derive the form of bifurcation
problem~\cite{Dionne1997}. This bifurcation problem describes the long-time
evolution of the twelve modes on the critical circle that are associated with
patterns that tile a plane in hexagonal fashion. The critical Fourier modes
associated with the  $21.8^\circ$ superlattice pattern are indicated in
Figure~\ref{fig:lattices}(b). We label these modes as follows: $(z_1,z_3,z_5)$
are complex amplitudes associated with wavevectors separated by $120^\circ$
and, together with their complex conjugates, they comprise the modes associated
with one hexagon, while $(z_2,z_4,z_6)$ and their complex conjugates are
associated with a second hexagon, rotated by approximately $21.8^\circ$
relative to the first.

The form of the
bifurcation problem associated with this hexagonal superlattice,
to cubic order in the amplitudes, is:
 \begin{align*}
 \frac{dz_1}{dt} &= \lambda z_1 + Q {\bar z}_3 {\bar z}_5
                    - \left(|z_1|^2 + B_{60}\left(|z_3|^2+|z_5|^2\right)
                                    + B_{22}|z_4|^2
                                    + B_{38}|z_6|^2
                                    + B_{82}|z_2|^2     \right)z_1\\
 \frac{dz_3}{dt} &= \lambda z_3 + Q {\bar z}_5 {\bar z}_1
                    - \left(|z_3|^2 + B_{60}\left(|z_5|^2+|z_1|^2\right)
                                    + B_{22}|z_6|^2
                                    + B_{38}|z_2|^2
                                    + B_{82}|z_4|^2     \right)z_3\\
 \frac{dz_5}{dt} &= \lambda z_5 + Q {\bar z}_1 {\bar z}_3
                    - \left(|z_5|^2 + B_{60}\left(|z_1|^2+|z_3|^2\right)
                                    + B_{22}|z_2|^2
                                    + B_{38}|z_4|^2
                                    + B_{82}|z_6|^2     \right)z_5,
 \end{align*}
with similar equations for $z_2$, $z_4$ and $z_6$, related by symmetry.  We have shortened
the labels of the angles to $22$ instead of $21.8$, {\em etc.} This
label indicates the angle between pairs of modes,
{\em e.g.}, there is an angle of
approximately $82^\circ$ between $z_1$ and $z_2$, while
$z_1$ and $z_4$ are separated by approximately $22^\circ$.
The
nonlinear coefficients $Q$ and $B_{60}$ are computed from the
governing PDEs by considering the problem of bifurcation on a
simple hexagonal lattice involving a subset of the modes, while
$B_{\theta}$ is computed for an
arbitrary $\theta\neq60^\circ$ on a rhombic lattice. We can then read
off $B_{22}$, $B_{38}$, {\em etc.}, as required.
The details are in Appendix~\ref{app:WNLT}.
Note that we have assumed that the
bifurcation to a stripe planform is supercritical so that we can rescale the
amplitudes to make the self--coupling coefficient $a=-1$.

The standard planforms, namely stripes, rhombs (associated with each of the angles $22^\circ$,
$38^\circ$, $82^\circ$),
hexagons, and superlattice patterns, correspond to equilibrium
solutions of these equations. The calculation of their linear
stability proceeds in a standard fashion and is summarized in~\cite{Dionne1997}.
In fact, due to the presence
of the quadratic term, all planforms bifurcate unstably, but
because $Q$ is typically very small for multi-frequency
forcing and sufficiently weak damping~\cite{Porter2002}, the planforms can be stabilized at small
amplitude by secondary bifurcations. The
superlattice patterns, which satisfy $z_1=z_2=\cdots=z_6$, come in two varieties that bifurcate together
and their relative stability is unresolved at cubic order. Specifically, hexagonal
superlattice patterns are associated with $z_j$ being real, while
triangular superlattice patterns are of the form $z_j=Re^{i\pi/3}$,
where $R$ is the real amplitude~\cite{Silber1998}.  Which of these two superlattice
patterns is favored in a given situation depends on higher order terms
in the bifurcation problem. We do
not calculate the coefficients of these terms and merely lump the two
types of superlattice patterns together in our bifurcation diagrams.

\section{Numerical experiments: selecting superlattice patterns}
\label{sec:Numericssuperlattice}

Motivated by experimental
observations~\cite{Epstein2006,Kudrolli1998,Arbell2002} of superlattice
patterns in the Faraday wave experiment with $6:7$ forcing, in this section
we use $6:7$ forcing, with some additional forcing at frequency~2 to drive
the difference frequency mode, in order to stabilise a $21.8^\circ$ hexagonal
superlattice pattern. We carry out the linear and weakly nonlinear calculations
to find parameter values for which hexagonal superlattice patterns are
predicted to be stable, and solve the model PDE numerically to confirm these
predictions. In addition, we make quantitative comparisons between the weakly
nonlinear predictions and the numerical solutions of the PDEs, comparing the
predicted amplitudes and ranges of stability of the patterns. The agreement is
not quantitative except at very small amplitude, and we develop an explanation
for this at the end of the section.

The PDE was solved numerically using the fourth-order Runge--Kutta
Exponential Time Differencing numerical method (ETD4RK)
of~\cite{Cox2002}. This pseudospectral method solves the linear part exactly,
excluding the parametric forcing term, which is included with the
nonlinear terms. This allows the use of a timestep based on
accuracy requirements, rather than numerical stability
limits. Timestepping takes place in spectral space, and the
timestep is chosen to be one-twentieth of the shortest of the periods
of the forcing function~$f(t)$, in order that the effect of the
time-dependent forcing is fully resolved by the fourth-order
method. The nonlinear terms are evaluated using
FFTW~\cite{Frigo2005}. The resolution was relatively low for the
examples in this section (up to $96\times56$ Fourier modes),
but we used up to $1536^2$ Fourier modes
for the largest quasipattern examples discussed below. At each
timestep, the upper half of the Fourier spectrum was removed in order
to dealias the cubic terms.

\begin{figure}
\hbox to \hsize{\hfil
  \mbox{\includegraphics[width=0.99\hsize]{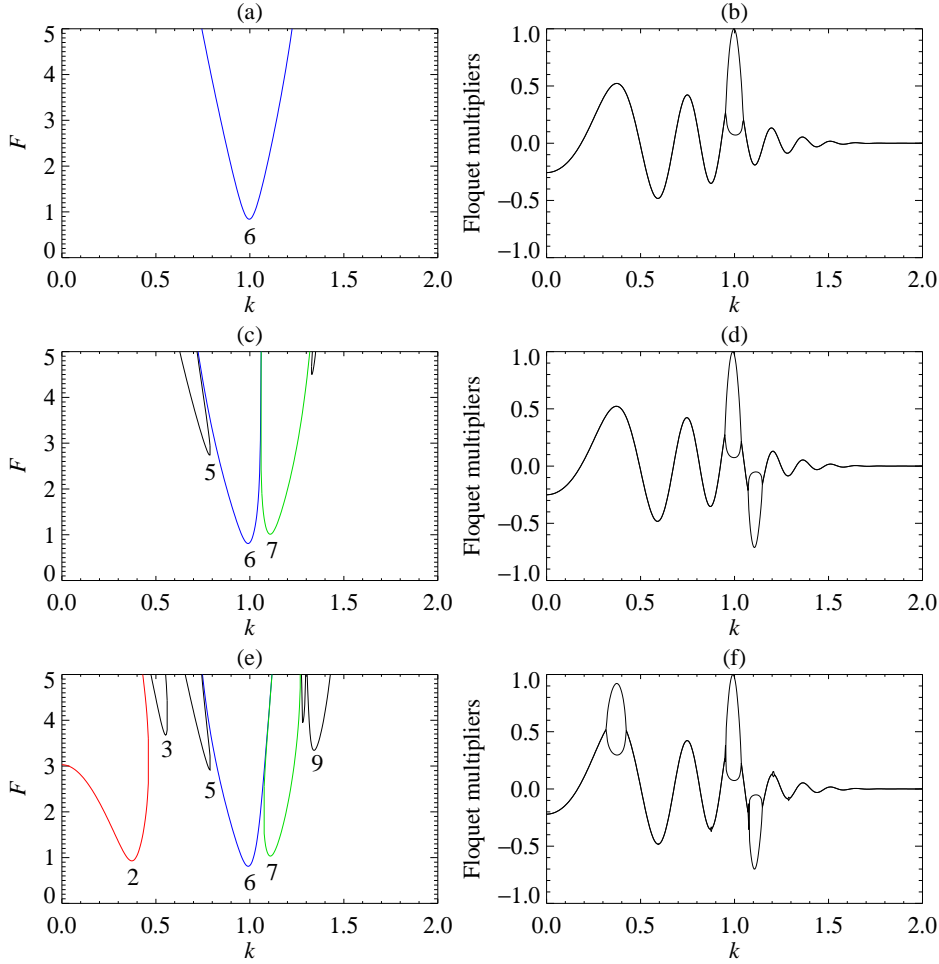}}
\hfil}
\caption{An example of the linear theory for one, two and three-frequency
forcing, with dispersion relation coefficients $\omega=2/3$, $\beta=-7/3$ and
$\delta=0$, and damping coefficients $\mu=-0.1$, $\alpha=0.01$ and
$\gamma=-0.1$.
(a,b) $6$~forcing, with $a_6=1$ and $F_c=0.83973$.
(c,d) $6:7$~forcing, with $(a_6,a_7)=(1,1)$, $(\phi_6,\phi_7)=(0,0)$
      and $F_c=0.80839$.
(e,f) $6:7:2$~forcing, with $(a_6,a_7,a_2)=(1,1,0.45)$,
      $(\phi_6,\phi_7,\phi_2)=(0,0,240^\circ)$, $F_c=0.80975$ and $k_c=0.9910$.
 (a,c,e) show neutral stability curves, whose minima define $F_c$ and the
 critical wavenumber (close to 1 in all cases). Curves corresponding to the
response to frequency~6 are blue, to frequency~7 are green, and to twice the
difference frequency are red, with the corresponding driving frequency
indicated. The minimum of the red curve in (e) is close to $k=\frac{1}{\sqrt{7}}$.
 (b,d,e) show the real parts of the Floquet multipliers at the critical
forcing.}
 \label{fig:lineartheorysl}
 \end{figure}

\begin{figure}
\hbox to \hsize{\hfil
  \mbox{\includegraphics[width=0.99\hsize]{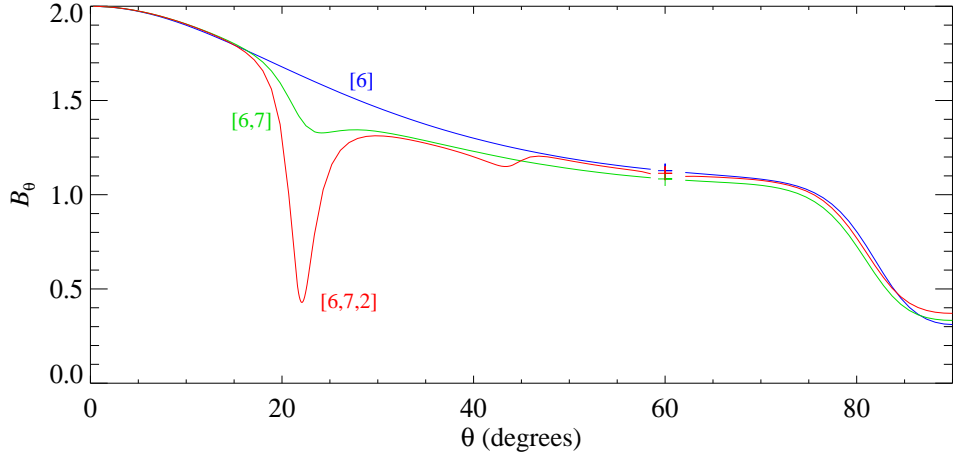}}
\hfil}
\caption{$B_{\theta}$, with linear parameters as in
figure~\ref{fig:lineartheorysl} and nonlinear coefficients $Q_1=2+i$,
$Q_2=1+2i$, $C=-1+30i$. The blue curve is with $6$~forcing only:
the dip at $90^\circ$ is because of $1:2$ resonance with frequency~6.
The green curve has $6:7$ forcing: note the dip starting at $22^\circ$,
corresponding to the difference frequency, even though this frequency is not
forced directly. The red curve has $6:7:2$ forcing: note the pronounced dip
at $22^\circ$, and the smaller dip around $43^\circ$, corresponding to
frequency~4 (which is in $1:2$ resonance with frequency~$2$).
$B_{60}$~is calculated separately, as described in Appendix~\ref{app:WNLT}.
The relevant coefficients are $B_{22}=0.46$, $B_{38}=1.23$, $B_{60}=1.11$ and
$B_{82}=0.63$.}
 \label{fig:Bthetasuperlattice}
 \end{figure}

\begin{figure}
\hbox to \hsize{\hfil
 \hbox to 0.3\hsize{\hfil (a)\hfil}\hfil
 \hbox to 0.3\hsize{\hfil (b)\hfil}\hfil
 \hbox to 0.3\hsize{\hfil (c)\hfil}\hfil}
\vspace{0.5ex}
\hbox to \hsize{\hfil
  \mbox{\includegraphics[width=0.3\hsize]{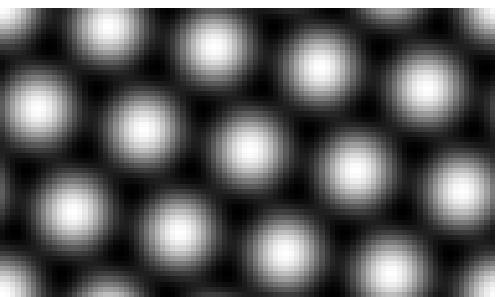}}\hfil
  \mbox{\includegraphics[width=0.3\hsize]{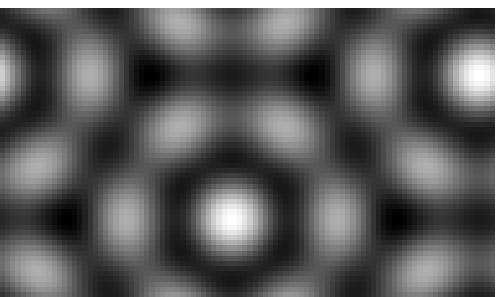}}\hfil
  \mbox{\includegraphics[width=0.3\hsize]{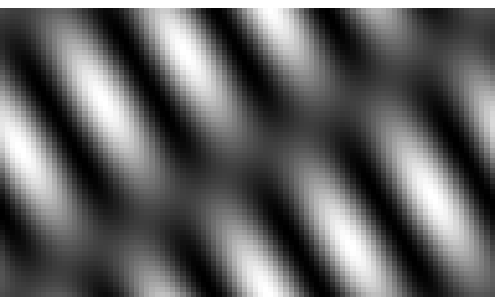}}\hfil}
\caption{With $6:7:2$ frequency forcing at (a,b)~1.004 and (c)~1.02
times critical, we find hexagons, a superlattice pattern, and
$22^\circ$~rectangles. The resolution was $96\times56$ Fourier modes, and
the domain is rectangular, $2\sqrt{7}\times2\sqrt{7/3}$ critical
wavelengths, big enough to fit two copies of the superlattice pattern.
The grey-scale represents the real part of $U(x,y,t)$ with $t$ equal to an
integer multiple of~$2\pi$.}
 \label{fig:hexsuperrect}
\end{figure}

\begin{figure}
\hbox to \hsize{\hfil
  \mbox{\includegraphics[width=0.99\hsize]{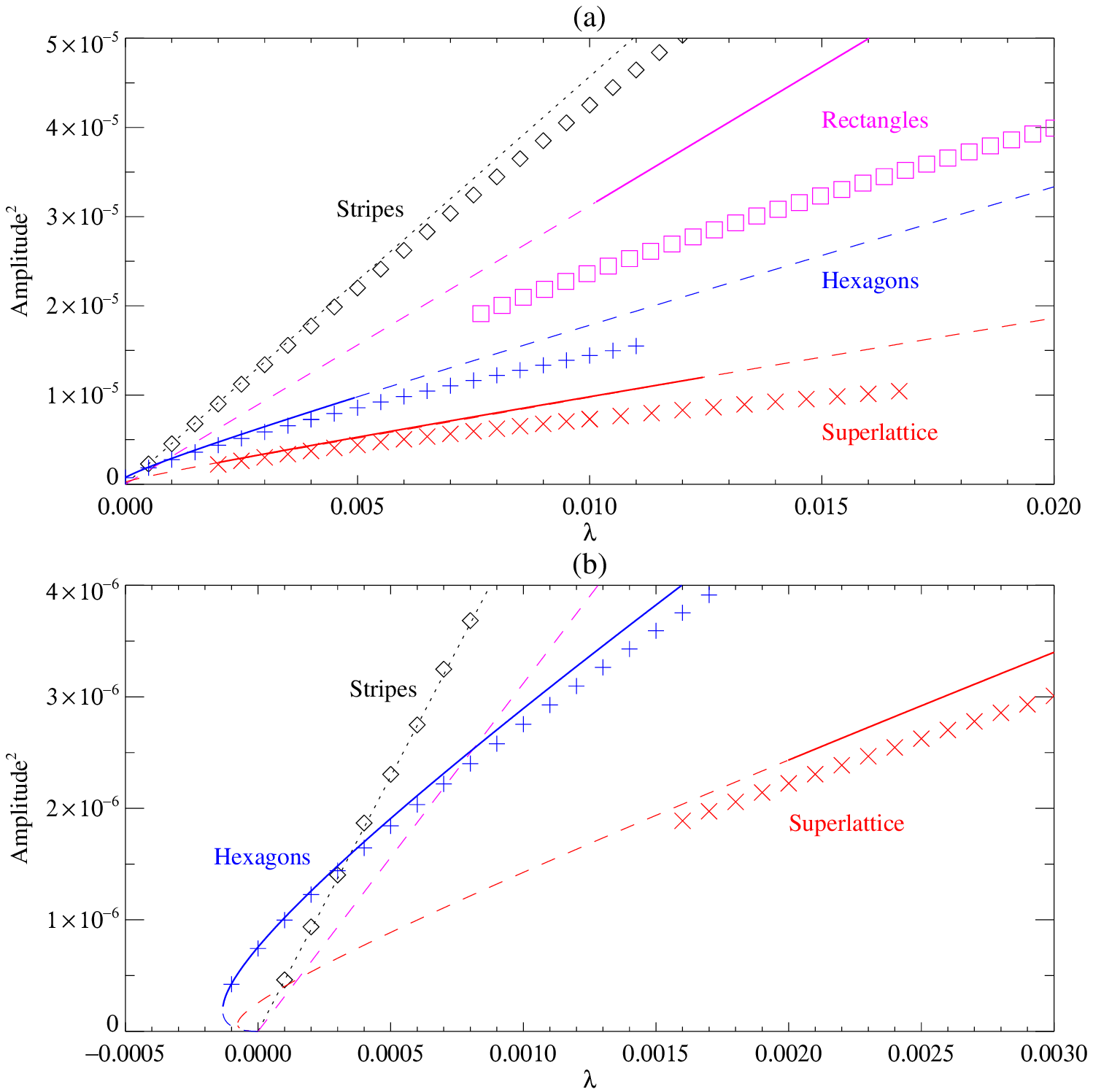}}
\hfil}
\caption{Amplitudes of hexagons (blue), the superlattice pattern (red),
$22^\circ$ rectangles (magenta) and stripes (black) as a function of $\lambda$
for $6:7:2$ forcing. Using weakly nonlinear theory, we denote stable
solutions with solid lines and unstable solutions with dashed and dotted lines.
We also denote numerically computed stable solutions of the PDEs by $+$~for
hexagons, $\times$ for the superlattice pattern, $\Box$~for rectangles, and
$\diamond$~for (unstable) stripes.
The amplitudes of the PDE solutions are computed by matching the time evolution
of the $k=1$ modes to the linear response functions.
(b)~is a detail of~(a), showing that the agreement between weakly nonlinear theory
and the PDEs improves very close to onset.}
 \label{fig:bifurcationsuperlattice}
\end{figure}

We choose parameters so that the mode driven subharmonically by the
$6$~frequency has wavenumber~$k=1$: ${\hat\Omega}(k=1)=3$. The wavenumber associated
with $21.8^\circ$ is $\frac{1}{\sqrt{7}}$, which we wish to correspond to the difference
frequency ($7-6$), so we set ${\hat\Omega}(\frac{1}{\sqrt{7}})=1$. This mode responds
subharmonically to driving at frequency~2. Furthermore, the
wavenumber associated with $81.8^\circ$ is part of this superlattice pattern, and
we can influence this if we set ${\hat\Omega}(\frac{4}{\sqrt{7}})=6$, in $1:2$ resonance with
the primary response. This yields $\omega=\frac{2}{3}$, $\beta=-\frac{7}{3}$
and $\delta=0$. We choose damping coefficients $\mu=-0.1$, $\alpha=0.01$ and
$\gamma=-0.15$, and nonlinear coefficients $Q_1=2+i$, $Q_2=1+2i$ and
$C=-1+30i$. Note that these parameters are not close to the Hamiltonian limit:
in fact, we have chosen the nonlinear coefficients so that the $1:2$
interaction {\em reduces} the cross-coupling coefficient in the range of angles close
to~$90^\circ$.

The linear theory for this problem is shown in figure~\ref{fig:lineartheorysl},
confirming that modes with $k$ close to 1 and $\frac{1}{\sqrt{7}}$ are neutral
and weakly damped respectively, with $6:7:2$ forcing
(figure~\ref{fig:lineartheorysl}f). For this example, we have set the phases of
the two main components of the forcing equal to zero, and (after some
experimentation) set the phase of the component that drives the difference mode
equal to~$240^\circ$.

The cross-coupling coefficient~$B_{\theta}$ is shown in
figure~\ref{fig:Bthetasuperlattice}: with $6$~forcing only, there is a dip in
the curve around $90^\circ$ owing to $1:2$ resonance
at~$82^\circ$. (Had we chosen parameters sufficiently close to a Hamiltonian limit, this feature would have
been a peak rather than a dip~\cite{Porter2004}.)
With $6:7$~forcing, the dip at $22^\circ$, corresponding to
the difference frequency, is visible, even though this frequency is not forced
directly. Finally, with $6:7:2$~forcing, we can control the depth of the dip
at~$22^\circ$. An additional feature at $43^\circ$ is visible, corresponding to
frequency~4, in $1:2$ resonance with frequency~$2$.

Solutions of the PDE with $6:7:2$~forcing are shown in
figure~\ref{fig:hexsuperrect}, confirming that hexagons, superhexagons and
$22^\circ$ rectangles are all stable solutions for different parameter values.

In figure~\ref{fig:bifurcationsuperlattice}, we make quantitative comparison
between the amplitudes and stability of these patterns as numerical solutions
of the PDE, and the values predicted by weakly nonlinear theory, for forcing up
to 1.02~times critical. In addition, we show the results of one-dimensional
simulations, which recover the unstable stripe pattern. In all cases,
numerical solutions of the PDE agree with the weakly nonlinear prediction,
provided we are close enough to onset (we have confirmed this by plotting the
data on a logarithmic scale). Secondary bifurcations, which delimit parameter
intervals where the patterns are stable, are also recovered, although the
agreement is only qualitatively correct.

It is notable that the agreement between the amplitudes predicted by weakly
nonlinear theory and measured from PDE simulations is not particularly good for
the multi-mode patterns, when compared to the much better agreement in the case of
stripes. The reason for this lack of quantitative agreement for the complex
patterns can be understood by going to higher order in the center manifold reduction that was
performed to go from (\ref{eq:resonanttriad}) to~(\ref{eq:ampsrhombs}). We
recall the framework: there are two weakly excited modes with amplitudes~$z_1$
and $z_2$, and a damped mode~$w$, which evolve according to:
 \begin{align}
 \dot{z}_1&= \lambda z_1+q_1 \bar{z}_2w+(a|z_1|^2+b|z_2|^2)z_1,\nonumber\\
 \dot{z}_2&= \lambda z_2+q_1 \bar{z}_1w+(a|z_2|^2+b|z_1|^2)z_2,\\
 \dot{w}&= \nu w+q_3z_1z_2,\nonumber
 \end{align}
where all coefficients are real, and $\nu<0$. We have discarded those
higher order terms that do not play a role in the centre manifold
reduction in order to emphasise the effect of the higher-order nonlinear terms
that appear as a result of the reduction. We express
$w$ on the centre manifold as a power series in $z_1$ and~$z_2$, and
perform a centre manifold reduction to find
 \begin{equation}
 w = -\frac{q_3}{\nu}z_1z_2
       + \frac{q_3(q_1q_3 -(a+b)\nu)}{\nu^3}\left(|z_1|^2+|z_2|^2\right)z_1z_2 + \cdots,
 \end{equation}
which results in
 \begin{align}
 \dot{z}_1&= \lambda z_1+(a|z_1|^2+\tilde{b}|z_2|^2)z_1
             + \frac{q_1q_3(q_1q_3-(a+b)\nu)}{\nu^3}\left(|z_1|^2+|z_2|^2\right)|z_2|^2z_1+\cdots,
 \nonumber\\
 \dot{z}_2&= \lambda z_2+(a|z_2|^2+\tilde{b}|z_1|^2)z_2
             + \frac{q_1q_3(q_1q_3-(a+b)\nu)}{\nu^3}\left(|z_1|^2+|z_2|^2\right)|z_1|^2z_2+\cdots,
 \label{eq:amplitudesquintic}
 \end{align}
where $\tilde{b}=b-q_1q_3/\nu$ as before. In order for the $w$~mode to
influence the coupling constant $B_\theta=\tilde{b}/a$, and so produce
interesting patterns, the quadratic coefficients $q_1$ and $q_3$ must be
non-zero (the modes must be in three-wave resonance) and $\nu$ must be
small (the mode must be weakly damped). However, if $\nu$ is small, the
$\nu^3$ in the denominator of the quintic terms imply that these
high-order terms will be important exactly where the most interesting patterns
will be found. Indeed, the graphs of amplitude against~$\lambda$ in
figure~\ref{fig:bifurcationsuperlattice} are
well fit by a quintic polynomial.

In contrast, the stripe pattern involves damped modes at $k=0$ and $k=2$, and
it can be seen from figure~\ref{fig:lineartheorysl}(f) that these modes are
well damped, so there is no reason for quantitative agreement not to
extend to larger values of~$\lambda$, and indeed it does.

One can estimate the range of validity of the cubic truncation
of~(\ref{eq:amplitudesquintic}) when $\lambda$ and $\nu$ are both small. In the
case of rhombs, the linear and cubic terms balance when
$|z_1|^2=|z_2|^2={\mathcal O}(\lambda\nu)$. The quintic terms are thus smaller than the
linear and cubic terms when $\lambda\ll\nu$, which is what one would expect:
the center manifold reduction is valid when all modes that are eliminated are
heavily damped compared to the modes that are retained.

Therefore, this codimension-one approach to finding interesting patterns has
the smallest range of validity exactly where the patterns are most likely to be
interesting. A proper treatment would require consideration of the
codimension-two problem $(\lambda,\nu)=(0,0)$, which is beyond the
scope of this paper.

A further complication in the Faraday wave situation (and in our model PDE) is
that in order for the interaction between the primary harmonic modes (driven by
the $m$~forcing) and the
weakly damped difference frequency modes ($n-m$) to take place, the subharmonic
mode~($n$) must be present in the forcing function: the quadratic
coefficients~$q_1$ and~$q_3$ increase with the strength of the subharmonic
forcing~$f_n$, so the interaction is strongest when $f_n$ is largest. This
implies that the subharmonic mode is itself only weakly damped, and will be
excited if~$f_n$ is increased beyond its critical value. Experimental evidence
suggests that the codimension-two point (or bicritical point), where the
primary harmonic ($m$) and subharmonic ($n$) modes are both neutral, is an
organising centre for the dynamics~\cite{Edwards1994,Kudrolli1998,Arbell2002}.
(There has been relatively little progress on the theoretical understanding of
the bicritical point, apart from a study in the case of a single
frequency~\cite{Wagner2003} and in a few particular cases for two-frequency
excitation~\cite{Porter2002,Porter2004a}.) Therefore, the problem should really
be treated as a codimension-three interaction between primary harmonic modes
and subharmonic modes, as well as weakly damped harmonic modes.

Notwithstanding these complications, it is clear that the idea that
pattern selection is being influenced by three-wave coupling to weakly damped
modes is fundamentally correct, and the codimension-one approach, while having
limited quantitative agreement with PDE simulations, is clearly providing the
correct qualitative interpretation of the observed patterns.


\section{Numerical experiments: 12-fold quasipatterns}
\label{sec:Numericsquasipatterns}

In the previous section, we demonstrated how to stabilise simple patterns by
driving the difference frequency. In this section, we show how this mechanism
can be used to predict parameter values for stable approximate 12-fold
quasipatterns, and we demonstrate how well a periodic pattern in a large domain
can approximate a quasipattern.

\begin{figure}
\hbox to \hsize{\hfil
  \mbox{\includegraphics[width=0.99\hsize]{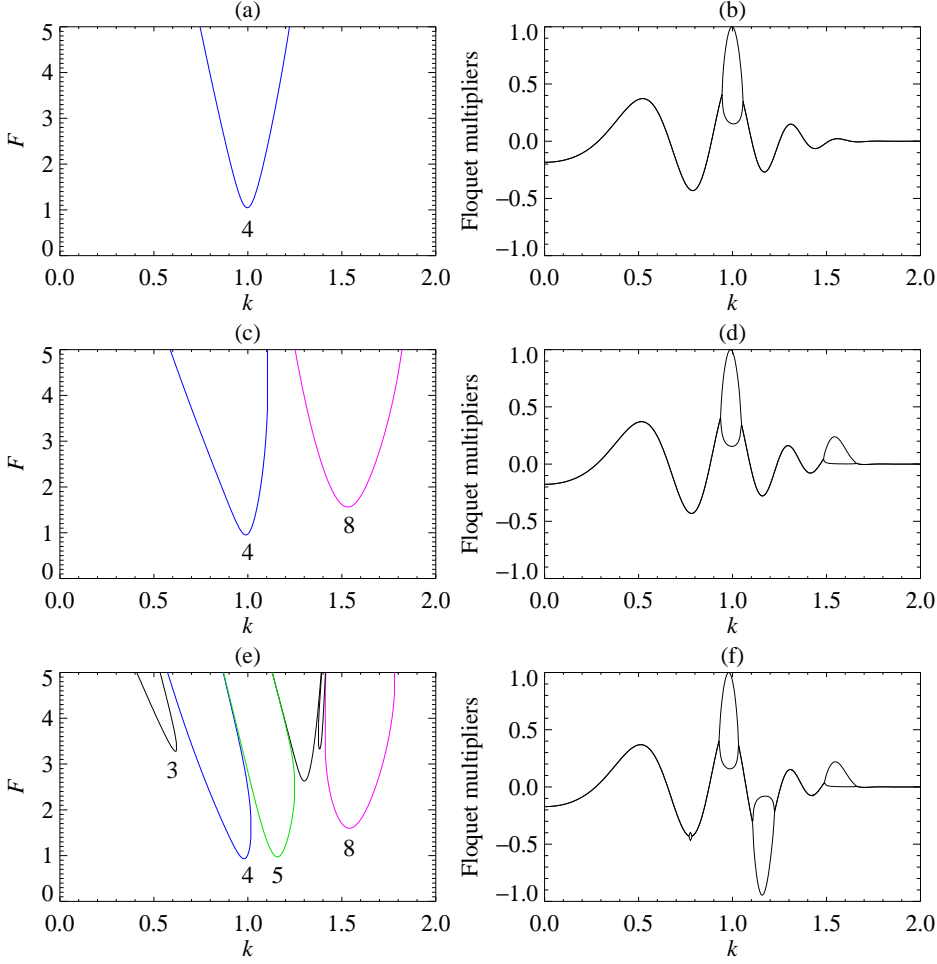}}
\hfil}
\caption{Linear theory for the quasipattern example. The dispersion relation
coefficients are $\omega=0.633975$, $\beta=-1.366025$ and $\delta=0$, and the
damping coefficients are $\mu=-0.2$, $\alpha=-0.2$ and $\gamma=-0.15$.
(a,b) $4$~forcing, with $a_4=0.57358$ and $F_c=1.04730$.
(c,d) $4:8$~forcing, with $(a_4,a_8)=(0.57358,1.6)$, $(\phi_4,\phi_8)=(0,0)$
      and $F_c=0.95214$.
(e,f) $4:5:8$~forcing, with $(a_4,a_5,a_8)=(0.57358,0.81915,1.6)$,
      $(\phi_4,\phi_5,\phi_8)=(0,0,0)$, $F_c=0.93159$ and $k_c=0.9798$.
 (a,c,e) show neutral stability curves, whose minima define $F_c$ and the
 critical wavenumber (close to 1 in all cases). Harmonic curves are blue and
subharmonic curves are green, with the corresponding driving frequency
indicated.
 (b,d,e) show the real parts of the Floquet multipliers.}
\label{fig:lineartheoryqp}
\end{figure}

In order to use triad interactions to encourage modes at $30^\circ$, we choose
$m=4$, $n=5$ forcing: $4:5$~forcing has been used in several experiments to
produce 12-fold quasipatterns~\cite{Kudrolli1998,Edwards1994,Arbell2000}. We
set ${\hat\Omega}(k=1)=2$ so that the subharmonic response to frequency~4 comes
at wavenumber~1, and we require that a wavenumber involved in $150^\circ$ mode
interactions ($k^2=2-\sqrt{3}$) correspond to the difference frequency:
${\hat\Omega}(k)=1$. One solution is $\omega=0.633975$, $\beta=-1.366025$ and
$\delta=0$. Twelve-fold quasipatterns also require modes at $90^\circ$ to be
favoured, and for this choice of parameters, ${\hat\Omega}(k=\sqrt{2})$
is~$3.37$. Although this is not particularly close to~$4$, we can use $1:2$
resonance (driving at frequency~8) to control the $90^\circ$ interaction. The
linear theory for these cases is in figure~\ref{fig:lineartheoryqp}, with
$\mu=-0.2$, $\alpha=-0.2$ and $\gamma=-0.15$. Note that the damping is
non-monotonic, and has a minimum at $k=1$.

\begin{figure}
\hbox to \hsize{\hfil
  \mbox{\includegraphics[width=0.99\hsize]{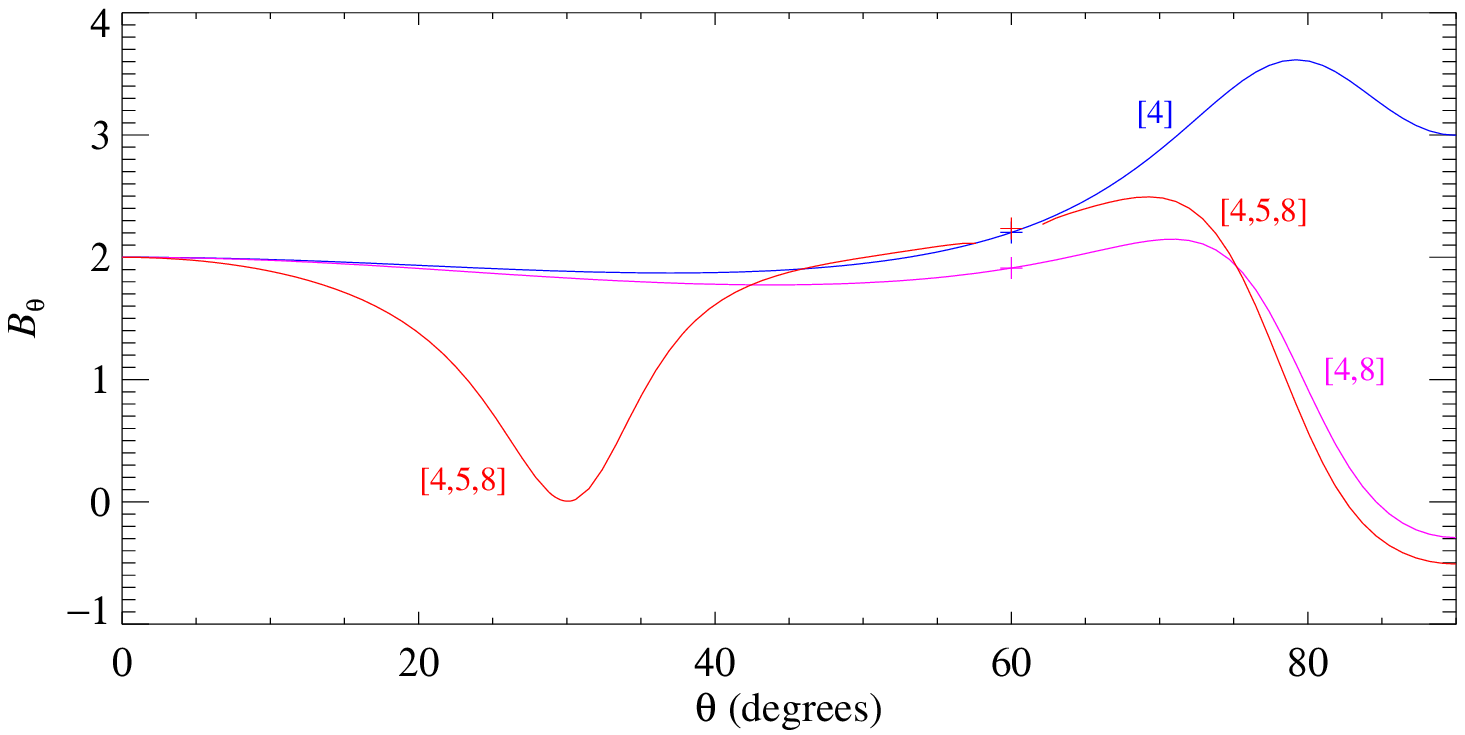}}
\hfil}
\caption{Cross-coupling coefficient $B_{\theta}$ for the parameters from
figure~\ref{fig:lineartheoryqp} and nonlinear coefficients $Q_1=1+i$,
$Q_2=-2+2i$, $C=-1+10i$. The blue curve is with $4$ forcing only: the peak
near $80^\circ$ is because of $1:2$ resonance with frequency~8. The magenta
curve has $4:8$ forcing, using the frequency~8 component to bring down the
curve close to~$90^\circ$. Bringing
in the 5~frequency (red curve) give a pronounced dip at $30^\circ$,
corresponding to the difference frequency~$(5-4)$, even though this mode is not
driven directly. The relevant coefficients are $B_{30}=-0.01$, $B_{60}=2.24$
and $B_{90}=-0.51$.}
 \label{fig:BthetaQP}
 \end{figure}

\begin{figure}
\hbox to \hsize{\hfil
  \mbox{\includegraphics[width=0.99\hsize]{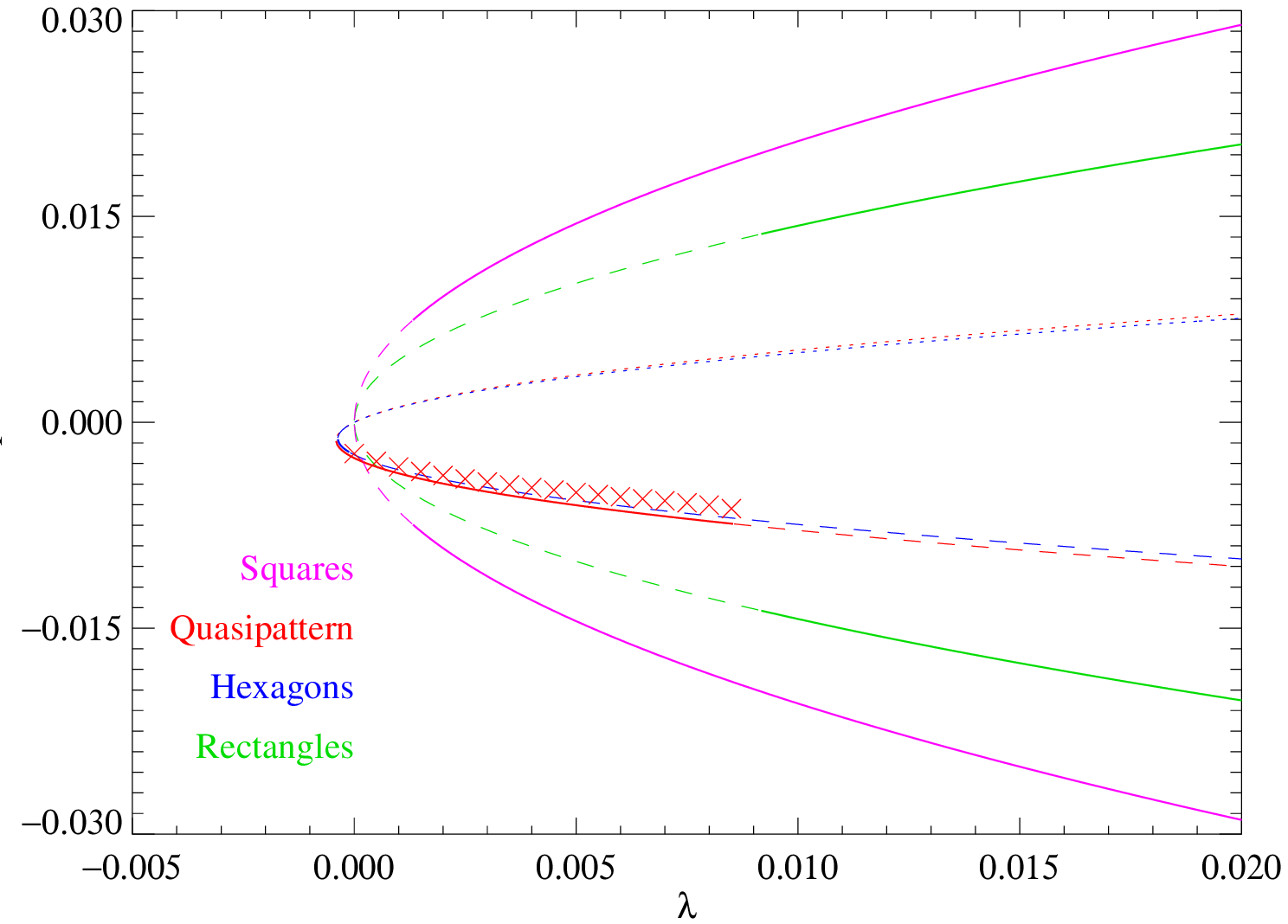}}
\hfil}
\caption{Bifurcation diagram with 3-frequency $4:5:8$ forcing based on a
12-amplitude cubic truncation, with parameter values as in
figure~\ref{fig:BthetaQP}. 12-fold quasipatterns (red) are predicted to be
stable up to $1.0085$ times critical; squares (magenta) are stable from
$1.0013$ times critical and $30^\circ$~rectangles (green) from $1.0092$ times
critical. The crosses show the amplitudes of stable approximate quasipattern
solutions of the PDE, calculated in a domain $2\sqrt{13}\times2\sqrt{13/3}$
critical wavelengths.}
 \label{fig:QPbifn}
 \end{figure}

The resulting $B_{\theta}$ curve with $4:5:8$ forcing and with
near-Hamiltonian choice of nonlinear coefficients (figure~\ref{fig:BthetaQP})
shows pronounced dips at $30^\circ$ and $90^\circ$ as required. In
figure~\ref{fig:QPbifn}, we show that within a 12-amplitude cubic truncation,
12-fold quasipatterns are stable between $0.9995$ and $1.0085$ times critical.
We have found stable approximate quasipatterns (marked by crosses on the
figure, and discussed in more detail below) in the same range. Squares are also
stable above $1.0013$ times critical. The agreement is good, better than the
examples in the previous section (figure~\ref{fig:bifurcationsuperlattice}),
since the difference frequency mode is still fairly well damped (compare the
damping for $k$ close to $0.378$ in figure~\ref{fig:lineartheorysl}f and
$k=0.518$ in figure~\ref{fig:lineartheoryqp}f).

\subsection{Choice of domain size for approximate quasipatterns}

Before presenting numerical solutions in large domains,
we discuss the issue of choice of domain for providing accurate
approximations to quasipatterns.

Just as hexagonal patterns can be approximated in rectangular
domains~\cite{Matthews1998}, there are many ways of choosing periodic domains
to allow accurate approximations of quasipatterns. Here we discuss three
plausible approaches to choosing domains for 12-fold quasipatterns. We show why
one approach, based on Pythagorean triplets, does not work at all, while two
other approaches both work well.

Reducible symmetry group representations  with square periodic domains have
been put forward as candidates for producing approximate
quasipatterns~\cite{Solari1997,Dawes2003}. In particular, Pythagorean triplets
have been identified as of particular interest in this case~\cite{Dawes2003}.
Consider a pair of integers~$(p,q)$ with $p>q>0$. Then $(p^2-q^2,2pq,p^2+q^2)$
forms a Pythagorean triplet ({\em i.e.}, $(p^2-q^2)^2+(2pq)^2=(p^2+q^2)^2)$,
and the vectors $\bfkone=(1,0)$ and $\bfktwo=(2pq,p^2-q^2)/(p^2+q^2)$ have the
same length. For example, with $p=7$, $q=4$, we have $\bfkone=(1,0)$ and
$\bfktwo=(\frac{56}{65},\frac{33}{65})$. If $\frac{p}{q}$ is a continued
fraction approximation of~$\sqrt{3}$ (table~\ref{table:sqrtthree}), then the
angle between these two vectors, namely
$\tan^{-1}\left((p^2-q^2)/(2pq)\right)$, tends to $30^\circ+\mathcal{O}(1/q^2)$
as the approximation to~$\sqrt{3}$ improves. Thus it might be thought that
square $(p^2+q^2)\times(p^2+q^2)$ domains might readily allow approximations to
12-fold quasipatterns.

Unfortunately, these Pythagorean domains do not provide good approximations to
quasipatterns. The reason is that the essential $60^\circ$~coupling is not
quite correct: consider the wavevector $\bfkone=(p^2+q^2,0)$ (in units of the
basic lattice vector), and, at $60^\circ$ on either side of~$\bfkone$ there are
the wavevectors $\bfkthr=(p^2-q^2,2pq)$ and $\bfkele=(p^2-q^2,-2pq)$. In order
for quadratic interactions between these three modes to occur, we need
$\bfkone=\bfkthr+\bfkele$. However, in this case,
$\bfkthr+\bfkele=(2(p^2-q^2),0)=\bfkone+(p^2-3q^2,0)$, which is close to, but
never equal to,~$\bfkone$: it can be shown that $p^2-3q^2=1$ or~$-2$, which is
small is compared to $|\bfkone|=p^2+q^2$. This small difference means that the
important nonlinear interactions between these three waves generate erroneous
long-wave modulations in square $(p^2+q^2)\times(p^2+q^2)$ domains, and if
modes with wavenumber close to zero are not heavily damped, these long-wave
modulations can dominate the pattern.

\begin{table}
\begin{center}
\begin{tabular}{|c|l|}
\hline $\strut$
$\frac{p}{q}\strut$ & $\frac{2}{1}$, $\frac{5}{3}$,
$\frac{7}{4}$, $\frac{19}{11}$, $\frac{26}{15}$, $\frac{71}{41}$,
$\frac{97}{56}$, $\frac{265}{153}$, $\frac{362}{209}$,
$\cdots{}\rightarrow\sqrt{3}$\\
\hline
\end{tabular}
\vspace{2mm}
\caption{Continued fraction approximations to $\sqrt{3}$.}
\label{table:sqrtthree}
\end{center}
\end{table}

\begin{table}
\begin{center}
\begin{tabular}{|c|l|l|l|l|}
\hline $\strut$
$p/q$ & $A=\mbox{Area}$ & First two wavevectors &
$(|\bfktwo|-1)A$ &
$(\angle_{12}-30^\circ)A$\\
\hline $\strut$
\strut$\frac{p}{q}$ & $\frac{4(p^2-3pq+3q^2)}{\sqrt{3}}$
      & $\bfkone=\frac{((p-q)\sqrt{3},3q-p)}{2\sqrt{p^2-3pq+3q^2}}$,
      &
      & \\
\strut &
      & $\bfktwo=\frac{((2q-p)\sqrt{3},p)}{2\sqrt{p^2-3pq+3q^2}}$
      &
      & \\
\strut$\frac{7}{4}$ & $30.02$
      & $\frac{\left(3\sqrt{3},5\right)}{\sqrt{52}}$,
        $\frac{\left(\sqrt{3}, 7\right)}{\sqrt{52}}$
      & $\phantom{-}0$
      & $\phantom{-}66.2$\\
\strut$\frac{19}{11}$ & $224.01$    
      & $\frac{\left(8\sqrt{3},14\right)}{\sqrt{338}}$,
        $\frac{\left(3\sqrt{3},19\right)}{\sqrt{338}}$
      & $\phantom{-}0$
      & $-132.3$\\
\strut$\frac{26}{15}$ & $418.00$
      & $\frac{\left(11\sqrt{3},19\right)}{\sqrt{724}}$,
        $\frac{\left(4 \sqrt{3},26\right)}{\sqrt{724}}$
      & $\phantom{-}0$
      & $\phantom{-}66.2$\\
\strut$\frac{71}{41}$ & $3120.0$
      & $\frac{\left(30\sqrt{3},52\right)}{\sqrt{5404}}$,
        $\frac{\left(11\sqrt{3},71\right)}{\sqrt{5404}}$
      & $\phantom{-}0$
      & $-132.3$\\
\strut$\frac{97}{56}$ & $5822.0$
      & $\frac{\left(41\sqrt{3},71\right)}{\sqrt{10084}}$,
        $\frac{\left(15\sqrt{3},97\right)}{\sqrt{10084}}$
      & $\phantom{-}0$
      & $\phantom{-}66.2$\\
\strut\vdots & & & & \\
\strut$\sqrt{3}$ & $\infty$
        & $\frac{\left(1,1\right)}{\sqrt{2}}$,
          $\frac{\left(\sqrt{3}-1,\sqrt{3}+1\right)}{2\sqrt{2}}$
        & $\phantom{-}$ & $\phantom{-}$ \\
\hline $\strut$
\strut$\frac{p}{q}$ & $2q\times2q$
      & $\bfkone=(1,0)$,
        $\bfktwo=\left(\frac{p}{2q},\frac{1}{2}\right)$
      &
      & \\
\strut$\frac{7}{4}$ & 64
      & $\bfkone=(1,0)$,
        $\bfktwo=\left(\frac{7}{8},\frac{1}{2}\right)$
      & $\phantom{-}0.498$
      & $-16.3$ \\
\strut$\frac{19}{11}$ & 484
      & $\bfkone=(1,0)$,
        $\bfktwo=\left(\frac{19}{22},\frac{1}{2}\right)$
      & $-1.001$
      & $\phantom{-}33.2$ \\
\strut$\frac{26}{15}$ & 900
        & $\bfkone=(1,0)$,
          $\bfktwo=\left(\frac{26}{30},\frac{1}{2}\right)$
        & $\phantom{-}0.500$
        & $-16.5$ \\
\strut$\frac{71}{41}$ & 6724
      & $\bfkone=(1,0)$,
        $\bfktwo=\left(\frac{71}{82},\frac{1}{2}\right)$
      & $-1.000$
      & $\phantom{-}33.1$ \\
\strut$\frac{97}{56}$ & 12544
        & $\bfkone=(1,0)$,
          $\bfktwo=\left(\frac{97}{112},\frac{1}{2}\right)$
        & $\phantom{-}0.500$
        & $-16.5$ \\
\strut\vdots & & & & \\
\strut$\sqrt{3}$ & $\infty$
        & $\bfkone=(1,0)$,
          $\bfktwo=\left(\frac{\sqrt{3}}{2},\frac{1}{2}\right)$
        & $\phantom{-}$ & $\phantom{-}$ \\
\hline
\end{tabular}
\vspace{2mm}
\caption{Domains that provide good approximations to 12-fold quasipatterns.
The first column gives the rational number $\frac{p}{q}$ that is a continued
fraction approximation to~$\sqrt{3}$, drawn from table~\ref{table:sqrtthree}.
The second and third columns give the area of a computational domain and
two of the
wavevectors that will make up an approximate quasipattern. The fourth and fifth
columns demonstrate that the errors in the length of $\bfktwo$ and in the
angle~$\angle_{12}$
between $\bfkone$ and $\bfktwo$ scale as $1/A$, or equivalently, as~$q^{-2}$.
The first set of rows refer to rectangular domains of size
$2\sqrt{p^2-3pq+3q^2}\times2\sqrt{(p^2-3pq+3q^2)/3}$, which allow superlattice
patterns that approximate quasipatterns. In these examples, all wavevectors
are the same length. The second set of rows refer to square
domains of size $2q\times2q$. In these domains, approximate quasipatterns have
wavevectors that have two slightly different lengths.
}
\label{table:domains}
\end{center}
\end{table}

A much better way of generating good approximations to 12-fold quasipatterns is
to choose $2q\times2q$ domains, with vectors $\bfkone=(1,0)$ and
$\bfktwo=(p,q)/2q$, with $\frac{p}{q}$ drawn from table~\ref{table:sqrtthree}.
Again, the angle between these vectors goes as $30^\circ+\mathcal{O}(1/q^2)$,
and, though the wavenumbers are not quite equal, we have
$|\bfktwo|\rightarrow1+\mathcal{O}(1/q^2)$. This approach works because with
this choice of wavevectors, we do have the correct three-wave coupling:
$\bfkthr=(q,p)/2q$, $\bfkele=(q,-p)/2q$ and so $\bfkthr+\bfkele=\bfkone$.

A third possibility is to approximate the quasipattern as a superlattice
pattern, using the 12-dimensional irreducible representations of~$D_6\sdp
T^2$~\cite{Dionne1997}. If we choose $\alpha=q$ and $\beta=p-q$ (in the
notation of~\cite[table~2]{Dionne1997}), with $\frac{p}{q}$ drawn from
table~\ref{table:sqrtthree}, then
 $\bfkone=((p-q)\sqrt{3},3q-p)$,
 $\bfktwo=((2q-p)\sqrt{3},p)$,
 $\bfkthr=((p-2q)\sqrt{3},p)$,
 $\bfkele=((q\sqrt{3},3q-2p)$ (all these should be divided
by their length, $\sqrt{p^2-3pq+3q^2}$). We have
 $\bfkthr+\bfkele=\bfkone$ as required, all wavevectors are the same length
(which is an advantage over the second alternative), and the angle between
$\bfkone$ and $\bfktwo$ goes as $30^\circ+\mathcal{O}(1/q^2)$. One
disadvantage of this approach is that numerical solutions of the PDE must be
carried out in $2\sqrt{p^2-3pq+3q^2}\times2\sqrt{(p^2-3pq+3q^2)/3}$ rectangular
domains in order to take advantage of spectral numerical methods. These domains
are big enough to contain two repeats of the pattern (as in
figure~\ref{fig:hexsuperrect}b) and so only half the Fourier coefficients are
used.

The last two methods are compared in table~\ref{table:domains}. The error
between the approximation and the 12-fold quasipattern is inversely
proportional to the area of the domain in both cases. The hexagonal
superlattice approximations have all wavevectors of the correct length, unlike
the square approximation. For this reason, the superlattice case is better for
computing bifurcation diagrams, as all modes bifurcate at the same value of the
forcing.

However, for similar domain areas, the wavevectors in the square case are about
four times closer to $30^\circ$ apart than the superlattice case. Moreover, the
square case is more amenable to efficient use of fast Fourier transforms, since
the number of modes can be chosen to be a power of two in each direction, while
the superlattice case leads to more awkward choices of numbers of modes.

For these reasons, we choose the rectangular $2\sqrt{13}\times2\sqrt{13/3}$
domain for the bifurcation diagram in figure~\ref{fig:QPbifn}, since the
quantitative comparison between numerical simulation and analysis is easier
when all modes have the same wavenumber and hence bifurcate at the same value
of the forcing (see below). We use the more convenient square ($8\times8$,
$30\times30$ and $112\times112$) domains for the remaining PDE simulations
described below. It would be interesting to see how sensitive the quasipattern
is to the exact choice of domain size. It would also be interesting to find
domains that provide particularly accurate approximations to 8-, 10-, 14-fold
and higher order quasipatterns, but these issues are beyond the scope of this
paper.

One consequence of the wavevectors being of unequal length in the square case
is that the two wavenumbers concerned have slightly different critical
forcings. In the  $30\times30$ and $112\times112$ cases, this difference is
negligible, but in the $8\times8$ case, the critical forcings for the
two wavenumber are appreciably different. We therefore make a small adjustment
to the domain size to make the two critical forcings the same, raised by a
factor of $1.00066$ above~$F_c$.

\begin{figure}
\hbox to \hsize{\hfil
 \hbox to 0.45\hsize{\hfil (a)\hfil}\hfil
 \hbox to 0.45\hsize{\hfil (b)\hfil}\hfil}
\vspace{0.5ex}
\hbox to \hsize{\hfil
  \mbox{\includegraphics[width=0.45\hsize]{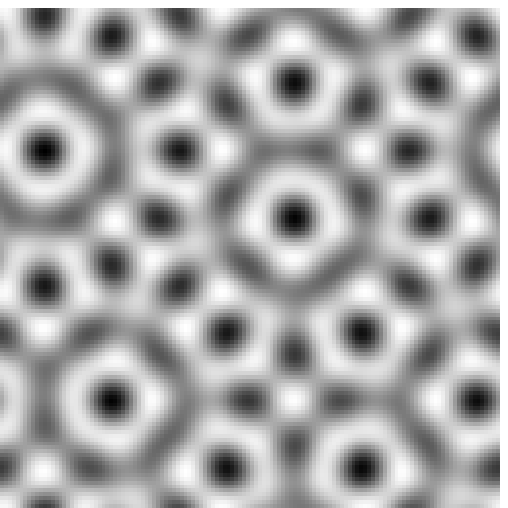}}\hfil
  \mbox{\includegraphics[width=0.45\hsize]{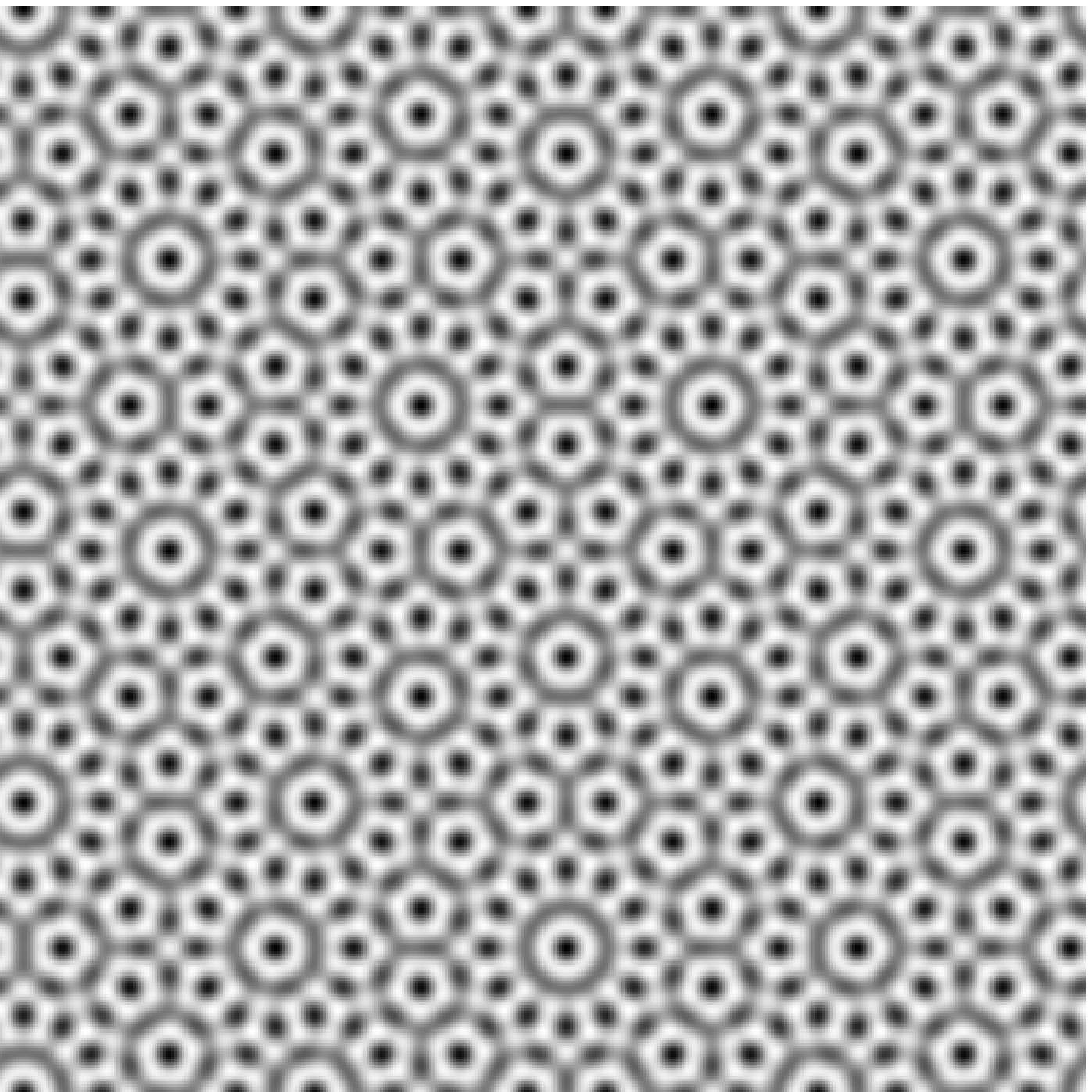}}\hfil}
\vspace{1ex}
\hbox to \hsize{\hfil (c)\hfil}
\vspace{0.5ex}
\hbox to \hsize{\hfil
   \mbox{\includegraphics[width=0.90\hsize]{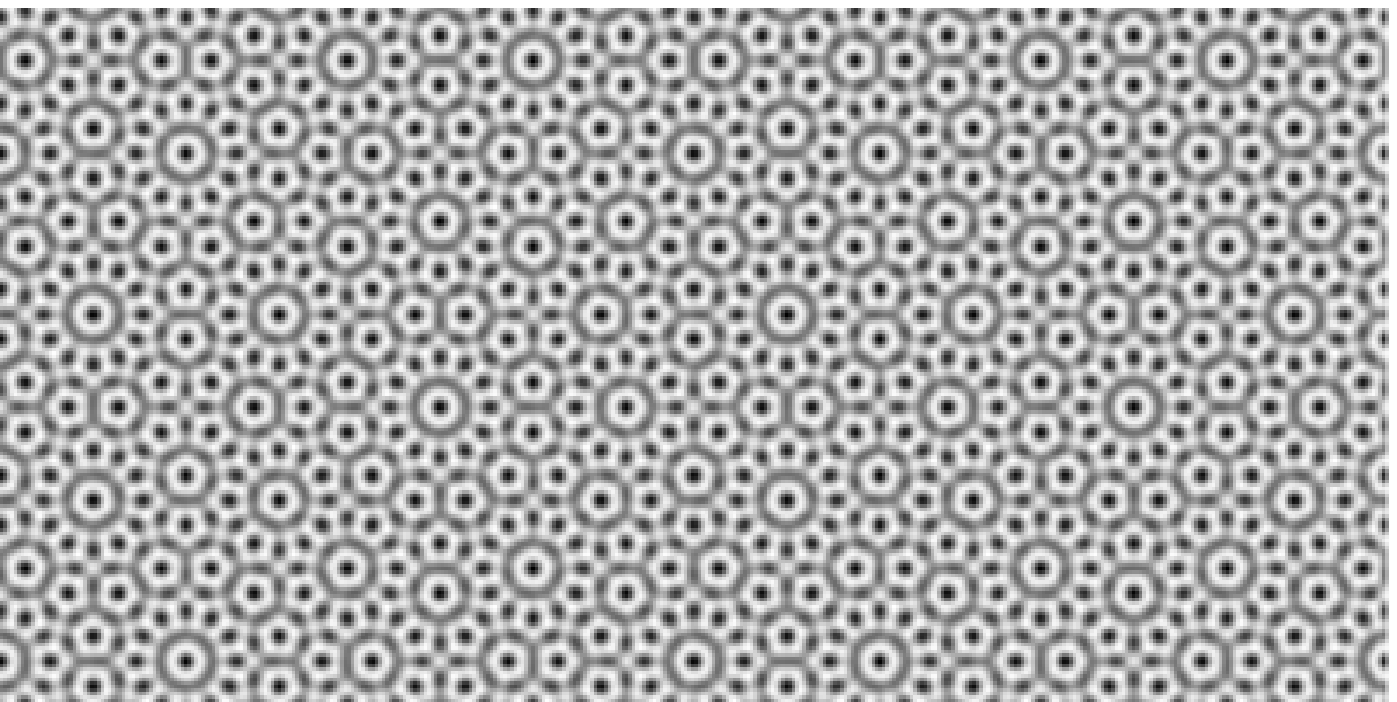}}\hfil}
\caption{Numerical solutions of the PDE with $4:5:8$ forcing at $1.003$ times
critical.
 (a) $8\times8$ domain.
 (b) $30\times30$ domain.
 (c) $112\times112$ domain (only a $60\times30$ section of the domain is
 shown).
 Parameter values are as in figure~\ref{fig:BthetaQP}.}
 \label{fig:QPexamples}
 \end{figure}

\subsection{Numerical examples of 12-fold quasipatterns}

Numerical solutions of the PDE~(\ref{eq:pde}) with $4:5:8$ forcing at $1.003$
times critical are shown in figure~\ref{fig:QPexamples}, in periodic domains
$8\times8$, $30\times30$ and $112\times112$ wavelengths with periodic boundary
conditions and using up to $1536^2$ Fourier modes. The solutions were followed
for at least $10\,000$ forcing periods in the largest domain. Most initial
conditions resulted in square patterns, but minor adjustments to the Fourier
amplitudes at an early stage of the calculation resulted in stable approximate
12-fold quasipatterns. Note, however, that the PDE solutions in
figure~\ref{fig:QPexamples} were not constrained to chose exactly the
wavevectors given in table~\ref{table:domains}.

The accuracy of the approximation improves with increasing domain size. The
modes involved in the $30\times30$ example are $(30,0)$ and $(26,15)$ and their
reflections, which are $29.98^\circ$ apart, and differ in length by $0.05\%$.
The larger $112\times112$ domain allows an improved approximation to the
quasipattern: the important wavevectors are $(112,0)$ and $(97,56)$, which are
$29.9987^\circ$ apart and differ in length by $0.004\%$. The amplitudes of
these modes differ by~$1\%$. We discuss other ways of evaluating the improved
approximation to quasiperiodicity in the next section.

\subsection{Fourier spectra of quasipatterns}

All the numerical PDE solutions presented here have been carried out in
periodic domains, so these solutions are only approximations to quasipatterns.
In the 12-fold examples (figures~\ref{fig:QPexamples} and \ref{fig:NPexamples}a
below), the most important twelve wavevectors in the pattern are not exactly
$30^\circ$ apart and do not have exactly the same length (see
table~\ref{table:domains}). On the other hand, it
is possible to generate true quasipatterns using twelve modes with $k=1$ evenly
spaced around the unit circle, but the asymptotic series in the weakly
nonlinear approximation are known to diverge~\cite{Rucklidge2003}. In this
section, we compare the detailed Fourier spectra of the approximate
quasipatterns to see how these differ from the spectra of true quasipatterns.

We make the comparison by computing the locations of modes generated by
nonlinear interactions up to a certain order, in the three cases of approximate
quasipatterns in figure~\ref{fig:QPexamples}, as well as in a true
quasipattern. To do this, we first define the {\em order} of a mode.

Quadratic interactions between the twelve modes with wavevectors $\bfkone$,
\dots, $\bfktwe$ generate new modes with
wavevectors $2\bfkone$, $\bfkone+\bfktwo$, $\bfkone+\bfkthr$ and so on.
Nonlinear interactions of~$N$ of the twelve modes
generate modes with wavevectors~$\bfkm$:
 \begin{equation}
 \bfkm=\sum_{j=1}^{12} m_j \bfk_j,
 \end{equation}
 where the $m_j$'s are non-negative integers adding up to~$N$. We define
$|\bfm|=\sum_{j=1}^{12}m_j$. In the case of a periodic domain, the set of all
possible~$\bfkm$ defines a lattice of wavevectors
(figure~\ref{fig:lattices}a,b), while in the quasipattern case, the set of all
possible~$\bfkm$ defines a {\em quasilattice}: examples of 12- and 14-fold
quasilattices with $|\bfm|\leq11$ ($|\bfm|\leq7$ in the 14-fold case) are shown
in figure~\ref{fig:lattices}(c,d).

If a wavevector~$\bfk$ is in the lattice or quasilattice, then $\bfk=\bfkm$ for
an infinite number of choices of vector~$\bfm$, since increasing $m_1$ and
$m_7$ (say) by the same amount does not change~$\bfkm$ but increases $|m|$
by~2. We define the {\em order} of wavevector $\bfk$ to be the smallest value
of~$|\bfm|$ with $\bfkm=\bfk$.

\begin{figure}
\hbox to \hsize{\hfil(a)\hfil}
\vspace{0.5ex}
\hbox to \hsize{\hfil
  \mbox{\includegraphics[width=0.8\hsize]{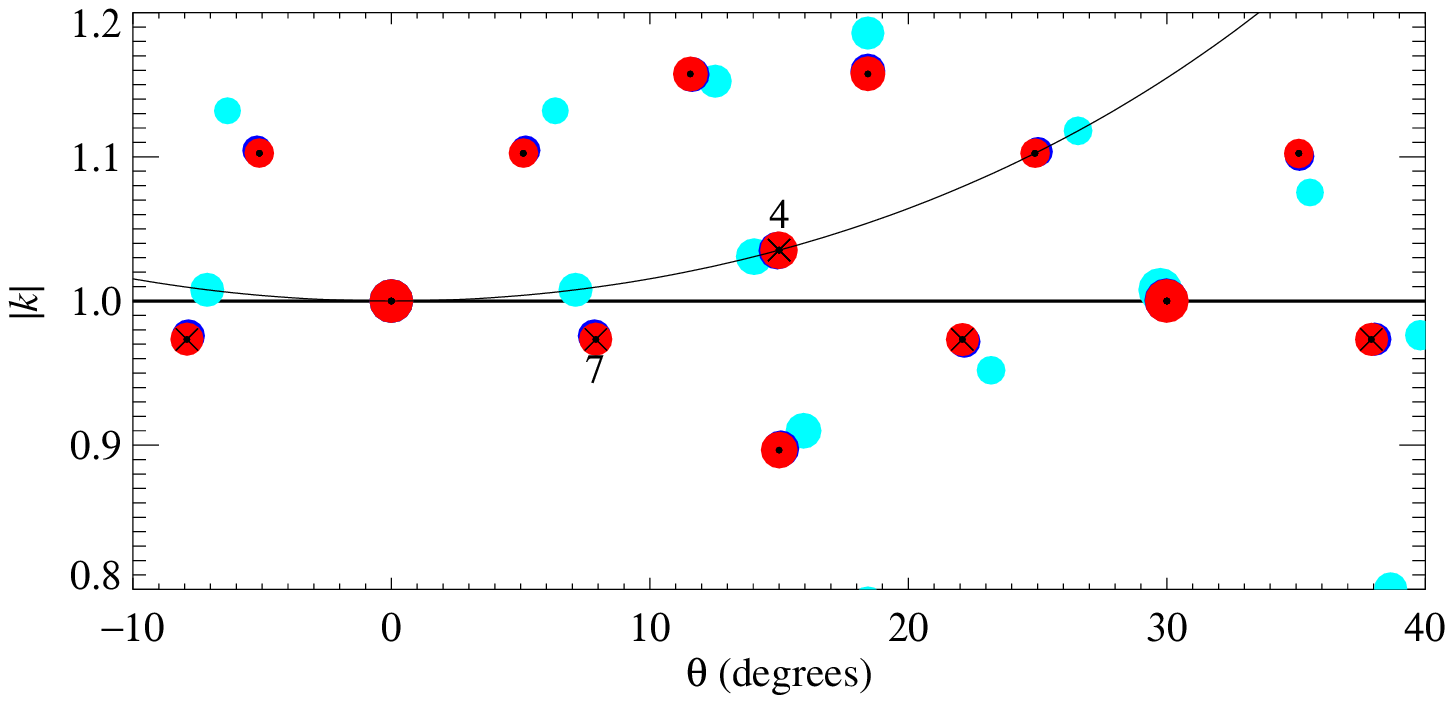}}\hfil}
\vspace{1ex}
\hbox to \hsize{\hfil(b)\hfil}
\vspace{0.5ex}
\hbox to \hsize{\hfil
  \mbox{\includegraphics[width=0.8\hsize]{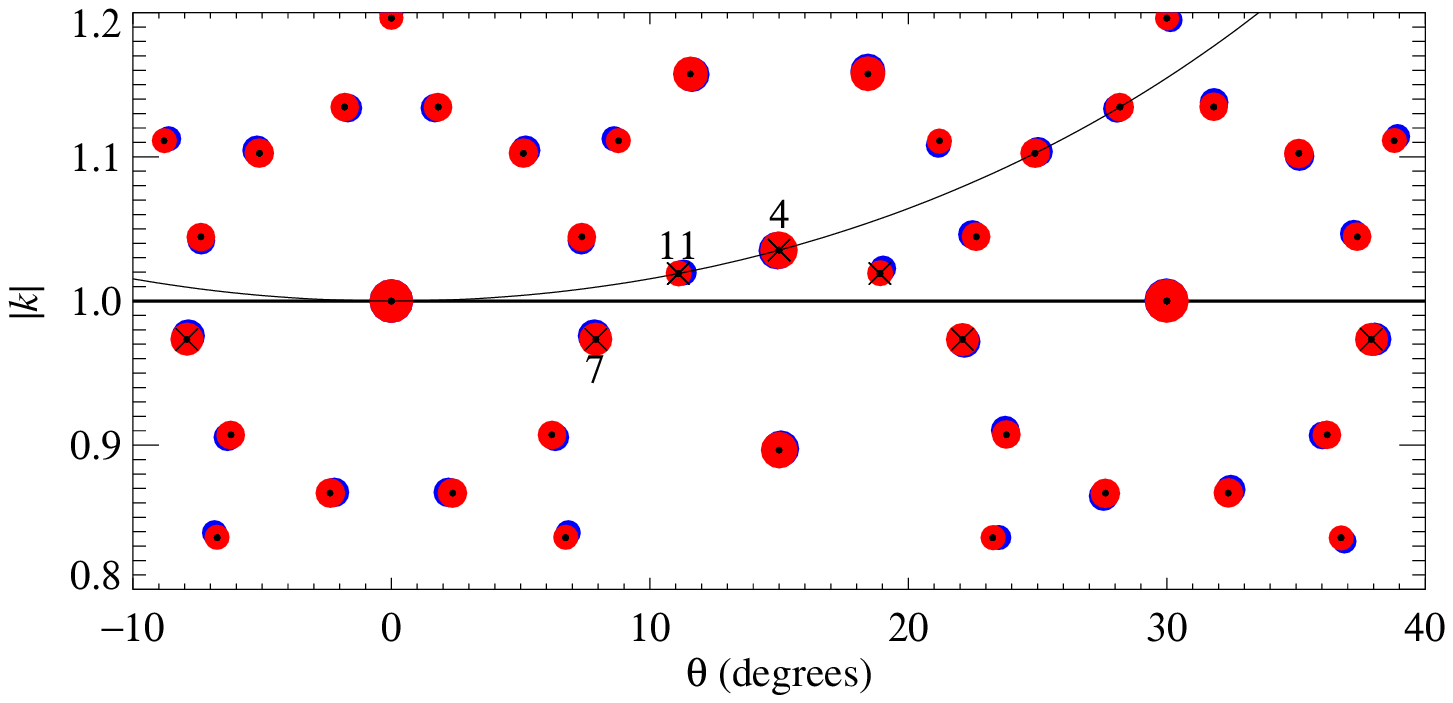}}\hfil}
\vspace{1ex}
\hbox to \hsize{\hfil(c)\hfil}
\vspace{0.5ex}
\hbox to \hsize{\hfil
  \mbox{\includegraphics[width=0.8\hsize]{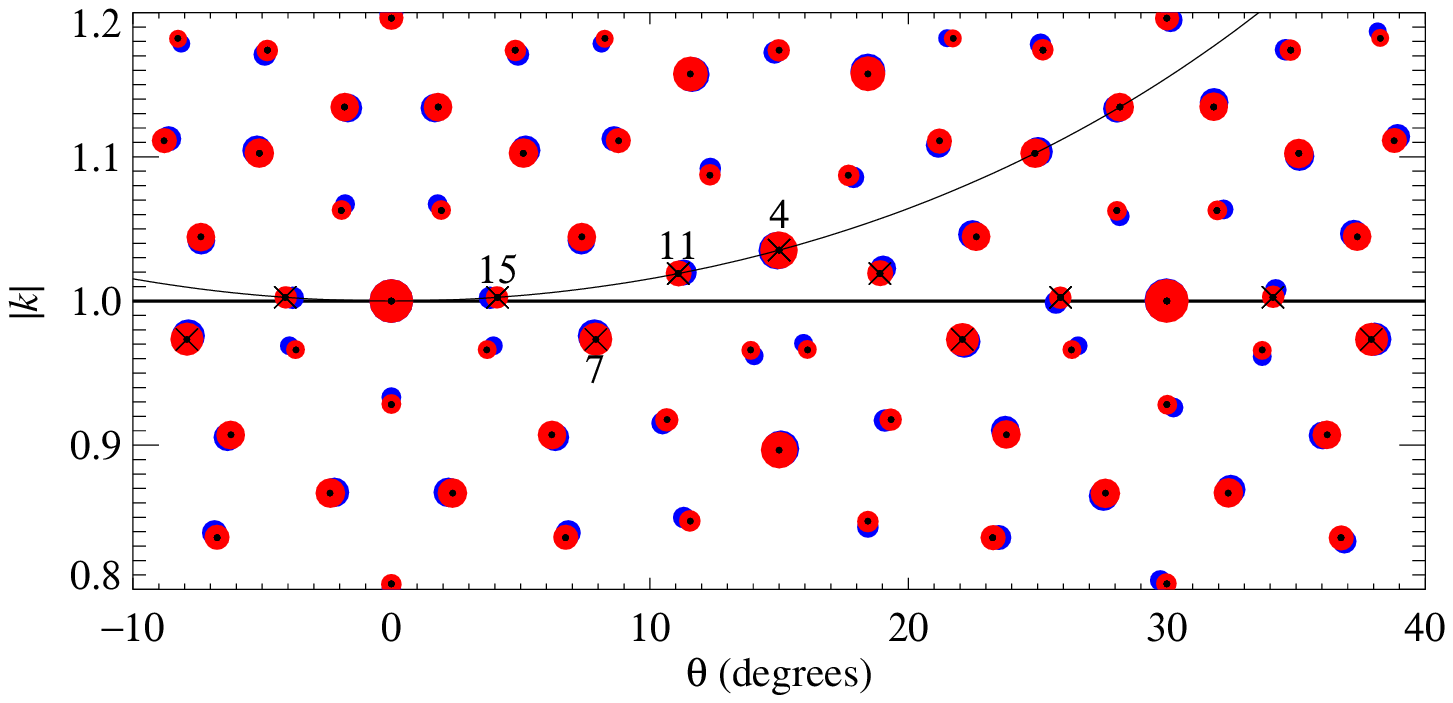}}\hfil}
\caption{Fourier spectra of $8\times8$ (cyan), $30\times30$ (blue),
$112\times112$ (red) approximate 12-fold quasipatterns, as well as the 12-fold
quasilattice (small black dots), up to (a)~7th, (b)~11th and (c)~15th order.
The size of the coloured marker indicates the amplitude of the corresponding
Fourier mode on a logarithmic scale, with the smallest markers having
$10^{17}$ times smaller amplitude than the largest. Only Fourier modes close to
the unit circle ($0.8\leq|\bfk|\leq1.2$) are shown, with wavevector angles
$-10^\circ\leq\theta\leq40^\circ$. The horizontal line $k=1$ represents the
unit circle, while the curved line represents $k_x=1$. Wavevectors that come
closest to the unit circle up to a particular order are labelled
with~$\times$.}
 \label{fig:QPFourierSpectra}
 \end{figure}

\begin{figure}
\hbox to \hsize{\hfil
 \hbox to 0.45\hsize{\hfil (a)\hfil}\hfil
 \hbox to 0.45\hsize{\hfil (b)\hfil}\hfil}
\vspace{0.5ex}
\hbox to \hsize{\hfil
  \mbox{\includegraphics[width=0.45\hsize]{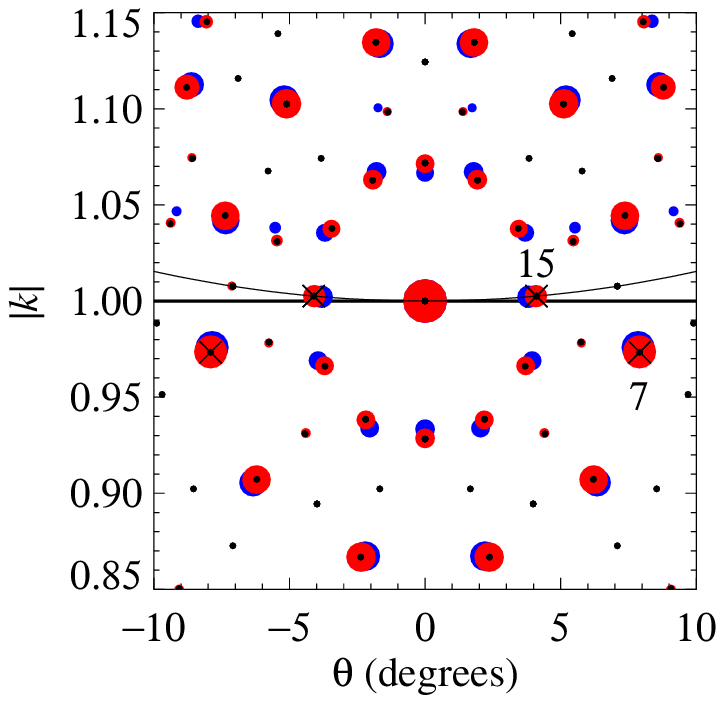}}\hfil
  \mbox{\includegraphics[width=0.45\hsize]{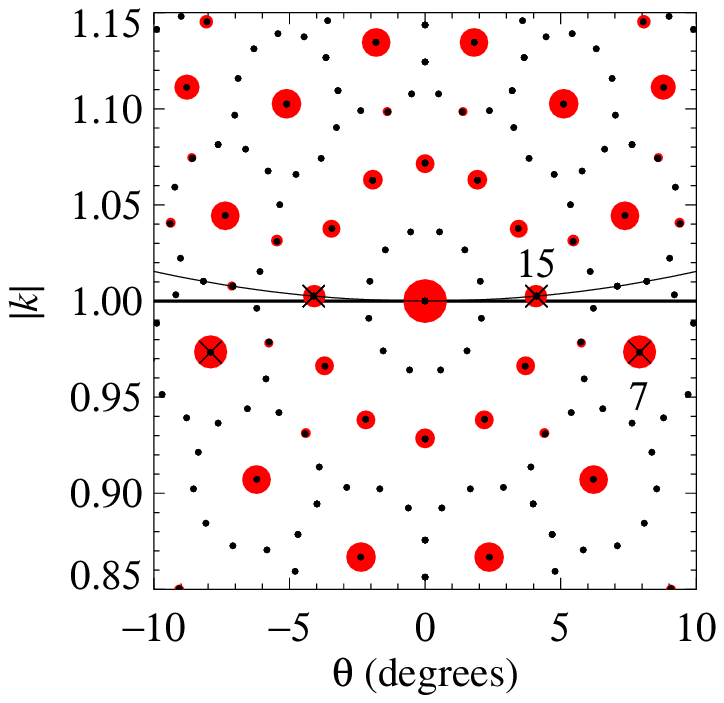}}\hfil}
\vspace{1ex}
\hbox to \hsize{\hfil
 \hbox to 0.45\hsize{\hfil (c)\hfil}\hfil
 \hbox to 0.45\hsize{\hfil (d)\hfil}\hfil}
\vspace{0.5ex}
\hbox to \hsize{\hfil
  \mbox{\includegraphics[width=0.45\hsize]{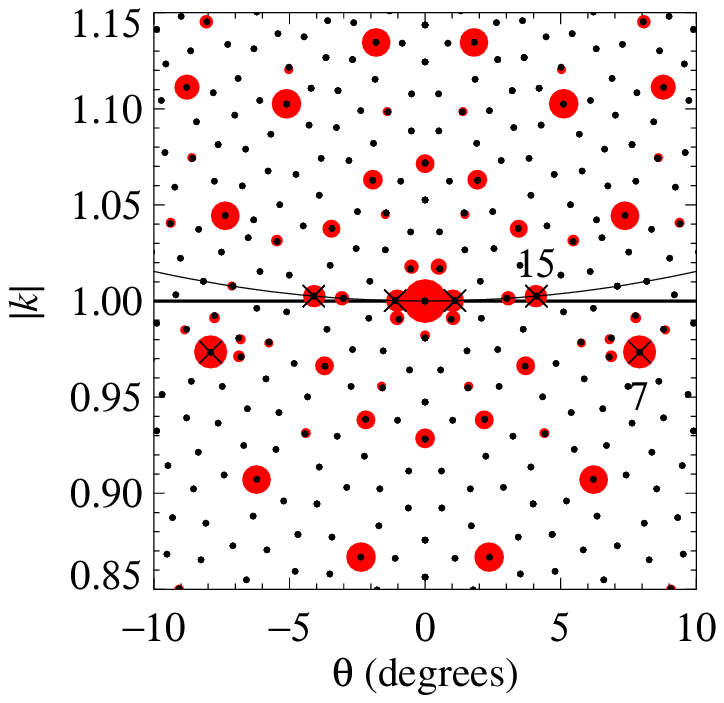}}\hfil
  \mbox{\includegraphics[width=0.45\hsize]{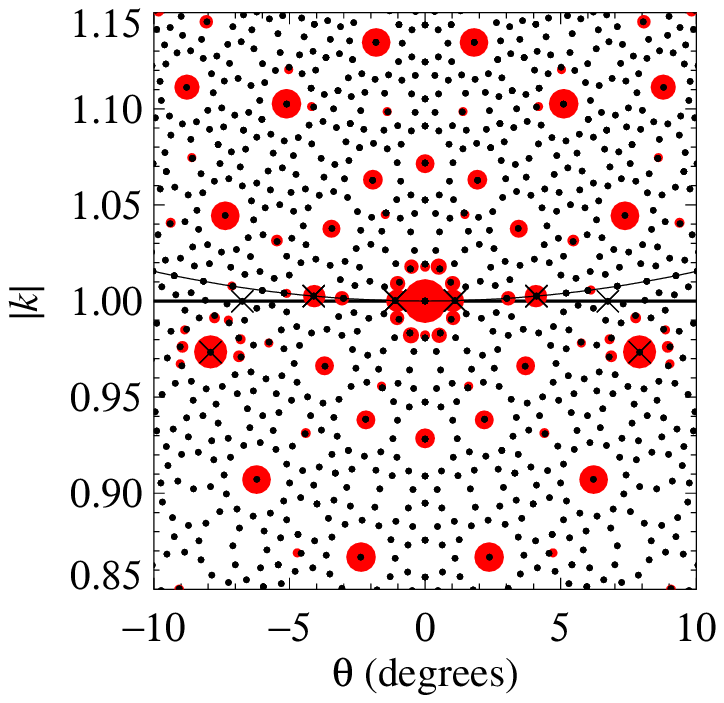}}\hfil}
\caption{Fourier spectra of $30\times30$ (blue) and $112\times112$ (red)
approximate 12-fold quasipatterns, as well as the 12-fold quasilattice (small
black dots), up to (a)~26th, (b)~39th, (c)~56th and (d)~94th order.
Here we show wavevectors within $10^\circ$ of $(1,0)$, and with
$0.85\leq|\bfk|\leq1.15$.}
 \label{fig:QPFourierSpectraDetail}
 \end{figure}

In figures~\ref{fig:QPFourierSpectra} and~\ref{fig:QPFourierSpectraDetail} we
compare the locations of wavevectors in the $8\times8$ (cyan), $30\times30$
(blue) and $112\times112$ (red) examples from the simulation results shown in
figure~\ref{fig:QPexamples}. The
$8\times8$ spectrum is only in figure~\ref{fig:QPFourierSpectra}(a) (up to 7th
order), and the $30\times30$ spectra are only up to 26th order.
The amplitudes of each Fourier mode is given by the size of the symbol (on a
logarithmic scale): the largest symbols are the modes with the largest
amplitudes, and the modes with the smallest symbols have amplitudes $10^{17}$
times smaller. Modes with amplitudes smaller than this are not plotted as these
are in the realm of round-off error (see below).

The success with which the periodic patterns approximate a true quasipattern
can be judged by the locations of the important modes in the pattern, as
compared to the locations of modes on the quasilattice, up to a given order. Up
to 7th order (figure~\ref{fig:QPFourierSpectra}a), the $30\times30$ and
$112\times112$ modes (blue and red) overlay each other almost exactly, and
correspond well with the quasilattice modes (small black dots at the centre of
the red markers). However, the $8\times8$ modes (cyan) deviate substantially
from the correct locations, and we conclude that the periodic pattern in an
$8\times8$ domain is a poor approximation to a quasipattern (in spite of
appearances in figure~\ref{fig:QPexamples}a).

At 11th order (figure~\ref{fig:QPFourierSpectra}b), the agreement between
$30\times30$, $112\times112$ and the true quasipattern is still good, while at
15th order (figure~\ref{fig:QPFourierSpectra}c), the $30\times30$ modes deviate
noticeably from the $112\times112$ and quasipattern modes. This is more
pronounced at 26th order (figure~\ref{fig:QPFourierSpectraDetail}a).

At 15th order (figure~\ref{fig:QPFourierSpectra}c), the agreement between
$112\times112$ and the true quasipattern is excellent: every red marker has a
black dot at its centre. The agreement is still very good at 26th order
(figure~\ref{fig:QPFourierSpectraDetail}a): the black dots are not quite in the
centres of the smallest red markers, and some black dots do not have
corresponding red markers. This means that those modes, present in the true
quasipattern at this order, have amplitudes in the $112\times112$ approximation
that are too small to be plotted. The situation at 39th order
(figure~\ref{fig:QPFourierSpectraDetail}b) is similar, but it isn't until 94th
order (figure~\ref{fig:QPFourierSpectraDetail}d) that the true quasipattern
modes miss the centres of the red markers entirely.

\begin{figure}
\hbox to \hsize{\hfil
  \mbox{\includegraphics[width=0.8\hsize]{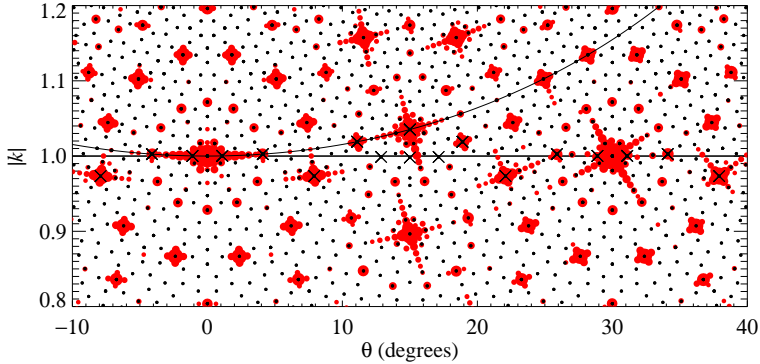}}\hfil}
\caption{Fourier spectra of the
$112\times112$ (red) approximate 12-fold quasipattern, as well as the 12-fold
quasilattice (small black dots). All modes with amplitude greater than
$10^{-17}$ times the maximum amplitude are plotted; the quasilattice is plotted
up to 56th order. The competing effects of the 12-fold order and the square
lattice in are apparent.}
 \label{fig:QPFourierSpectraLattice}
 \end{figure}

In fact, a curious situation arises at 56th and 94th order: it is apparent that
modes close to $\bfk=(1,0)$, and in particular modes on the line $k_x=1$, have
amplitudes that are higher than might be expected, since usually the amplitudes
of modes decreases with their order. These modes are discussed in more
detail below. However, it should be noted that plotting only modes up to a
certain order masks the effect of the underlying lattice in the numerical
solutions. In figure~\ref{fig:QPFourierSpectraLattice}, we show all modes in
the $112\times112$ down to round-off error, and quasilattice modes up to 56th
order. The underlying square numerical lattice is clearly seen in the strings
of red markers emanating from each large-amplitude mode. These give an
impression that the large-amplitude modes could be considered to be clusters of
modes in Fourier space.

\begin{figure}
\hbox to \hsize{\hfil (a)\hfil}
\vspace{0.5ex}
\hbox to \hsize{\hfil
  \mbox{\includegraphics[width=0.8\hsize]{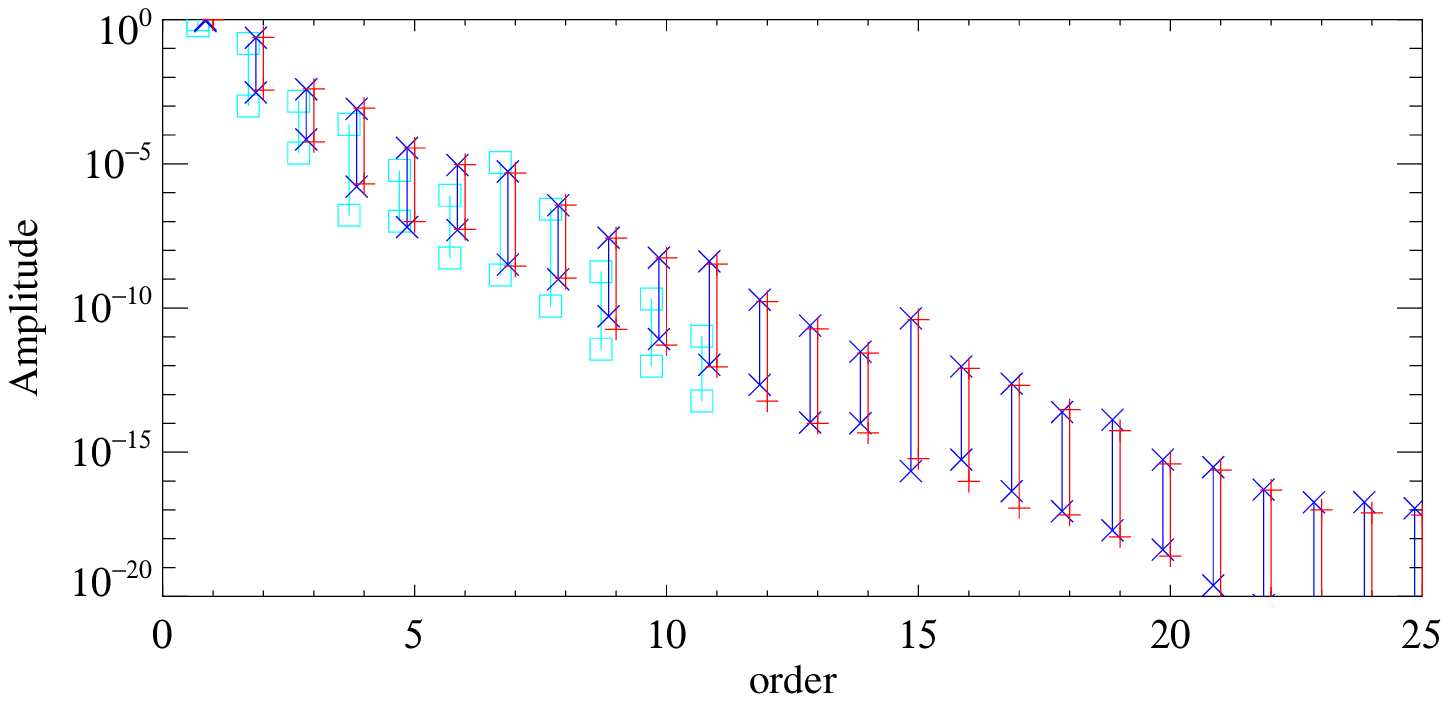}}\hfil}
\vspace{1ex}
\hbox to \hsize{\hfil (b)\hfil}
\vspace{0.5ex}
\hbox to \hsize{\hfil
  \mbox{\includegraphics[width=0.8\hsize]{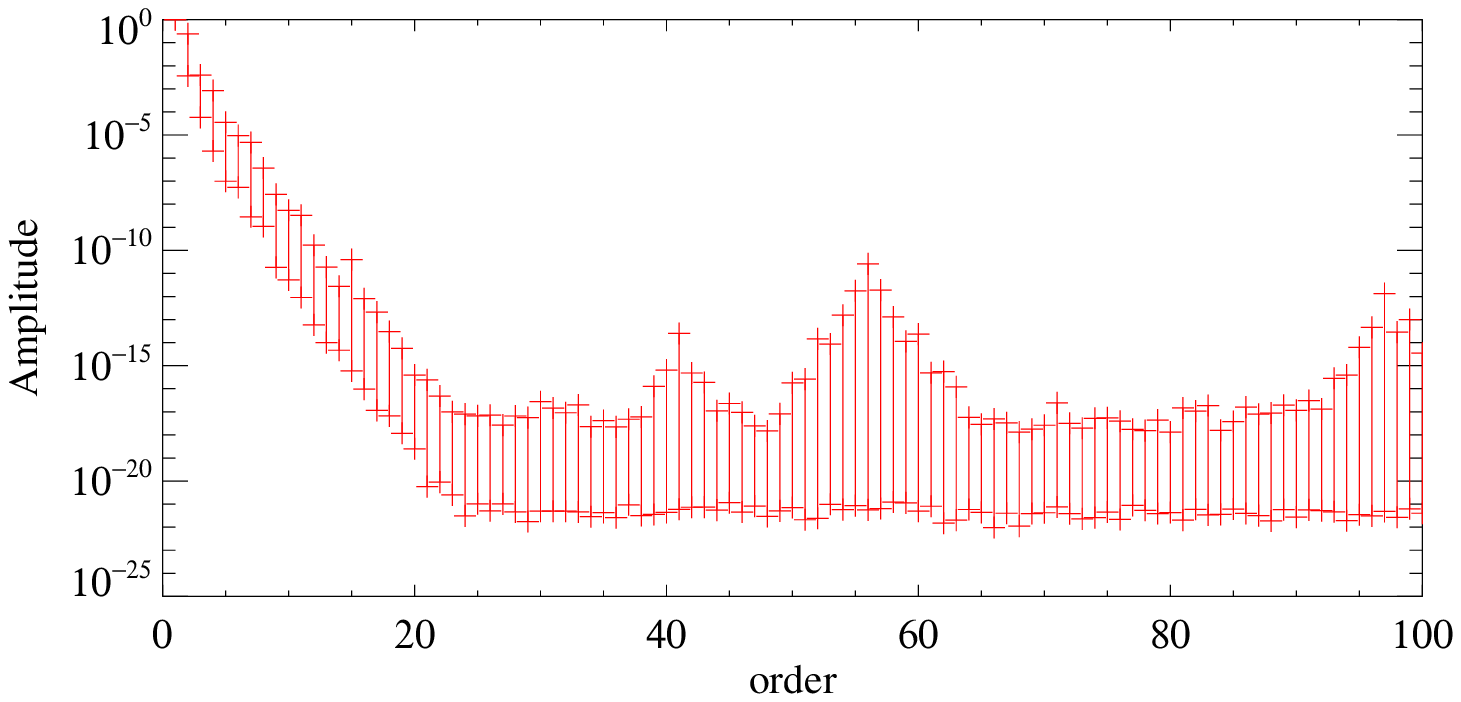}}\hfil}
\caption{Fourier spectra of $8\times8$ (cyan), $30\times30$ (blue) and
$112\times112$ (red) approximate 12-fold quasipatterns, as a function of
order, (a)~up to 25th order, (b)~up to 100th order ($112\times112$ only). The
vertical lines indicate the range of amplitudes of modes at that order, scaled
to the maximum amplitude at order~1.}
\label{fig:QPFourierSpectraOrder}
\end{figure}

In figure~\ref{fig:QPFourierSpectraOrder}, we show the range of amplitudes of
the modes as a function of order, for the $8\times8$, $30\times30$ and
$112\times112$  approximate quasipatterns. We note that the $8\times8$ example
deviates significantly from the other two at all orders, while the $30\times30$
and $112\times112$ examples are in fairly good agreement until 19th order or
so, another indication of how the approximation improves with the domain size.
We note that direct comparisons between the $8\times8$ case and the other two
cases are not strictly valid, as the domain size in the
$8\times8$ case had to be adjusted slightly to allow for the different
wavenumbers in the pattern, as discussed above.

Figure~\ref{fig:QPFourierSpectraOrder}(b) shows the spectrum of the
$112\times112$ example at all orders up to 100. The range of amplitudes at a
given order decays exponentially with order up to about 23rd order, but then
levels off at the level of the round-off error: about $10^{-17}$. On top of
this pattern, there is a peak in amplitude at 15th order, and broad peaks
around 41st, 56th and 97th order. We attribute these to the presence of weakly
damped modes, with $\bfk$ close to unity, that appear around these orders. With
weak damping, these modes will amplify the numerical noise in the PDE
solutions, and nonlinear coupling implies that the amplified noise will feed in
to nearby modes.

\begin{figure}
\hbox to \hsize{\hfil
  \mbox{\includegraphics[width=0.8\hsize]{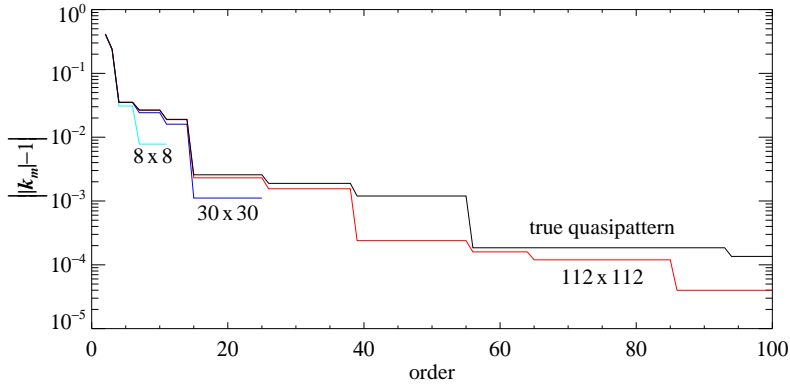}}\hfil}
\caption{Minimum value of $\big||\bfkm|-1\big|$, as a function of
order~$|\bfm|$, for $8\times8$ (cyan), $30\times30$ (blue),
$112\times112$ (red) approximate 12-fold quasipatterns, and for a true 12-fold
quasipattern (black)~\cite{Rucklidge2003}. There are drops in this minimum
quantity at 4th, 7th, 11th, 15th, 26th, 39th and 56th order in all cases. Modes
that are responsible for these drops are identified in
figures~\ref{fig:QPFourierSpectra} and~\ref{fig:QPFourierSpectraDetail}.
In the $30\times30$ case, the Pythagorean mode $(0.6,0.8)$ (which is of unit
length) is generated at 26th order.}
 \label{fig:QPFourierSpectraClosest}
 \end{figure}

This is shown in figure~\ref{fig:QPFourierSpectraClosest}, where we plot
$\big||\bfkm|-1\big|$ as a function of order~$|\bfm|$ for the different
examples. We note the marked drop in this quantity in particular at 15th and
39th order. While the drops in $\big||\bfkm|-1\big|$ do not line up exactly
with the peaks in amplitude, we suspect that it is this marked change in the
damping of modes appearing at these orders that is responsible for the
amplification of noise at around the same order.

Of course, for the numerical patterns in periodic domains, there is a lower
bound to $\big||\bfkm|-1\big|$ that does not depend on~$|\bfm|$ (this lower 
bound occurs within figure~\ref{fig:QPFourierSpectraClosest}), while for the
true 12-fold quasipattern, there is no lower bound~\cite{Rucklidge2003}. If
$|\bfkm|\neq1$, then
 \begin{equation}
 \big||\bfkm|-1\big| > \frac{K}{|\bfm|^2},
 \end{equation}
where $K$~is an order-one constant. For some particular
choices of wavevectors, this asymptotic limit is achieved: for example, if
 \begin{equation}\label{eq:vectors12}
 \bfkm = p \bfkfou + (q-1)\bfknin + (q+1)\bfkele = (1,p-\sqrt{3}q),
 \end{equation}
where $p$ and $q$ are integers, then $|\bfm|=p + 2q$ and
$|\bfkm|^2=1+(p-\sqrt{3}q)^2$. When $\frac{p}{q}$ is a
continued fraction
approximation to~$\sqrt{3}$ (table~\ref{table:sqrtthree}), then
$\big||\bfkm|-1\big|\sim\frac{1}{2}(p-\sqrt{3}q)^2\leq\frac{K_2}{q^2}$, so
$\big||\bfkm|-1\big|\leq\frac{K'}{|\bfm|^2}$, where $K_2$ and $K'$ are order-one
constants~\cite{Rucklidge2003}. These particular vectors are not always the
closest ones that can be found at a given order, but they demonstrate that
modes approach the unit circle arbitrarily closely (and so are arbitrarily
weakly damped) as the order of the mode increases. For the fractions listed in
table~\ref{table:sqrtthree}, we have $|\bfm|=p+2q=4$, $11$, $15$, $41$, $56$,
$153$, \dots, but in fact there are step decreases in the minimum of
$\big||\bfkm|-1\big|$ at $|\bfm|=4$, $7$, $11$, $15$, $26$, $39$, $56$, $94$. This is
the reason for the choices of orders in figures~\ref{fig:QPFourierSpectra}
and~\ref{fig:QPFourierSpectraDetail}.

In summary, having identified which sizes result in the most accurate
approximations to 12-fold quasipatterns, we find that in the largest example
($112\times112$), the locations of the Fourier modes of the approximation
deviate significantly from those of the true quasipattern only beyond 26th
order. At this point, the amplitudes of the modes have reached the level of
numerical round-off, so going any larger than $112\times112$ would not lead
to any significant improvement in the approximation to a true quasipattern for
these parameter values. The small divisors reveal themselves by amplifying the
amplitudes of the Fourier modes at (or close to) the order at which the small
divisor appears, but they do not appear to cause the amplitudes of the Fourier
modes to become excessively large, at least at this level of forcing, and up to 
the maximum order (153) available in this domain.


\section{Turbulent crystals: quasipatterns using $1:2$ resonance}
\label{sec:Numericsturbulentcrytals}

\begin{figure}
\hbox to \hsize{\hfil
  \mbox{\includegraphics[width=0.99\hsize]{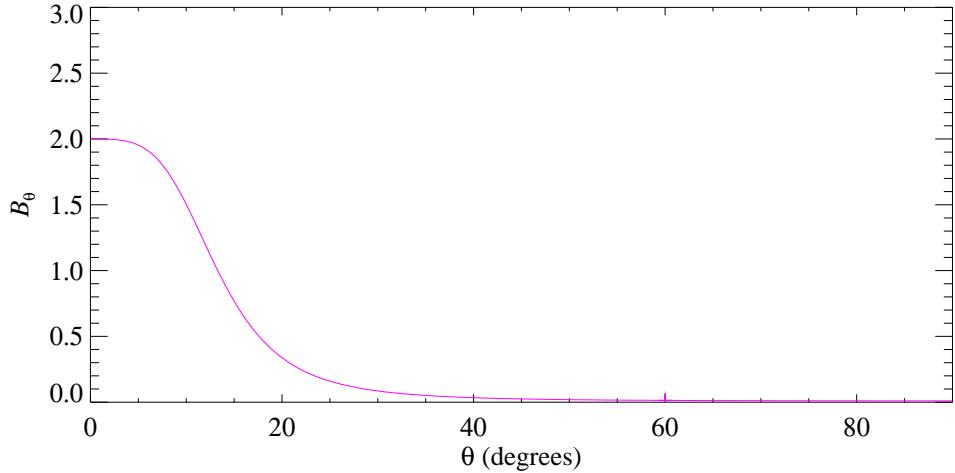}}
\hfil}
\caption{Cross-coupling coefficient $B_{\theta}$ for single frequency forcing,
with $1:2$ resonance in space and time: $\omega=\frac{1}{3}$,
$\beta=-\frac{1}{6}$, $\delta=0$, $\mu=-0.005$, $\alpha=0.001$, $\gamma=0$,
$Q_1=3+4i$, $Q_2=-6+8i$, $C=-1+10i$, $F_c=0.024002$ and $k_c=0.9999$.
Note that $B_{\theta}$
drops
away sharply as $\theta$ increases, and is close to zero for
$\theta\geq30^\circ$. The relevant coefficients are
$B_{30}=0.088$, $B_{60}=0.014$ and $B_{90}=0.010$.}
\label{fig:BthetaNP}
\end{figure}

In order to have $1:2$ resonance in space and time with single frequency
forcing ($m=1$), we impose ${\hat\Omega}(1)=\frac{1}{2}$ and
${\hat\Omega}(2)=1$, which leads to $\omega=\frac{1}{3}+4\delta$ and
$\beta=-\frac{1}{6}+5\delta$. We choose $\delta=0$, small values for the
damping coefficients~$\mu$, $\alpha$ and~$\gamma$, and order~one values for the
nonlinear coefficients close to the Hamiltonian limit. The resulting
cross-coupling curve is shown in figure~\ref{fig:BthetaNP}: as explained in
section~\ref{sec:Background}, the $1:2$ resonance in space and time has
enhanced the self-coupling coefficient (by about four orders of magnitude
compared to the previous cases), and so the cross-coupling coefficient
$B_{\theta}$ drops away sharply, and is close to zero for $\theta\geq30^\circ$.

\begin{figure}
\hbox to \hsize{\hfil
  \mbox{\includegraphics[width=0.99\hsize]{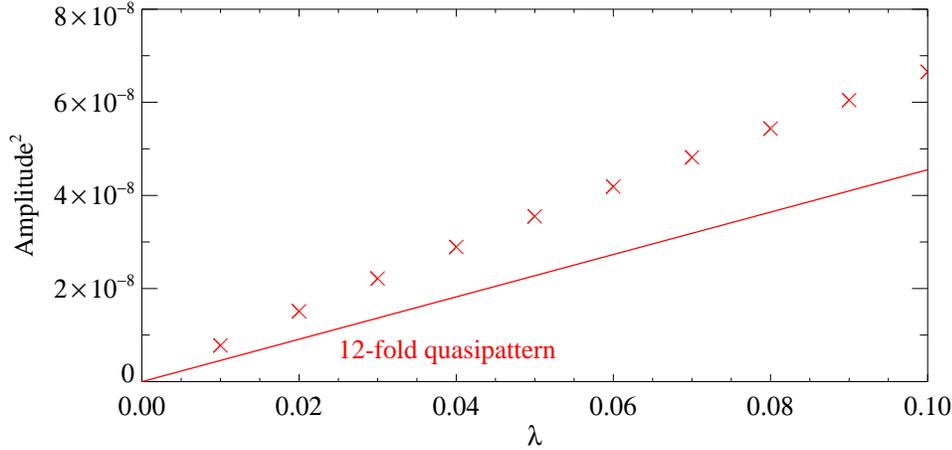}}
\hfil}
\caption{Bifurcation diagram showing the weakly nonlinear
predicted amplitude (solid line) and amplitudes of approximate quasipattern
solutions of the PDEs in an $8\times8$ domain (crosses).}
 \label{fig:NPbifn}
 \end{figure}

\begin{figure}
\hbox to \hsize{\hfil
 \hbox to 0.45\hsize{\hfil (a)\hfil}\hfil
 \hbox to 0.45\hsize{\hfil (b)\hfil}\hfil}
\vspace{0.5ex}
\hbox to \hsize{\hfil
  \mbox{\includegraphics[width=0.45\hsize]{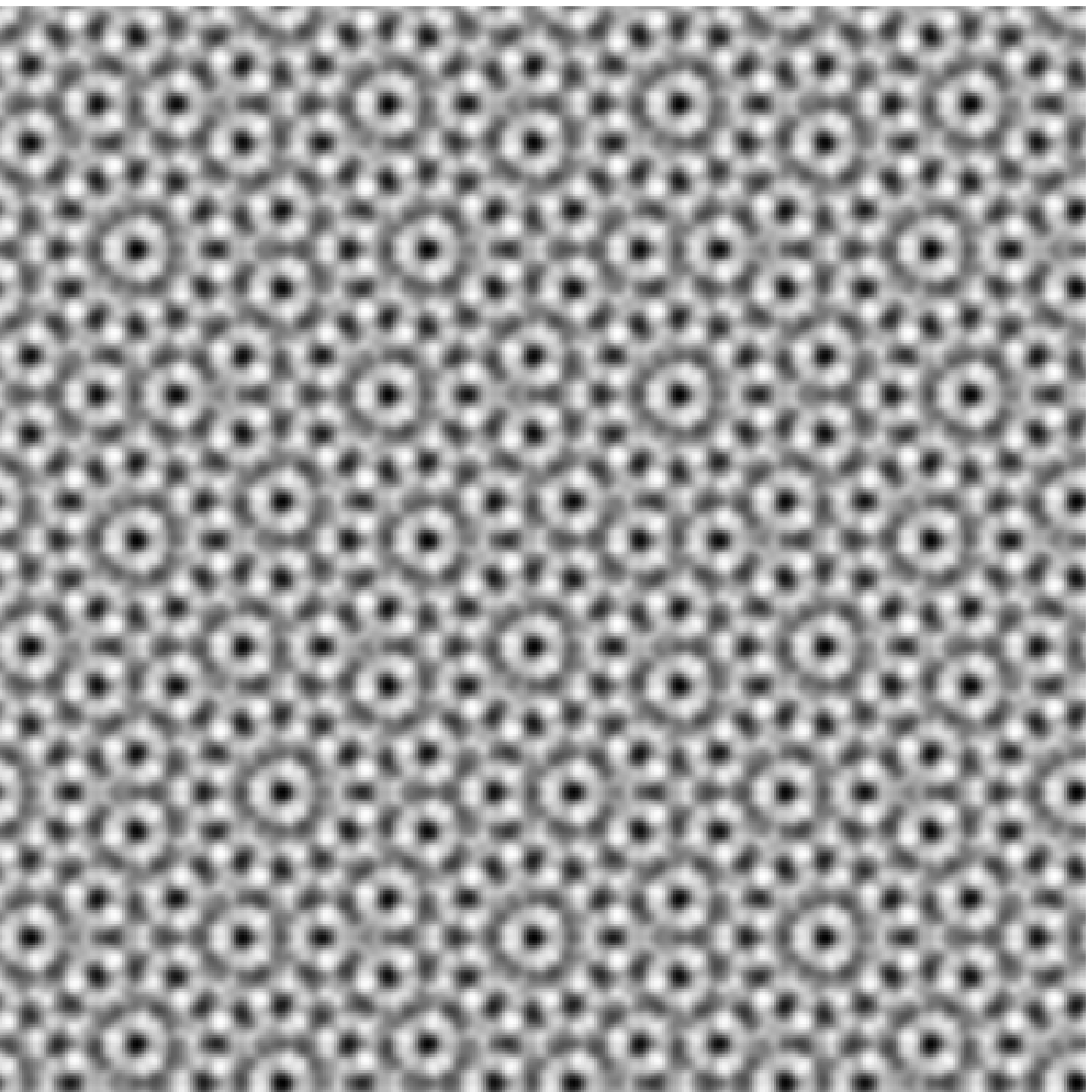}}\hfil
  \mbox{\includegraphics[width=0.45\hsize]{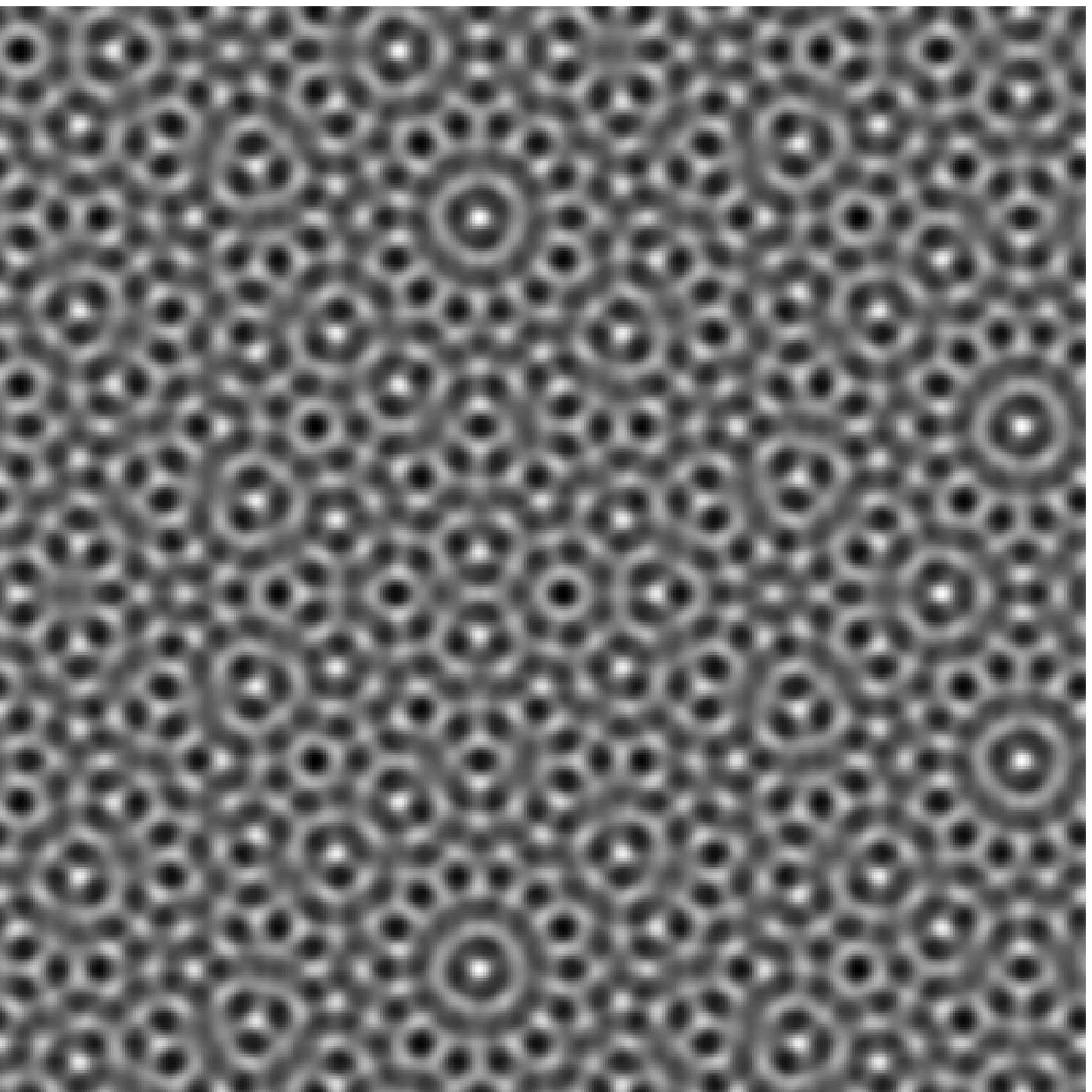}}\hfil}
\vspace{1ex}
\hbox to \hsize{\hfil (c)\hfil}
\vspace{0.5ex}
\hbox to \hsize{\hfil
   \mbox{\includegraphics[width=0.95\hsize]{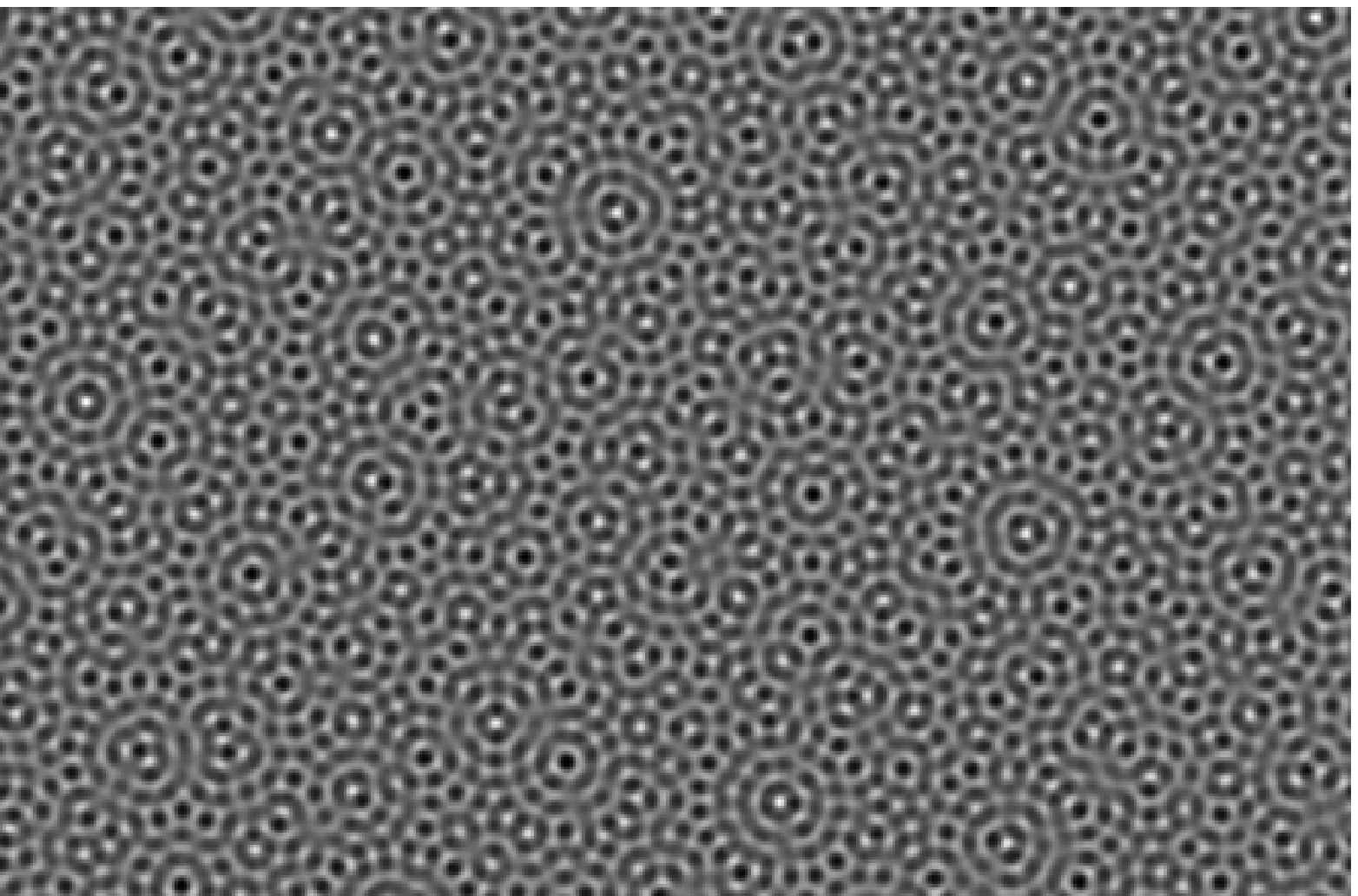}}\hfil}
\caption{For the parameter values from figure~\ref{fig:BthetaNP}, we find three
different approximate quasipatterns depending on the amplitude of the forcing
and the size of the domain.
 (a) 1.1 times critical, $30\times30$ domain: a 12-fold quasipattern.
 (b) 1.3 times critical, $30\times30$ domain: a 14-fold quasipattern.
 (c) 1.3 times critical, $60\times60$ domain: a 20-fold quasipattern
(only $\frac{2}{3}$ of the domain is shown). An animation of the transition
from (a) to (b), also showing details of the Fourier spectrum, can be found
online.}
 \label{fig:NPexamples}
 \end{figure}

\begin{figure}
\hbox to \hsize{\hfil
  \mbox{\includegraphics[width=0.99\hsize]{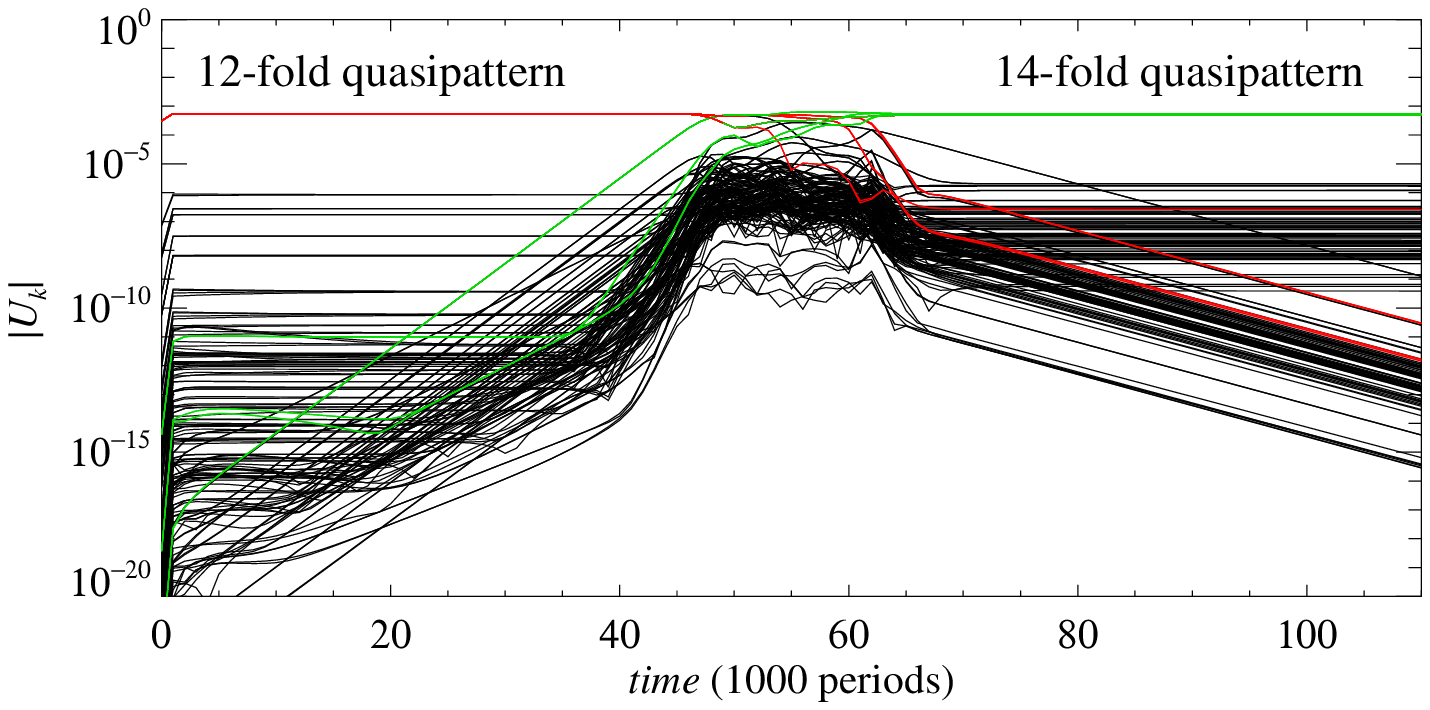}}
\hfil}
\caption{Amplitudes of Fourier modes as a function of time, at 1.3 times
critical in a $30\times30$ domain. The initial condition was the 12-fold
quasipattern from 1.1 times critical (red). This is unstable and, after an
extended transient of $70\,000$ periods, it is replaced by the 14-fold
quasipattern (green). Amplitudes of other Fourier modes close to $k=1$ are
shown in black. An animation of this transition can be found online.}
 \label{fig:NPquasitransition}
 \end{figure}

Within the restrictions of a 12-mode expansion, 12-fold
quasipatterns are stable (figure~\ref{fig:NPbifn}). Indeed, at $1.1$
times the critical forcing, in a $30\times30$ domain, the numerical solution of
the PDE with random initial conditions is a stable 12-fold approximate
quasipattern (figure~\ref{fig:NPexamples}a). As above, the primary modes that
make up the pattern are $(30,0)$ and $(26,15)$ and their reflections, in units
of basic lattice vectors. The amplitudes of the 12~modes differ by $0.5\%$. The
initial condition was not in any invariant subspace, and the PDE was integrated
for $160\,000$ periods of the forcing. The agreement between the weakly
nonlinear predictions and the computed amplitudes is not good
(figure~\ref{fig:NPbifn}), which we expect since the $k=2$ mode is weakly
damped (see discussion above).

However, there is no feature in the cross-coupling curve
(figure~\ref{fig:BthetaNP}) to indicate that modes at~$30^\circ$ should enjoy a
special status. When the forcing is changed to $1.3$~times critical with this
12-fold quasipattern as the initial condition, we find that it is unstable, and
is replaced (after a transient of $70\,000$ periods) by a stable approximate
14-fold quasipattern (figures~\ref{fig:NPexamples}b
and~\ref{fig:NPquasitransition}, and animation online). In this case, the 14
modes are $(30,0)$, $(27,13)$, $(19,23)$, $(7,29)$ and their reflections,
differing in length by $0.5\%$ and having angles within $1.5^\circ$ of
$360^\circ/14$. The amplitudes differ by about $10\%$. Both the 12-fold and
14-fold quasipatterns are stable at 1.2~times critical, in a $30\times30$
domain.

\begin{figure}
\hbox to \hsize{\hfil
  \mbox{\includegraphics[width=0.99\hsize]{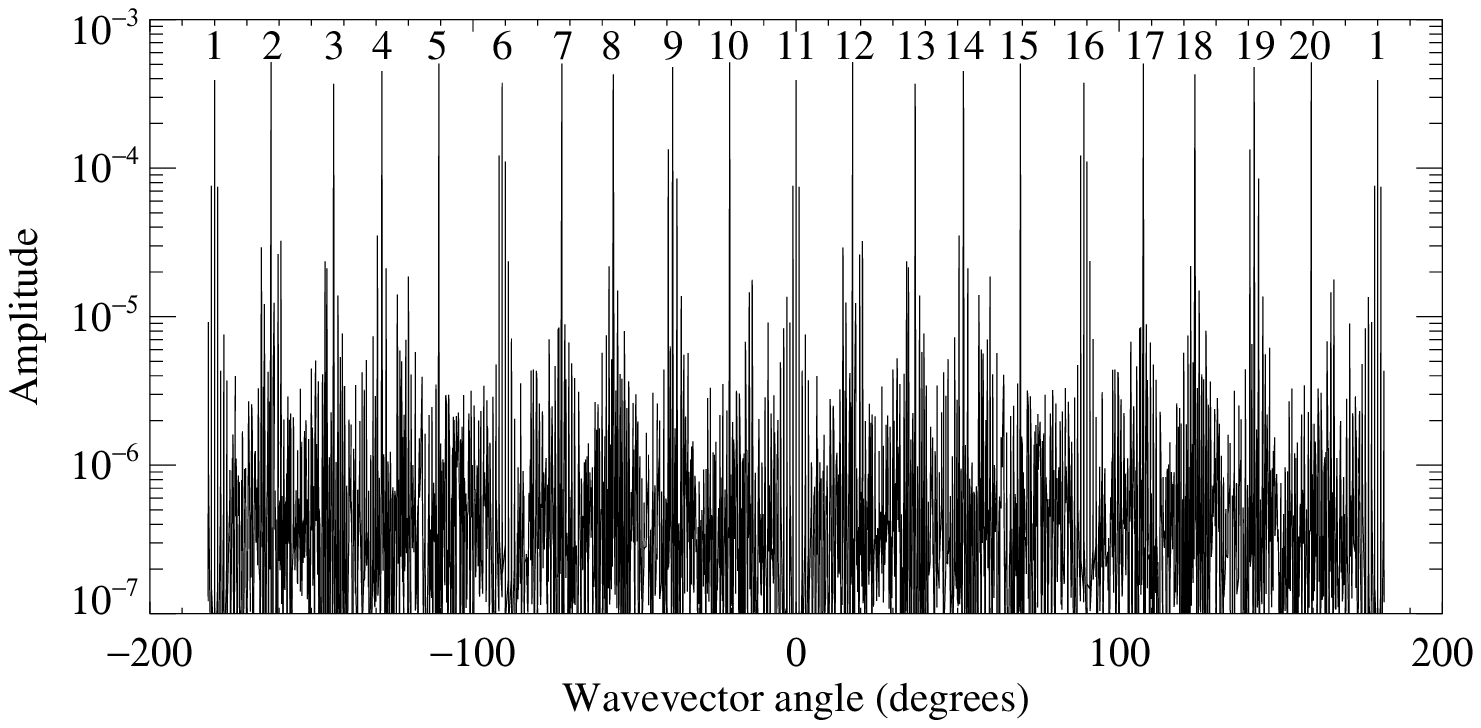}}
\hfil}
\caption{Amplitudes of Fourier modes of the 20-fold quasipattern from
figure~\ref{fig:NPexamples}(c), as a function of wavevector orientation,
for $0.95\leq|\bfk|\leq1.05$, showing twenty peaks roughly evenly distributed.}
 \label{fig:NPstructurefunction}
 \end{figure}

More complex quasipatterns are also possible:
calculations done in larger $60\times60$ and $90\times90$ domains at 1.3~times
critical, starting with random initial conditions, both yield an approximate  20-fold
quasipattern (figure~\ref{fig:NPexamples}c). However, this is not a particularly
accurate approximation to a 20-fold quasipattern: in the $60\times60$ case,
the 20~modes are
 $(60,0)$,
 $(57,18)$,
 $(48,36)$,
 $(37,47)$,
 $(21,56)$,
 $(1,60)$,
 $(-18,57)$,
 $(-33,50)$,
 $(-47,37)$,
 $(-56,21)$ and their $180^\circ$ rotations, which differ in length by $0.4\%$
and which have angles within $2^\circ$ of $360^\circ/20$. The amplitudes of the
20 modes differ by up to $40\%$. However, figure~\ref{fig:NPstructurefunction}
shows an examination of the Fourier spectrum of figure~\ref{fig:NPexamples}(c):
there are 20~peaks close to $k=1$, of similar amplitudes and arranged roughly
evenly around the unit circle. The $90\times90$ example is similar.
We speculate that in other domains and at other forcings, 16-fold and 18-fold
approximate quasipatterns could also be observed.

The cross-coupling $B_\theta$ curve suggests that modes that are more than
about $30^\circ$ apart do not influence each other (at least in an amplitude
equation truncated at cubic order). This suggests that patterns containing many
modes at essentially arbitrary angles might be expected -- such patterns have
been termed {\em turbulent crystals} by Newell and Pomeau~\cite{Newell1993}. It
is hard to see why one quasipattern should be favoured over another, on indeed,
why quasipatterns (with wavevectors evenly distributed around the $k=1$ circle)
should be favoured over more complex patterns.

We have not attempted to compare the locations of the high-order modes in these
14- and 20-fold approximate quasipatterns with those in the true quasipatterns,
as these higher quasilattices are very densely populated
(figure~\ref{fig:lattices}d), and the approximate solutions are not close
enough to the true quasipatterns.

As an aside, there is an interesting connection that can be made between the
Fourier spectra of these high-order quasipatterns and the fractal dynamics of
the complex ODE $d^2\zeta/dt^2=-\zeta^{n-1}$, where $n$~is an even
integer~\cite{Grinevich2007}. Solutions of this equation lie on multiply
branched Riemann surfaces, with branch points occurring densely at points in a
quasilattice of order~$n$.

The rotational degeneracy of the plane (in the absence of
boundary conditions) implies that any
mode with $|\bfk|=1$ is linearly excited, and it is an open question as to why
patterns and quasipatterns, with a finite number of modes evenly distributed
around the unit circle, should be the preferred patterns close to onset in so
many examples of pattern-forming systems. (Of course, pattern-forming systems
are by definition those that produce regular patterns close to onset!) Newell
and Pomeau~\cite{Newell1993} wrote down the evolution equation of $N$ modes,
ignoring quadratic terms and truncating at cubic order:
 \begin{equation}
 \frac{dA_j}{dt} = \lambda A_j - \sum_{k=1}^{N} B_{\theta_{jk}} |A_k|^2A_j,
 \end{equation}
where $A_j$ is the complex amplitude of mode~$j$, $\lambda$ is the growth rate,
and $\theta_{jk}$ is the angle between the wavevectors of modes $j$ and~$k$.
Truncated in this way, the phases of the amplitudes do not enter the dynamics:
Newell and Pomeau~\cite{Newell1993} attribute this to the translation symmetry
of the underlying problem, but Melbourne~\cite{Melbourne1998} points out that
this phase-invariance is in fact a normal form symmetry and hence not exact.

This system of ODEs is variational, and evolves to minimise a free energy
 \begin{equation}
 \label{eq:ZVLF}
 \mathcal{F} = -\lambda\sum_{j=1}^{N}|A_j|^2
               + \frac{1}{2}\sum_{j,k=1}^{N} B_{\theta_{jk}} |A_j|^2|A_k|^2.
 \end{equation}
Newell and Pomeau~\cite{Newell1993} claim that in the case that the state that
minimizes the free energy~$\mathcal{F}$ has many modes, with the magnitudes of
all amplitudes equal but with arbitrary phases, then the pattern will resemble
a spatially random field (a turbulent crystal). This case will be realised when
$B_{\theta}<1$ over a wide range of~$\theta$ (as in figure~\ref{fig:BthetaNP},
for example). We therefore propose the quasipattern solutions in
figure~\ref{fig:NPexamples} (12-, 14- and 20-fold quasipatterns) as examples of
turbulent crystals.

\begin{figure}
\hbox to \hsize{\hfil
  \mbox{\includegraphics[width=0.8\hsize]{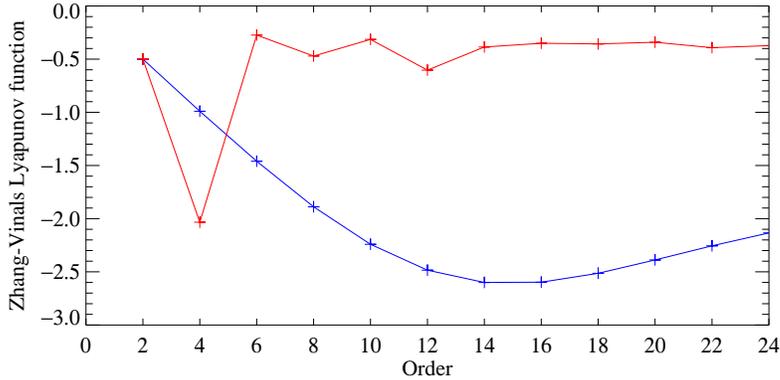}}\hfil}
\caption{Value of the free energy~$\mathcal{F}$ (\ref{eq:ZVLF}) for $N$-fold
patterns and quasipatterns, derived from the $B_\theta$ curves in
figure~\ref{fig:BthetaNP} with single-frequency forcing (blue) and
figure~\ref{fig:BthetaQP} with multi-frequency forcing (red).}
 \label{fig:ZVLF}
\end{figure}

It is interesting to consider what determines the number of modes in these
turbulent crystals. One argument, explored in more detail by Zhang and
Vi\~nals~\cite{Zhang1997}, is that the preferred pattern at onset should be the
global minimum of the free energy~$\mathcal{F}$. Assuming equal amplitudes, the
free energy of an $N$-fold pattern or quasipattern depends on $B_{\theta}$
evaluated at~$360^\circ/N$, $720^\circ/N$ {\em etc.} Figure~\ref{fig:ZVLF}
shows $\mathcal{F}$ derived from the $B_\theta$ curves in
figures~\ref{fig:BthetaNP} and \ref{fig:BthetaQP}, with single and
multi-frequency forcing respectively.

In the single-frequency case (blue curve in figure~\ref{fig:ZVLF}), where
$B_{\theta}$ is close to zero for $\theta\geq30^\circ$
(figure~\ref{fig:BthetaNP}), there is a broad minimum around 14-fold or 16-fold
quasipatterns. However, 12-fold, 14-fold and 20-fold examples were found
(figure~\ref{fig:NPexamples}), depending on the forcing strength and domain
size. No doubt other patterns could also be found with more exploration.

In the multi-frequency case (red curve in figure~\ref{fig:ZVLF}), where
$B_{\theta}$ has pronounced dips at $30^\circ$ and $90^\circ$
(figure~\ref{fig:BthetaQP}), there are local minima at $N=4$ (squares), $N=8$
and $N=12$. In numerical experiments, most initial conditions found squares,
though 12-fold quasipatterns were also stable (figures~\ref{fig:QPbifn}
and~\ref{fig:QPexamples}). We have not looked for 8-fold quasipatterns.

These results suggest that these free energy arguments provide a useful
qualitative tool for understanding pattern selection, but reality is often more
complicated than the arguments might suggest.


\section{Conclusions}
\label{sec:Conclusions}

We have introduced a new model PDE~(\ref{eq:pde}) for investigating pattern
formation and pattern design in parametrically forced systems. The PDE is
intended to play the same role for the Faraday wave experiment that the
Swift--Hohenberg equation~\cite{Swift1977} plays for convection: while the
model cannot be derived from the fluid mechanics, it has qualitatively correct
linear behaviour and the right type of nonlinear interactions in order to
provide useful illumination of the processes that are going on. The model
produces superlattice patterns (section~\ref{sec:Numericssuperlattice}) and
quasipatterns (sections~\ref{sec:Numericsquasipatterns}
and~\ref{sec:Numericsturbulentcrytals}) in response to single and
multi-frequency forcing, for the same reasons that these complex patterns are
found in the Faraday wave experiment -- confirming that the mechanisms are
generic. The ease of calculating weakly nonlinear coefficients and of computing
large-scale numerical solutions has allowed a quantitative exploration of the
agreement between the theoretical understanding of the pattern selection
mechanism and the patterns that are actually found.

Of course, the model does not capture every detail of the physics of the 
Faraday wave experiment. In particular, the dispersion relations have different
structures, and the model does not include the mean flow effects that are
important in nearly inviscid Faraday experiments~\cite{Knobloch2002}. The
latter could be addressed by coupling the model to a conserved quantity or to a
mean-flow equation (as in~\cite{Tsimring1997,Greenside1985}) or by taking the
negative Laplacian of the right-hand side of the PDE (as
in~\cite{Cox2003,Dawes2003}).

The Zhang--Vi\~nals~\cite{Zhang1996} equations do not have these drawbacks:
these are derived from the Navier--Stokes equations in the limit of infinite
depth and zero viscosity. One might ask what is gained by looking at a simpler
PDE that is even further from the physics. There are two advantages of the new
model: one is that it is very simple: the dispersion relation can be controlled
easily for studying any resonant interaction or response to multi-frequency
forcing; in addition, the weakly nonlinear theory can be computed very easily.
A second advantage is that is is very well suited to the use of efficient
numerical methods such as Exponential Time Differencing~\cite{Cox2002}: the
linear terms are diagonal in Fourier space, and the nonlinear terms do not
involve any derivatives. In contrast, the Zhang--Vi\~nals equations are
considerably more complicated and the weakly nonlinear computations are more
involved. Numerical solutions are also more time-consuming, as the linear term
is not diagonal in spectral space, and most of the nonlinear terms involve
products of derivatives, resulting in more Fourier transforms for their
evaluation. As a result of the relatively low cost of the calculations, we have
been able to follow branches of solutions in detail, and to go to much larger
domains and for much longer times than previous calculations. Of course, in the
end it would be desirable to work directly with the Navier--Stokes equations,
but for these, the weakly nonlinear theory is very
challenging~\cite{Skeldon2007} and there are as yet no large-scale numerical
simulations.

Like the Zhang--Vi\~nals equations, the new model includes explicit time
dependence. In contrast, other approaches, based on developing a description of
the slow evolution of the amplitude of an underlying pattern, use the
Ginzburg--Landau equation with additional complex conjugate terms to capture
the effect of the time-dependent
forcing~\cite{Coullet1992c,Conway2007,Conway2007a,Halloy2007}. As a result, any
complex patterns that are found must be interpreted in terms of slow,
long-wavelength amplitude modulations of an underlying pattern, which
complicates any effort to make quantitative comparison between theoretical
ideas and the behaviour of the real system.

The significance of three-wave coupling to weakly damped modes and its role in
pattern selection has long been
recognised~\cite{Mermin1985,Newell1993,Edwards1994,Zhang1997}. We have put this
idea to a quantitative test by using it to choose forcing functions that
stabilise a desired pattern in large domain calculations. However, the
codimension-one approach to this idea, where the weakly damped modes are slaved
to the pattern-forming modes, does not provide quantitative predictions of
amplitudes of patterns, and of parameter regimes where the desired patterns
should be stable, except for very close to onset. The reason is that
computation of the cross-coupling coefficient~$B_{\theta}$ is only valid when
all modes are strongly damped compared to the pattern-forming modes. This poses
difficulties because the most interesting patterns occur where the
pattern-forming modes are coupled to weakly damped modes, and this is where the
theory used to calculate properties of these patterns is of limited validity.
As a result, parameters had to be chosen very close to onset in order to find
stable numerical examples of the desired patterns in parameter regimes where
they were predicted to be stable. A codimension-two approach would extend the
range of validity of the theory, and will be the subject of future work.

We have investigated two mechanisms for the formation of quasipatterns. One
mechanism uses three-wave interactions involving a damped mode associated with
the difference of the two frequencies in the forcing to select a particular
angle ($30^\circ$~in the example presented here). Using different primary
frequencies, or altering the dispersion relation, would allow other angles, or
combinations of angles, to be selected. The advantage is that a forcing
function can be designed to produce a particular pattern: the mechanism is
quite selective, and requires some fine-tuning of the parameters.

The second mechanism uses $1:2$ resonance in space and time to magnify the
self-interaction coefficient and thereby, on rescaling, diminish the
cross-coupling coefficient~$B_{\theta}$ for angles greater than
about~$30^\circ$. This can lead to the formation of turbulent
crystals~\cite{Newell1993}. The mechanism is robust (the patterns are found
well above onset), and requires only single frequency forcing. A dispersion
relation that supports $1:2$ resonance in space and time is needed. Within this
framework, an inherent complication is that it is not clear why regular 8, 10,
12 or 14-fold quasipatterns, or indeed any other combination of modes, should
be favoured. Indeed we have found that 12-, 14- and 20-fold approximate
quasipatterns can be stabilised by altering the level of the forcing or the
domain size, without changing other parameters, and we have reported the
transition between two different types of quasipattern. The Lyapunov function
approach~\cite{Zhang1996} cannot make this distinction, and would predict that
14- or 16-fold quasipatterns should be found at onset for these parameter
values (figure~\ref{fig:ZVLF}). It remains an open question as to why one
turbulent crystal should be favoured over another.

The existence of 14-fold (and higher) quasipatterns has been suggested
before~\cite{Zhang1996,Rucklidge2003,Topaz2004,Steurer2007}, but we have
presented here the first examples of spontaneously formed 14-fold and 20-fold
approximate quasipatterns that are stable solution of a PDE
(figure~\ref{fig:NPexamples}), with preliminary results
in~\cite{Rucklidge2007}. Examples where 14-fold symmetry is imposed externally
have been reported in optical experiments~\cite{Pampaloni1995}. The Fourier
spectra of 12-fold and 14-fold quasipatterns are both dense
(figure~\ref{fig:lattices}c,d), but those of 14-fold quasipatterns are much
denser, owing to the difference between quadratic and cubic irrational
numbers~\cite{Rucklidge2003}. This difference may have profound consequences
for their mathematical treatment.

We have identified what domain sizes result in the most accurate approximations
to 12-fold quasipatterns, based on square and on hexagonal domains
(table~\ref{table:domains}), and produced exceptionally clean examples of
approximate quasipatterns in relatively large computational domains. Comparing
the Fourier spectra of the approximate quasipatterns as a function of domain
size, we have identified at what order the locations of Fourier modes in the
approximate quasipatterns deviate from those of the true quasipatterns. In the
largest example ($112\times112$), the locations of the Fourier modes deviate
significantly only beyond 26th order
(figure~\ref{fig:QPFourierSpectraDetail}a), at which point the amplitudes of
the modes have reached the level of numerical round-off
(figure~\ref{fig:QPFourierSpectraOrder}). This suggests that going any larger
than $112\times112$ would not lead to any significant improvement in the
approximation to a true quasipattern, at least for these parameter values.

We have compared the amplitudes of the Fourier modes of the approximate
quasipatterns and the leading order weakly nonlinear prediction
(figures~\ref{fig:QPbifn} and~\ref{fig:NPbifn}), and found quantitative
agreement very close to onset, but only qualitative agreement  at larger
amplitude, which is what would be expected from the problem of eliminating
weakly damped modes, as discussed above. We have not extended this comparison
to higher order since the weakly nonlinear calculations are too difficult for
this parametrically forced problem. An extension of this work would be to
devise a PDE without parametric forcing that also produces stable quasipattern
solutions: this would allow high order weakly nonlinear calculations (as
in~\cite{Rucklidge2003}) and very large domain numerical solutions, and so allow
a comparison between computed mode amplitudes (as a function of order) and the
weakly nonlinear theory. Standard weakly nonlinear theory produces amplitudes
that diverge at high order because of the presence of small
divisors~\cite{Rucklidge2003}, while the PDE solutions have amplitudes that
decay exponentially with order -- although the small divisors in this problem
do make themselves felt by amplifying the magnitudes of the Fourier modes at (or
close to) the order at which the small divisor appears
(figure~\ref{fig:QPFourierSpectraOrder}). Such a PDE could be based on (for
example) the Swift--Hohenberg
equation~\cite{Frisch1995,Lifshitz1997,Muller1994}, but the Swift--Hohenberg
equation itself does not allow the weakly damped modes that are necessary to
stabilise quasipatterns.

Other numerical studies of quasipatterns as solutions of a PDE have not made a
systematic study of the effect of domain size. Zhang and
Vi\~nals~\cite{Zhang1996,Zhang1998} report approximate 8-fold quasipatterns in
a $64\times64$ domain in their quasipotential model of the nearly inviscid
Navier--Stokes equations, for parameter values close to the $1:2$ resonance in
space and time. The modes involved were separated by  $41^\circ$, $42^\circ$,
$48^\circ$ and $49^\circ$~\cite{Zhang1998}, so the approximation was not
particularly accurate; our careful choice of domain size allowed much closer
approximation. M\"uller~\cite{Muller1994} developed a model based on two
coupled Swift--Hohenberg equations, with parameters chosen so that the two
unstable modes had wavenumbers that would favour 8-fold or 12-fold
quasipatterns. Numerical simulations in a $10\times10$ domain in the second
case found approximate 12-fold quasipatterns. The modes involved are not
stated, but we estimate them to be $\bfkone=(10,-1)$, $\bfktwo=(9,4)$,
$\bfkthr=(6,8)$ in units of the fundamental lattice vector. These have lengths
$10.05$, $9.85$ and $10.00$ respectively, and they are separated by
$28.48^\circ$ and $29.17^\circ$. With $\bfkele=(4,-9)$, we have
$\bfkthr+\bfkele=\bfkone$, so the $60^\circ$~resonance condition is satisfied.
Frisch and Sonnino\cite{Frisch1995} present a similar model and report 10-fold
quasipatterns. Lifshitz and Petrich~\cite{Lifshitz1997} found a 12-fold
approximate quasipattern in a roughly $30\times30$ domain, in a model based on
a single Swift--Hohenberg equation with a degenerate double minimum in its
marginal stability curve. The modes involved appear to be the same as those in
the $30\times30$ examples discussed in section~\ref{sec:Numericsquasipatterns}.

While we have not discussed the possibility of long-wave instabilities of
quasipatterns, the Fourier spectra of the $112\times112$
example~(figures~\ref{fig:QPFourierSpectraDetail}(d)
and~\ref{fig:QPFourierSpectraLattice}) suggests that long-wave modes that are
close to the primary wavevectors in a direction {\em tangent} to the critical
circle are forced by high-order nonlinear interactions. This is also apparent
from the locations of modes responsible for the small
divisors~\cite{Rucklidge2003}. Therefore, any treatment of the long-wave
stability of quasipatterns should take into account the presence of these
modes. This is a delicate question. The only study of the sideband
instabilities of quasipatterns~\cite{Echebarria2001} focusses on instabilities
associated with modes that are perpendicular to the unit circle, using coupled
Ginzburg--Landau equations for each of the primary mode directions in the
quasipattern. This approach could be extended to include instabilities
associated with modes that are tangent to the unit circle by looking at coupled
Newell--Whitehead--Segel equations, along the lines suggested
by~\cite{Gunaratne1994}, although high-order nonlinear interaction may not be
captured in a long-wave analysis truncated at cubic order.


 \section*{Acknowledgments}
 We are grateful for support from National Science Foundation (DMS-0309667) and
from the Engineering and Physical Sciences Research Council (GR/S45928/01). We
are also grateful to many people who have helped shape these ideas: Jessica
Conway, St\'ephan Fauve, Jay Fineberg, Rebecca Hoyle, G\'erard Iooss, Edgar
Knobloch, Paul Matthews, Ian Melbourne, Werner Pesch, Michael Proctor, Jeff
Porter, Hermann Riecke, Anne Skeldon, Jorge Vi\~nals and Gene Wayne. We thank
Michael Proctor for pointing out the effect of higher order terms on the centre
manifold, discussed in section~\ref{sec:Numericssuperlattice}. Finally, we are
grateful to the Isaac Newton Institute for Mathematical Sciences, where part of
this work was carried out.


\begingroup

\def\url#1{}

\bibliographystyle{siam}
\bibliography{rs}

\endgroup


\appendix

\section{Weakly nonlinear theory}
\label{app:WNLT}

In this appendix, we present the weakly nonlinear theory for the
PDE~(\ref{eq:pde}). We will describe the calculation in terms of a harmonic
primary bifurcation; the subharmonic case is similar, with the main differences
being that the period is $4\pi$ rather then~$2\pi$, and that the quadratic
coefficient $Q$ is identically zero.

We start by writing $U=u+iv$, where $u(x,y,t)$ and $v(x,y,t)$ are real
functions, and so
 \begin{align*}
 \frac{\partial u}{\partial t}&=
   \left(\mu+\alpha\nabla^2+\gamma\nabla^4\right)u
  -\left(\omega+\beta\nabla^2+\delta\nabla^4\right)v\\
  &\qquad{}
  +Q_{1r}(u^2-v^2) - Q_{1i}(2uv) + Q_{2r}(u^2+v^2)
  +C_r(u^2+v^2)u - C_i(u^2+v^2)v\\
 \frac{\partial v}{\partial t}&=
   \left(\omega+\beta\nabla^2+\delta\nabla^4\right)u
  +\left(\mu+\alpha\nabla^2+\gamma\nabla^4\right)v\\
  &\qquad{}
  +Q_{1i}(u^2-v^2) + Q_{1r}(2uv) + Q_{2i}(u^2+v^2)
  +C_r(u^2+v^2)v + C_i(u^2+v^2)u\\
  &\qquad{}+f(t)u.
 \end{align*}
We define differential operators $\mathcal L$ and $\mathcal M$:
 \begin{equation*}
 {\mathcal L}=\frac{\partial}{\partial t}-\left(\mu+\alpha\nabla^2+\gamma\nabla^4\right)
 \qquad\hbox{and}\qquad
 {\mathcal M}=\left(\omega+\beta\nabla^2+\delta\nabla^4\right),
 \end{equation*}
so the PDEs for $u$ and $v$ are
 \begin{align*}
 {\mathcal L} u &=          - {\mathcal M}v + \hbox{NL}_u,\\
 {\mathcal L} v &= \phantom{-}{\mathcal M}u + \hbox{NL}_v + f(t)u.
 \end{align*}
The nonlinear terms $\hbox{NL}_u$ and $\hbox{NL}_v$ are:
 \begin{align*}
 \hbox{NL}_u &=
   Q_{1r}(u^2-v^2) - Q_{1i}(2uv) + Q_{2r}(u^2+v^2)
   +C_r(u^2+v^2)u - C_i(u^2+v^2)v,\\
 \hbox{NL}_v &=
   Q_{1i}(u^2-v^2) + Q_{1r}(2uv) + Q_{2i}(u^2+v^2)
   +C_r(u^2+v^2)v + C_i(u^2+v^2)u.
 \end{align*}

\subsection{Linear theory}

We seek solutions of the form $u=\eikdotx p_1(t)$ and $v=\eikdotx q_1(t)$, where $p_1$ and $q_1$ are
periodic functions of period~$T$, and define
 \begin{equation*}
 {\hat\gamma}_1=2\left(-\mu+\alpha k^2-\gamma k^4\right),
 \qquad
 {\hat\Omega}_1=\omega-\beta k^2+\delta k^4,
 \qquad
 \Omega_1=\sqrt{\left(\frac{{\hat\gamma}_1}{2}\right)^2 + \left({\hat\Omega}_1\right)^2},
 \end{equation*}
we get
 \begin{align*}
 {\mathcal L}_1 p_1 &=          - {\mathcal M}_1 q_1,\\
 {\mathcal L}_1 q_1 &= \phantom{-}{\mathcal M}_1 p_1 + f(t)p_1,
 \end{align*}
or
 \begin{equation*}
 {\mathcal L}_1^2 p_1 = - {\mathcal M}_1{\mathcal L}_1 q_1
                      = - {\mathcal M}_1^2 p_1 - f(t){\mathcal M}_1 p_1,
 \end{equation*}
where ${\mathcal L}_1$ and ${\mathcal M}_1$ act on $p_1(t)$ and $q_1(t)$ as
 \begin{equation*}
 {\mathcal L}_1=\frac{d}{dt}+\frac{{\hat\gamma}_1}{2}
 \qquad\hbox{and}\qquad
 {\mathcal M}_1={\hat\Omega}_1.
 \end{equation*}
The linearised PDE reduces to a damped Mathieu equation for $p_1$:
 \begin{equation*}
 \left(\frac{d}{dt}+\frac{{\hat\gamma}_1}{2}\right)^2 p_1 + {\hat\Omega}_1^2 p_1 + f(t){\hat\Omega}_1 p_1=0
 \end{equation*}
or
 \begin{equation*}
 {\ddot p}_1 + {\hat\gamma}_1 {\dot p}_1 + \left(\Omega_1^2 + {\hat\Omega}_1 f(t)\right) p_1=0
  ={\mathbf L}p_1.
 \end{equation*}
The adjoint equation is:
 \begin{equation*}
 {\ddot {\tilde p}}_1 - {\hat\gamma}_1{\dot {\tilde p}}_1 + \left(\Omega_1^2 + {\hat\Omega}_1f(t)\right){\tilde p}_1=0
  ={\mathbf{\tilde L}}{\tilde p}_1,
 \end{equation*}
with respect to an inner product
 \begin{equation*}
 \langle g,h\rangle=\frac{1}{T}\int_0^T\, g(t)h(t)\,dt,
 \end{equation*}
with $T=2\pi$ (harmonic case) or $T=4\pi$ (subharmonic case),
so $\langle g,{\mathbf L}h\rangle= \langle {\mathbf{\tilde L}}g,h\rangle$

For a given value of~$k$, seeking periodic solutions of ${\mathbf L}p=0$ yields
an eigenvalue problem whose eigenvalue is the amplitude of the forcing function
$f(t)$. We use the method of~\cite{Besson1996} to solve this eigenvalue problem
with multi-frequency forcing~$f(t)$, providing the critical forcing amplitude.
Minimising this critical forcing amplitude over $k$ yields the critical
wavenumber $k_c$, critical forcing function $f_c(t)$, and critical
eigenfunction $p_1(t)$. The corresponding $q_1(t)$ is determined by solving
${\mathcal L}_1q_1={\hat\Omega}p_1+f_c(t)p_1$.

\subsection{Rhombs}

We consider $f$ close to $f_c$,
writing $f(t)=f_c(t) (1+\epsilon^2F_2)$,
and seek small-amplitude rhombic solutions associated with  two wavevectors
$\bfkone$ and $\bfktwo$ at the critical wavenumber:
$k_1=k_2=k_c$, separated by an angle~$\theta$. We formally expand the solution as
 \begin{align*}
 u&=\epsilon u_1 + \epsilon^2 u_2 + \epsilon^3 u_3 + \cdots\\
{}&=\epsilon\left(
           z_1(T_2)\eikonedotx + z_2(T_2)\eiktwodotx + \cc
           \right)p_1(t)\\
  &\quad{}
  + \epsilon^2\Big(
        \left(z_1^2\etwoikonedotx + z_2^2\etwoiktwodotx + \cc\right)p_2(t)
      + \left(|z_1|^2 + |z_2|^2\right)p_3(t)\\
  &\quad\qquad{}
      + \left(z_1z_2\eikonepktwodotx + \cc\right)p_4(t)
      + \left(z_1{\bar z}_2\eikonemktwodotx + \cc\right)p_5(t)
           \Big)
  + {\mathcal O}(\epsilon^3),
 \end{align*}
with a similar expression for $v$ in terms of $v_1$, $v_2$ and $v_3$, and
$q_1$, ..., $q_5$, where $T_2$ is a slow time, varying on a scale of
$\epsilon^{-2}$, and the functions $p_2(t)$, ..., $q_5(t)$ are to be
determined. The form of this solution is chosen by knowing in advance the
structure of the nonlinear terms and the modes to be generated by them.

Substituting these expressions for $u$ and $v$ into the PDE and ordering
in powers of~$\epsilon$, we recover,
at leading order in~$\epsilon$,  the linear theory.
At second order in~$\epsilon$, we split the PDE into terms that go as
$\etwoikonedotx$, $\etwoiktwodotx$, terms without spatial dependence,
and terms that go as $\eikonepktwodotx$ and $\eikonemktwodotx$.

The terms like $\etwoikonedotx$ lead to equations for~$p_2$ and~$q_2$:
 \begin{align*}
 {\mathcal L}_2 p_2 &=          - {\mathcal M}_2 q_2 + \hbox{NL}^{(2)}_{p2},\\
 {\mathcal L}_2 q_2 &= \phantom{-}{\mathcal M}_2 p_2 + \hbox{NL}^{(2)}_{q2} + f_c(t)p_2,
 \end{align*}
where the linear operators are:
 \begin{align*}
 {\mathcal L}_2&=\frac{d}{dt}-\left(\mu-4\alpha k_c^2+16\gamma k_c^4\right)
                =\frac{d}{dt}+\frac{{\hat\gamma}_2}{2},\\
 {\mathcal M}_2&=\left(\omega-4\beta k_c^2+16\delta k_c^4\right)={\hat\Omega}_2,
 \end{align*}
and the nonlinear terms
$\hbox{NL}^{(2)}_{p2}$ and $\hbox{NL}^{(2)}_{q2}$ are:
 \begin{align*}
 \hbox{NL}^{(2)}_{p2} &=
   Q_{1r}(p_1^2-q_1^2) - Q_{1i}(2p_1q_1) + Q_{2r}(p_1^2+q_1^2),\\
 \hbox{NL}^{(2)}_{q2} &=
   Q_{1i}(p_1^2-q_1^2) + Q_{1r}(2p_1q_1) + Q_{2i}(p_1^2+q_1^2).
 \end{align*}
The function $q_2$ is eliminated by operating with ${\mathcal L}_2$, resulting
in a second-order non-constant coefficient inhomogeneous linear ODE for~$p_2$:
 \begin{equation*}
 \left({\mathcal L}_2^2 + {\mathcal M}_2^2 + {\mathcal M}_2f(t)\right)p_2 =
   {\mathcal L}_2 \hbox{NL}^{(2)}_{p2} - {\mathcal M}_2 \hbox{NL}^{(2)}_{q2}.
 \end{equation*}
or
 \begin{equation*}
 {\ddot p_2} + {\hat\gamma}_2 {\dot p_2} + \left(\Omega_2^2 + {\hat\Omega}_2f_c(t)\right)p_2 =
   \left(\frac{d}{dt} + \frac{{\hat\gamma}_2}{2}\right)\hbox{NL}^{(2)}_{p2} - {\hat\Omega}_2 \hbox{NL}^{(2)}_{q2}.
 \end{equation*}
This can be solved numerically for $p_2$ using Fourier transform methods, and
$q_2$ can then be found. Terms that go as  $\etwoiktwodotx$ result in the same
equation.

Terms without spatial dependence, and terms with spatial dependence $\eikonepktwodotx$ and $\eikonemktwodotx$,
result in similar equations for $p_3$, $p_4$ and $p_5$, but with linear operators:
 \begin{align*}
 {\mathcal L}_3&=\frac{d}{dt}- \mu = \frac{d}{dt} + \frac{{\hat\gamma}_3}{2},
 \qquad\hbox{and}\qquad
 {\mathcal M}_3=\omega={\hat\Omega}_3,\\
 {\mathcal L}_4&=\frac{d}{dt}-\left(\mu-4\cos^2\left(\frac{\theta}{2}\right)\alpha k_c^2+16\cos^4\left(\frac{\theta}{2}\right)\gamma k_c^4\right)
                = \frac{d}{dt} + \frac{{\hat\gamma}_4}{2},\\
 {\mathcal M}_4&=\left(\omega-4\cos^2\left(\frac{\theta}{2}\right)\beta k_c^2+16\cos^4\left(\frac{\theta}{2}\right)\delta k_c^4\right)
                = {\hat\Omega}_4,\\
 {\mathcal L}_5&=\frac{d}{dt}-\left(\mu-4\sin^2\left(\frac{\theta}{2}\right)\alpha k_c^2+16\sin^4\left(\frac{\theta}{2}\right)\gamma k_c^4\right)
                = \frac{d}{dt} + \frac{{\hat\gamma}_5}{2},\\
 {\mathcal M}_5&=\left(\omega-4\sin^2\left(\frac{\theta}{2}\right)\beta k_c^2+16\sin^4\left(\frac{\theta}{2}\right)\delta k_c^4\right)
                = {\hat\Omega}_5,
 \end{align*}
and nonlinear terms:
 \begin{align*}
 \hbox{NL}^{(2)}_{p3} &=  \hbox{NL}^{(2)}_{p4} =  \hbox{NL}^{(2)}_{p5} = 2\hbox{NL}^{(2)}_{p2},\\
 \hbox{NL}^{(2)}_{q3} &=  \hbox{NL}^{(2)}_{q4} =  \hbox{NL}^{(2)}_{q5} = 2\hbox{NL}^{(2)}_{q2}.
 \end{align*}
Note that ${\hat\gamma}_4$,  ${\hat\gamma}_5$, ${\hat\Omega}_4$ and ${\hat\Omega}_5$ depend on~$\theta$, the angle
between the chosen wavevectors, whereas ${\hat\gamma}_1$, ${\hat\gamma}_2$, ${\hat\gamma}_3$, ${\hat\Omega}_1$,
${\hat\Omega}_2$ and ${\hat\Omega}_3$ do not.

At third order in~$\epsilon$, the problem has the following structure:
 \begin{align*}
 {\mathcal L} u_3 + \frac{\partial u_1}{\partial T_2} &=          - {\mathcal M}v_3 + \hbox{NL}^{(3)}_u,\\
 {\mathcal L} v_3 + \frac{\partial v_1}{\partial T_2} &= \phantom{-}{\mathcal M}u_3 + \hbox{NL}^{(3)}_v + f_c(t)u_3 + F_2f_c(t)u_1.
 \end{align*}
The nonlinear terms $\hbox{NL}^{(3)}_u$ and $\hbox{NL}^{(3)}_v$ are:
 \begin{align*}
 \hbox{NL}^{(3)}_u &=
   2Q_{1r}(u_1 u_2 - v_1 v_2) - Q_{1i}(2u_1v_2+2u_2v_1) + 2Q_{2r}(u_1u_2+v_1v_2)\\
  &\qquad{}+C_r(u_1^2+v_1^2)u_1 - C_i(u_1^2+v_1^2)v_1,\\
 \hbox{NL}^{(3)}_v &=
   2Q_{1i}(u_1 u_2 - v_1 v_2) + Q_{1r}(2u_1v_2+2u_2v_1) + 2Q_{2i}(u_1u_2+v_1v_2)\\
  &\qquad{}+C_r(u_1^2+v_1^2)v_1 + C_i(u_1^2+v_1^2)u_1.
 \end{align*}
Eliminating~$v_3$, we obtain
\begin{equation*}
\left({\mathcal L}^2+{\mathcal M}^2+{\mathcal M}f_c\right) u_3 =
  -{\mathcal L}\frac{\partial u_1}{\partial T_2}
  +{\mathcal M}\frac{\partial v_1}{\partial T_2}
  -{\mathcal M}F_2f_cu_1
  +{\mathcal L}\hbox{NL}^{(3)}_u - {\mathcal M}\hbox{NL}^{(3)}_v.
\end{equation*}
The operator on the left is the singular operator from the linearised problem,
so the equation can only be solved for $u_3$ if a solvability condition is
applied to the terms that are proportional to $\eikonedotx$ and $\eiktwodotx$
and complex conjugates. If we take the inner product between ${\tilde p}_1$ and
the $\eikonedotx$ component of the above, we find \begin{equation*}
 \tau\frac{\partial z_1}{\partial T_2} =
        \sigma F_2 z_1
        + \left({\hat A} |z_1|^2 + ({\hat B}_{\hbox{indep}}+{\hat B}_{\hbox{res}}(\theta))|z_2|^2\right)z_1,
\end{equation*}
with a similar equation for $z_2$, where
\begin{equation*}
 \tau=\left\langle {\tilde p}_1,
      2 \left({\dot p}_1 + \frac{{\hat\gamma}_1}{2}p_1\right)
      \right\rangle
\qquad\hbox{and}\qquad
 \sigma=\left\langle {\tilde p}_1,
      -{\hat\Omega}_1 f_c p_1
      \right\rangle
\end{equation*}
(using ${\mathcal L}_1p_1=-{\mathcal M}_1q_1$) and
 \begin{align*}
 {\hat A} &= \Big\langle {\tilde p}_1,
        \Big(\frac{d}{dt}+\frac{{\hat\gamma}_1}{2}\Big)
        \Big(2Q_{1r}(p_1 p_2 + p_1 p_3 - q_1 q_2 - q_1 q_3) \\
   & \qquad\qquad\qquad\qquad
         {} - 2Q_{1i}(p_1 q_2 + p_1 q_3 + q_1 p_2 + q_1 p_3) \\
   & \qquad\qquad\qquad\qquad
         {} + 2Q_{2r}(p_1 p_2 + p_1 p_3 + q_1 q_2 + q_1 q_3) \\
   & \qquad\qquad\qquad\qquad
         {} + 3C_r(p_1^2+q_1^2)p_1
            - 3C_i(p_1^2+q_1^2)q_1\Big)\\
    & \qquad {} - {\hat\Omega}_1
        \Big(2Q_{1r}(p_1 q_2 + p_1 q_3 + q_1 p_2 + q_1 p_3) \\
   & \qquad\qquad\qquad
         {} + 2Q_{1i}(p_1 p_2 + p_1 p_3 - q_1 q_2 - q_1 q_3) \\
   & \qquad\qquad\qquad
         {} + 2Q_{2i}(p_1 p_2 + p_1 p_3 + q_1 q_2 + q_1 q_3) \\
   & \qquad\qquad\qquad
         {} + 3C_r(p_1^2+q_1^2)q_1
            + 3C_i(p_1^2+q_1^2)p_1\Big)
              \Big\rangle, \\
{\hat B}_{\hbox{indep}} &= \Big\langle {\tilde p}_1,
        \Big(\frac{d}{dt}+\frac{{\hat\gamma}_1}{2}\Big)
        \Big(2Q_{1r}(p_1 p_3 - q_1 q_3)
            - 2Q_{1i}(p_1 q_3 + q_1 p_3) \\
   & \qquad\qquad\qquad\qquad
         {} + 2Q_{2r}(p_1 p_3 + q_1 q_3)
            + 6C_r(p_1^2+q_1^2)p_1
            - 6C_i(p_1^2+q_1^2)q_1\Big)\\
    & \qquad {} - {\hat\Omega}_1
        \Big(2Q_{1r}(p_1 q_3 + q_1 p_3)
            + 2Q_{1i}(p_1 p_3 - q_1 q_3)
            + 2Q_{2i}(p_1 p_3 + q_1 q_3) \\
   & \qquad\qquad\qquad
         {} + 6C_r(p_1^2+q_1^2)q_1
            + 6C_i(p_1^2+q_1^2)p_1\Big)
      \Big\rangle,\\
{\hat B}_{\hbox{res}}(\theta) &= \Big\langle {\tilde p}_1,
        \Big(\frac{d}{dt}+\frac{{\hat\gamma}_1}{2}\Big)
        \Big(2Q_{1r}(p_1 p_4 + p_1 p_5 - q_1 q_4 - q_1 q_5) \\
   & \qquad\qquad\qquad\qquad
         {} - 2Q_{1i}(p_1 q_4 + p_1 q_5 + q_1 p_4 + q_1 p_5) \\
   & \qquad\qquad\qquad\qquad
         {} + 2Q_{2r}(p_1 p_4 + p_1 p_5 + q_1 q_4 + q_1 q_5)\Big)\\
    & \qquad {} - {\hat\Omega}_1
        \Big(2Q_{1r}(p_1 q_4 + p_1 q_5 + q_1 p_4 + q_1 p_5) \\
   & \qquad\qquad\qquad
         {} + 2Q_{1i}(p_1 p_4 + p_1 p_5 - q_1 q_4 - q_1 q_5) \\
   & \qquad\qquad\qquad
         {} + 2Q_{2i}(p_1 p_4 + p_1 p_5 + q_1 q_4 + q_1 q_5)\Big)
      \Big\rangle.
 \end{align*}
For convenience, we have separated the parts of the cross-coupling coefficient
that do not depend on the angle between the modes (${\hat B}_{\hbox{indep}}$)
from those that do (${\hat B}_{\hbox{res}}(\theta)$). We discuss below how
these coefficients are then scaled.

\subsection{Hexagons}

As with rhombs, we look for $f$ close to $f_c$, writing $f(t)=f_c(t)
(1+\epsilon^2F_2)$, but now we chose three wavevectors, $\bfkone$, $\bfktwo$
and $\bfkthr$ oriented at $120^\circ$ to each other, with
$\bfkone+\bfktwo+\bfkthr=\mathbf{0}$. We look for small-amplitude solutions
with equal amplitudes of the three waves, and write
 \begin{align*}
 u&=\epsilon u_1 + \epsilon^2 u_2 + \epsilon^3 u_3 + \cdots\\
{}&=\epsilon z(T_1,T_2)\left(
           \eikonedotx + \eiktwodotx + \eikthrdotx + \cc
           \right)p_1(t)\\
  &\quad{}
  + \epsilon^2\Big(
        z^2\left(\etwoikonedotx + \etwoiktwodotx + \etwoikthrdotx + \cc\right)p_2(t)
      + 3|z|^2 p_3(t)\\
  &\quad\qquad{}
      + |z|^2 \left(\eikonemktwodotx +\eiktwomkthrdotx +\eikthrmkonedotx + \cc\right){\tilde p}_5(t)\\
  &\quad\qquad{}
      + {\bar z}^2 \left(
           \eikonedotx + \eiktwodotx + \eikthrdotx + \cc
           \right)p_6(t)
           \Big)
  + {\mathcal O}(\epsilon^3),
 \end{align*}
with a similar expression for $v$ in terms of $q_1$, ..., $q_6$, where $T_1$
and $T_2$ are slow times, varying on scales $\epsilon^{-1}$ and
$\epsilon^{-2}$, and the functions $p_2(t)$, ..., $q_6(t)$ are to be
determined. The form of the expression is chosen by knowing in advance the
structure of the nonlinear terms.

Substituting these expressions for $u$ and $v$ into the PDE and ordering in
powers of~$\epsilon$, at leading order in~$\epsilon$ we recover the linear
theory. At second order in~$\epsilon$, we split the PDE into terms that go as
$\etwoikonedotx+\etwoiktwodotx+\etwoikthrdotx+\cc$, terms without spatial
dependence, terms that go as $\eikonemktwodotx +\eiktwomkthrdotx
+\eikthrmkonedotx + \cc$, and finally terms that go as $\eikonedotx +
\eiktwodotx + \eikthrdotx + \cc$, which have to be considered specially.

Terms like $\etwoikonedotx+\etwoiktwodotx+\etwoikthrdotx+\cc$ and terms without
spatial dependence give the same equations for $p_2$, $q_2$, $p_3$ and $q_3$ as
in the case of rhombs. In particular, the inhomogeneous nonlinear terms are the
same. Terms that go as  $\eikonemktwodotx +\eiktwomkthrdotx +\eikthrmkonedotx +
\cc$ result in equations for ${\tilde p}_5$ and ${\tilde q}_5$ that are the
equations for $p_5$ and $q_5$ evaluated for $\theta=120^\circ$.

Terms that go as  $\eikonedotx + \eiktwodotx + \eikthrdotx + \cc$ require the
use of two time-scales and a solvability condition. The linear operators are
the same as for the initial linear problem:
 \begin{align*}
 {\mathcal L}_1 p_6 &=          - {\mathcal M}_1 q_6 + \hbox{NL}^{(2)}_{p6} \phantom{{}+ f_c(t)p_6}
- \frac{\partial z/\partial T_1}{{\bar z}^2} p_1,\\
 {\mathcal L}_1 q_6 &= \phantom{-}{\mathcal M}_1 p_6 + \hbox{NL}^{(2)}_{q6} + f_c(t)p_6
- \frac{\partial z/\partial T_1}{{\bar z}^2} q_1.
 \end{align*}
The nonlinear terms are
$\hbox{NL}^{(2)}_{p6}$ and $\hbox{NL}^{(2)}_{q6}$ are:
 \begin{equation*}
 \hbox{NL}^{(2)}_{p6} = 2\hbox{NL}^{(2)}_{p2}
 \qquad\hbox{and}\qquad
 \hbox{NL}^{(2)}_{q6} = 2\hbox{NL}^{(2)}_{q2}.
 \end{equation*}
This can be reduced to a
second-order non-constant coefficient inhomogeneous linear ODE for~$p_6$:
 \begin{equation*}
 \left({\mathcal L}_1^2 + {\mathcal M}_1^2 + {\mathcal M}_1f(t)\right)p_6 =
   {\mathcal L}_1 \hbox{NL}^{(2)}_{p6} - {\mathcal M}_1 \hbox{NL}^{(2)}_{q6}
   +  \frac{\partial z/\partial T_1}{{\bar z}^2}
      \left({\mathcal M}_1 q_1 - {\mathcal L}_1 p_1\right)
 \end{equation*}
Since the operator on the LHS is the singular linear operator~${\mathbf L}$, we
must apply a solvability condition:
 \begin{equation*}
 \langle {\tilde p}_1,{\mathbf L}p_6\rangle = 0 =
 \langle {\tilde p}_1,
         {\mathcal L}_1 \hbox{NL}^{(2)}_{p6} - {\mathcal M}_1
         \hbox{NL}^{(2)}_{q6} \rangle
 - 2 \frac{\partial z/\partial T_1}{{\bar z}^2}
     \langle {\tilde p}_1,
             {\mathcal L}_1 p_1
     \rangle
 \end{equation*}
since ${\mathcal M}_1 q_1=-{\mathcal L}_1 p_1$. We define
 \begin{equation*}
 \tau=\left\langle{\tilde p}_1,
                    2\left(\frac{dp_1}{dt}+\frac{{\hat\gamma}_1}{2}p_1\right)
      \right\rangle
 \end{equation*}
as before and
 \begin{equation*}
 {\hat\epsilon}=\left\langle{\tilde p}_1,
                    {\mathcal L}_1 \hbox{NL}^{(2)}_{p6} -
                    {\mathcal M}_1 \hbox{NL}^{(2)}_{q6}
                \right\rangle,
 \end{equation*}
and obtain an equation for the slow ($T_1$) evolution of the amplitude~$z$:
 \begin{equation*}
 \tau \frac{\partial z}{\partial T_1} = {\hat\epsilon} {\bar z}^2.
 \end{equation*}
Once the solvability condition has been imposed, the ODE can be solved for $p_6$ and
$q_6$. The computed solution~$p_6$ contains an arbitrary amount of~$p_1$; the solution
is made unique by specifying that $\langle{\tilde p}_1,p_6\rangle=0$.

At third order in~$\epsilon$, the problem has the following structure:
 \begin{align*}
 {\mathcal L} u_3 + \frac{\partial u_2}{\partial T_1} + \frac{\partial u_1}{\partial T_2} &=
                 - {\mathcal M}v_3 + \hbox{NL}^{(3h)}_u,\\
 {\mathcal L} v_3 + \frac{\partial v_2}{\partial T_1} + \frac{\partial v_1}{\partial T_2} &=
        \phantom{-}{\mathcal M}u_3 + \hbox{NL}^{(3h)}_v + f_c(t)u_3 + F_2f_c(t)u_1.
 \end{align*}
We only need keep track of terms proportional to $\eikonedotx$ in our
derivation of the bifurcation problem, so the $\partial u_2/\partial T_1$ and
$\partial v_2/\partial T_1$ terms yield $p_6\partial {\bar z}^2/\partial T_1$
and $q_6\partial {\bar z}^2/\partial T_1$. The $\eikonedotx$ components of the
nonlinear terms $\hbox{NL}^{(3h)}_u$ and $\hbox{NL}^{(3h)}_v$ are specified
below.

Eliminating~$v_3$, we obtain
\begin{align*}
\left({\mathcal L}^2+{\mathcal M}^2+{\mathcal M}f_c\right) u_3 &=
  -{\mathcal L}\left(\frac{\partial u_2}{\partial T_1} + \frac{\partial u_1}{\partial T_2}
                     - \hbox{NL}^{(3h)}_u \right)\\
  &\qquad{}
  +{\mathcal M}\left(\frac{\partial v_2}{\partial T_1} + \frac{\partial v_1}{\partial T_2}
  - F_2f_cu_1 - \hbox{NL}^{(3h)}_v \right).
\end{align*}
The equation can only be solved for $u_3$ if a solvability condition
is applied to the terms that are proportional to $\eikonedotx$,
$\eiktwodotx$ and $\eikthrdotx$, and complex conjugates. If we take the inner product
between ${\tilde p}_1$ and the
$\eikonedotx$ component of the above, we find
\begin{equation*}
 \tau\frac{\partial z}{\partial T_2} =
        \sigma F_2 z
        + \left({\hat A}+2{\hat B}_{60}\right) |z|^2 z,
\end{equation*}
where $\tau$ and $\sigma$ are unchanged from the rhombic calculations, and
 \begin{align*}
 {\hat A}+2{\hat B}_{60} &= \Big\langle {\tilde p}_1,
        \Big(\frac{d}{dt}+\frac{{\hat\gamma}_1}{2}\Big)\hbox{NL}^{(3h)}_u -
        {\hat\Omega}_1 \hbox{NL}^{(3h)}_v\\
       & \qquad\qquad {}+\left(-{\mathcal L}_1 p_6 + {\mathcal M}_1 q_6\right)2\frac{{\hat\epsilon}}{\tau}
   \Big\rangle,
\end{align*}
where we have used
\begin{equation*}
 \frac{\partial{\bar z}^2}{\partial T_1}  = 2{\bar z}\frac{\partial{\bar z}}{\partial T_1}
                                          = 2\frac{{\hat\epsilon}}{\tau} |z|^2 z
\end{equation*}
and
 \begin{align*}
 \hbox{NL}^{(3h)}_u &=
   2Q_{1r}(p_1p_2 + 3p_1p_3 + 2p_1{\tilde p}_5 + 2p_1p_6
         - q_1q_2 - 3q_1q_3 - 2q_1{\tilde q}_5 - 2q_1q_6)\\
& \qquad
{} - 2Q_{1i}(p_1q_2 + 3p_1q_3 + 2p_1{\tilde q}_5 + 2p_1q_6
         + q_1p_2 + 3q_1p_3 + 2q_1{\tilde p}_5 + 2q_1p_6)\\
& \qquad
{} + 2Q_{1r}(p_1p_2 + 3p_1p_3 + 2p_1{\tilde p}_5 + 2p_1p_6
         + q_1q_2 + 3q_1q_3 + 2q_1{\tilde q}_5 + 2q_1q_6)\\
& \qquad
{} + 15C_r(p_1^2+q_1^2)p_1 - 15C_i(p_1^2+q_1^2)q_1,\\
 \hbox{NL}^{(3h)}_v &=
   2Q_{1r}(p_1q_2 + 3p_1q_3 + 2p_1{\tilde q}_5 + 2p_1q_6
         + q_1p_2 + 3q_1p_3 + 2q_1{\tilde p}_5 + 2q_1p_6)\\
& \qquad
{} + 2Q_{1i}(p_1p_2 + 3p_1p_3 + 2p_1{\tilde p}_5 + 2p_1p_6
         - q_1q_2 - 3q_1q_3 - 2q_1{\tilde q}_5 - 2q_1q_6)\\
& \qquad
{} + 2Q_{2i}(p_1p_2 + 3p_1p_3 + 2p_1{\tilde p}_5 + 2p_1p_6
         + q_1q_2 + 3q_1q_3 + 2q_1{\tilde q}_5 + 2q_1q_6)\\
& \qquad
{} + 15C_r(p_1^2+q_1^2)q_1 + C_i(p_1^2+q_1^2)p_1.
 \end{align*}
From the value of $\hat A$ calculated for rhombs, we can recover ${\hat
B}_{60}$, which is effectively the cross-coupling coefficient for modes with
wavevectors at~$60^\circ$.

\subsection{Reconstitution}

At this stage, the pattern formation problem on a hexagonal lattice would take
the form:
 \begin{align*}
 \tau\frac{\partial z_1}{\partial T_1} &= {\hat \epsilon} {\bar z}_2 {\bar z}_3,\\
 \tau\frac{\partial z_1}{\partial T_2} &=
        \sigma F_2 z_1
        + \left({\hat A} |z_1|^2 + {\hat B}_{60}|z_2|^2+{\hat B}_{60}|z_3|^2\right)z_1,
 \end{align*}
where $z_1$, $z_2$ and $z_3$ are amplitudes of $\eikonedotx$, $\eiktwodotx$ and
$\eikthrdotx$. Similar equations are found for $\partial z_2/\partial T_1$ {\em
etc.} Recall that the small factor~$\epsilon$ has been used so that the
amplitude of the original amplitude~$U$ is explicitly small: $U=\epsilon
z_1(T_1,T_2)\eikonedotx p_1(t)+\dots$

There is more than one way to combine these equations into a single ODE.
Properly, we should consider only the case where the coefficient of the
quadratic term $\hat\epsilon$ is itself small (order~$\epsilon$). This occurs
either for small values of $Q_1$ and~$Q_2$, or
near a codimension-one line in $(Q_1,Q_2)$ space:
 \begin{align*}
 0 &=  {\hat\epsilon}=\left\langle{\tilde p}_1,
                          {\mathcal L}_1 \hbox{NL}^{(2)}_{p6} -
                          {\mathcal M}_1 \hbox{NL}^{(2)}_{q6}\right\rangle\\
   &= 2Q_{1r}\left\langle{\tilde p}_1,
                          {\mathcal L}_1 (p_1^2-q_1^2) -
                          {\mathcal M}_1 (2p_1q_1)\right\rangle +
      2Q_{1i}\left\langle{\tilde p}_1,
                          {\mathcal L}_1 (-2p_1q_1) -
                          {\mathcal M}_1 (p_1^2-q_1^2)\right\rangle \\
   &\qquad\qquad {}+  2Q_{2r}\left\langle{\tilde p}_1,
                          {\mathcal L}_1 (p_1^2+q_1^2)\right\rangle +
      2Q_{2i}\left\langle{\tilde p}_1,
                        - {\mathcal M}_1 (p_1^2+q_1^2)\right\rangle.
 \end{align*}
Alternatively, we note that for many of these multi-frequency forced problems,
${\hat\epsilon}$ is small anyway~\cite{Porter2002,Porter2004a}.

Having decided that ${\hat\epsilon}$ is small,
we define a fast time scale~$t^*$, related to the original time scale~$t$
by an order-one factor $\sigma/\tau$:
 \begin{equation*}
 \frac{d}{dt^*}=\epsilon\frac{\tau}{\sigma} \frac{\partial}{\partial T_1} +
               \epsilon^2\frac{\tau}{\sigma} \frac{\partial}{\partial T_2},
 \end{equation*}
where $\epsilon$ is the original small parameter. We now scale the $z$'s by
$1/\epsilon$, so that the original amplitude~$U$ is implicitly small:
$U=z_1(T_1,T_2)\eikonedotx p_1(t)+\dots$, and obtain:
 \begin{equation*}
 \frac{dz_1}{dt^*} = \lambda z_1 + \frac{{\hat \epsilon}}{\sigma} {\bar z}_2 {\bar z}_3
                 + \frac{{\hat A}}{\sigma}\left(|z_1|^2 + B_{60}|z_2|^2+B_{60}|z_3|^2\right)z_1,
 \end{equation*}
where $\lambda=\epsilon^2 F_2$, so that the forcing amplitude is $(1+\lambda)$
times the critical amplitude, and $B_{60}={\hat B}_{60}/{\hat A}$.

The advantage of reconstituting in this way is that the quadratic and cubic
terms appear at the same order. The disadvantage is that the regime of validity
($\lambda\ll1$, ${\hat\epsilon}\ll1$, $z\ll1$) is not made explicit. In
particular, this validity condition will only be satisfied for hexagons when
the coefficient of the quadratic term is small -- which is precisely the limit
required for the quadratic and cubic terms to be of the same order.

Finally, we scale the amplitudes once more by a factor of $\sqrt{|\sigma/{\hat
A}|}$, rename the time variable back to~$t$, and obtain:
 \begin{equation*}
 \frac{dz_1}{dt} = \lambda z_1 + Q {\bar z}_2 {\bar z}_3
                   + s \left(|z_1|^2 + B_{60}|z_2|^2+B_{60}|z_3|^2\right)z_1,
 \end{equation*}
with similar equations for $z_2$ and $z_3$,
where
 \begin{equation*}
 Q =\frac{{\hat \epsilon}}{\sigma}
    \sqrt{\left|\frac{\sigma}{{\hat A}}\right|}
\qquad\hbox{and}\qquad
s=\hbox{sgn}\left(\frac{{\hat A}}{\sigma}\right)
 \end{equation*}
(usually, $s=-1$).

Repeating the same reconstitution for the rhombic lattice results in
 \begin{align*}
 \frac{dz_1}{dt} &= \lambda z_1
                    + s \left(|z_1|^2 + B_{\theta}|z_2|^2\right)z_1\\
 \frac{dz_2}{dt} &= \lambda z_2
                    + s \left(|z_2|^2 + B_{\theta}|z_1|^2\right)z_2
 \end{align*}
where $B_{\theta}=({\hat B}_{\hbox{indep}} + {\hat B}_{\hbox{res}}(\theta))/{\hat A}$.

The equations for other lattices can be found from combinations of the above,
as can candidate equations for quasipatterns.

\end{document}